\documentclass[%
 reprint,
 superscriptaddress,
nofootinbib,
 amsmath,amssymb,
 aps,
floatfix,
]{revtex4-2}

\usepackage{graphicx}% Include figure files
\usepackage{subfigure}
\usepackage{dcolumn}% Align table columns on decimal point
\usepackage{bm}% bold math
\usepackage[utf8]{inputenc} % allow utf-8 input
\usepackage[T1]{fontenc}    % use 8-bit T1 fonts
\usepackage{hyperref}       % hyperlinks
\usepackage{url}            % simple URL typesetting
\usepackage{booktabs}       % professional-quality tables
\usepackage{amsfonts}       % blackboard math symbols
\usepackage{nicefrac}       % compact symbols for 1/2, etc.
\usepackage{microtype}      % microtypography
\usepackage{xcolor}         % colors
\usepackage{amsmath}
\usepackage{xspace}
\usepackage{comment}
\usepackage[capitalise]{cleveref}
\crefname{appendix}{Appendix}{Appendices}
\Crefname{appendix}{Appendix}{Appendices}
\usepackage{amssymb}

\usepackage[printonlyused]{acronym}
\newacro{GW}{gravitational wave}
\newacro{CP}{conformal prediction}
\newacro{SNR}{signal-to-noise ratio}
\newacro{FAR}{false alarm rate}
\newacro{IFAR}{inverse false alarm rate}
\newacro{TP}{true positive}
\newacro{FP}{false positive}
\newacro{FN}{false negative}
\newacro{TN}{true negative}
\newacro{TPR}{true positive rate}
\newacro{FPR}{false positive rate}
\newacro{ML}{machine learning}
\newacro{CNN}{convolutional neural network}
\newacro{ROC}{receiver operating characteristic}
\newacro{AUC}{area under the curve}
\newacro{MLP}{multi-layer perceptron}
\newacro{NN}{neural network}
\newacro{KNN}{K-nearest neighbours}
\newacro{LR}{logistic regression}
\newacro{BNS}{binary neutron star}
\newacro{BBH}{binary black hole}
\newacro{NSBH}{neutron star black hole}
\newacro{CBC}{compact binary coalescence}
\newacro{LVK}{LIGO-Virgo-KAGRA}
\newacro{SHAP}{SHapley Additive exPlanations}
\newacro{MDC}{mock data challenge}
\newacro{LLPIC}{low-latency production integration challenge}
\newacro{GWTC}{gravitational wave transient catalogue}

\newcommand{\pastro}{\ensuremath{p_{\rm astro}}\xspace}
\newcommand{\maxpastro}{maximum-\ensuremath{p_{\rm astro}}\xspace}
\newcommand{\maxifar}{maximum-$\log_{10}$IFAR\xspace}
\newcommand{\logIFAR}{$\log_{10}({\rm IFAR})$\xspace}

\newcommand{\sklearn}{\texttt{scikit-learn}\xspace}

\newcommand{\pycbc}{\texttt{PyCBC}\xspace}
\newcommand{\gstlal}{\texttt{GstLAL}\xspace}
\newcommand{\spiir}{\texttt{SPIIR}\xspace}
\newcommand{\mbta}{\texttt{MBTA}\xspace}
\newcommand{\cwb}{\texttt{cWB}\xspace}

\newcommand{\bilby}{\texttt{Bilby}\xspace}

\newcommand{\gwtcTwo}{GWTC-2.0\xspace}
\newcommand{\gwtcTwoPointOne}{GWTC-2.1\xspace}
\newcommand{\gwtcThree}{GWTC-3.0\xspace} % this bugs if writing \gwtc3
\newcommand{\gwtcFour}{GWTC-4.1\xspace}
\newcommand{\gwtcFive}{GWTC-5.0\xspace}

\makeatletter
\AtBeginDocument{
   \def\ltx@label#1{\cref@label{#1}}%add braces
   \def\label@in@display@noarg#1{\cref@old@label@in@display{#1}}%remove braces
\def\label@in@mmeasure@noarg#1{%
    \begingroup%
      \measuring@false%
      \cref@old@label@in@display{#1}%remove braces for multline, see https://tex.stackexchange.com/q/737204/2388
    \endgroup}%  
 } %
\makeatother

\begin{document}
\preprint{APS/123-QED}

\title{Combining gravitational wave search pipelines to find subthreshold signals in GWTC-5.0}

\author{Ann-Kristin Malz}
\email{ann-kristin.malz@ligo.org}
\affiliation{Department of Physics, Royal Holloway, University of London}
\author{Samuel Russell}
\affiliation{Department of Physics, Royal Holloway, University of London}
\author{Gregory Ashton}
\affiliation{Department of Physics, Royal Holloway, University of London}
\affiliation{School of Mathematical Sciences, University of Southampton}
\author{Nicolo Colombo}
\affiliation{Department of Computer Science, Royal Holloway University of London}

\date{\today}

\begin{abstract}

The detection of transient gravitational wave signals relies on independent search algorithms that analyse detector data and assign significance measures to candidate events. 
However, varying performance complicates their interpretation. 
We use supervised machine learning combined with conformal prediction, a framework to quantify uncertainties, to merge multi-pipeline information into well-calibrated confidence scores. 
We demonstrate that this approach is robust across different classifier architectures and remains stable when trained on different simulated datasets. 
When applied to events across the GWTC catalogue up to and including the second part of the fourth observing run, the framework identifies several subthreshold candidates with elevated confidence, including the binary neutron star candidate GW200311\_103121. 
We examine the reliability of these up-rankings, finding evidence that high-confidence predictions correspond to signal-like events.
This framework enables simplified systematic candidate assessment for gravitational wave catalogues and real-time alerts by providing a single, well-calibrated confidence measure per candidate. 

\end{abstract}

\maketitle

\section{Introduction}\label{sec:intro}
The field of \ac{GW} astronomy is rapidly expanding with the growing number of detections. 
With the latest data release, \gwtcFive~\citep{GWTC5}, the cumulative catalogue of events now contains almost 400 signals generated by \acp{CBC} and observed by the \ac{LVK} detectors~\citep{LIGO, Virgo, KAGRA}. 
The majority of these signals are caused by \ac{BBH} mergers, with a small number of \ac{NSBH} and only two \ac{BNS} detections. 
\ac{BNS} signals are of particular interest because they can produce electromagnetic counterparts that enable multi-messenger astronomy~\citep{LIGOScientific:2017vwq}, allowing for independent measurements of cosmological parameters \citep{LIGOScientific:2017adf} and providing crucial insights into the properties of ultra-dense matter \citep{LIGOScientific:2018cki}.

\ac{GW} signals are extracted from noisy detector data~\citep{LIGOScientific:2019hgc} using advanced search algorithms (pipelines), which assign a \ac{FAR} to each candidate by comparing its detection statistic against an estimated background \citep{GWTC5method}. 
The pipeline outputs can then be combined with an astrophysical model to calculate a Bayesian \pastro~\citep{Farr:2013yna}, which quantifies the probability that the candidate is an astrophysical signal~\citep{GWTC5}. 

Within the \ac{LVK} collaboration, multiple search pipelines are currently routinely used, providing complementary sensitivity and improved robustness. 
Four of them (\gstlal~\citep{Messick:2016aqy, Sachdev:2019vvd, Tsukada:2023edh, Sakon:2022ibh, Joshi:2025nty,  Cannon:2020qnf, Hanna:2019ezx}, \mbta~\citep{Allene:2025saz, Aubin:2020goo, Adams:2015ulm}, \pycbc~\citep{Allen:2005fk, DalCanton:2014hxh, Usman:2015kfa, Nitz:2017svb, Allen:2004gu, Davies:2020tsx}, and \spiir~\citep{Chu:2020pjv, Luan:2011qx}) rely on matched filtering, where models of \ac{CBC} signals are used, while \cwb~\citep{Mishra:2024zzs, Klimenko:2005xv, Klimenko:2008fu, Klimenko:2015ypf, Klimenko:2004qh, Klimenko:2011hz} performs an unmodelled, coherence-based search. 
Additional information is detailed in \citet{GWTC5method}. 
There are also several other search pipelines, such as the \ac{ML} based \texttt{Aframe}~\citep{Marx:2024wjt} and \texttt{MLy}~\citep{Skliris:2020qax} pipelines, as well as independent searches \citep{Nitz:2021zwj, Mehta:2023zlk}.

The use of multiple pipelines creates challenges when their outputs for a particular candidate do not agree. 
This can arise from differences in search methodology, tuning, and configuration choices, as well as parameter-space-dependent sensitivities and varying susceptibility to noise artifacts. 
Combining these outputs into a single robust significance measure per candidate is therefore essential.

Currently, the maximum \ac{IFAR} across pipelines, where \ac{IFAR}=1/\ac{FAR}, is used to classify candidates as significant, and those with $\pastro \geq 0.5$ from at least one pipeline are included in the \ac{GWTC} as astrophysical signals~\citep{GWTC3, GWTC4, GWTC5}.
This maximum across pipelines approach is straightforward and computationally efficient. 
Alternative methods have been explored, such as producing a unified \pastro~\citep{Banagiri:2023ztt}. 

Nevertheless, differing pipeline outputs carry additional information about the nature of a candidate. 
The maximum significance approach neglects this, as it only accounts for information from the most significant pipeline candidate.
A signal, for instance, should be consistently detected across pipelines with compatible properties, whereas a noise artifact may only trigger a single pipeline sensitive to that particular noise type. 
By exploiting these correlations, we can therefore learn additional information about candidate events.

Several other \ac{ML} classifier approaches have also been proposed (\citet{Tsukamoto:2025vuu}, \citet{Moustakidis:2026prep}) to learn such correlations and improve the characterisation of candidate events.
Most \ac{ML} algorithms produce uncalibrated predictions without associated uncertainties, yet reliable uncertainty quantification is essential when assessing whether a \ac{GW} candidate is of astrophysical origin. 
In \citet{Ashton:2025jhn}, we therefore applied \ac{CP}~\cite{vovk2005algorithmic, angelopoulos2021gentle} to quantify the uncertainties of the \ac{ML} algorithm and provide well-calibrated conditional confidence scores for each candidate. 
The conditional confidence~\citep{Ashton:2024wae} is a score in $[0,1]$ quantifying how signal-like a candidate appears.

The application of \ac{CP} in astrophysics is recent but expanding, with work covering redshift estimation~\citep{2024ApJ...974..159J, Soriano:2026mwu}, stellar spectra classification~\citep{2024sccc.conf...43P}, black hole mass estimation~\citep{Yong:2023lcp}, neutron star equation of state inference~\citep{Mendes:2026fsj}, and measurement error propagation~\citep{2025MNRAS.539.1372S}, as well as work in high-energy physics~\citep{Araz:2025vuw} and a series of applications in gravitational-wave astronomy~\citep{Ashton:2024wae, Malz:2024zjd, Ashton:2025jhn}, including the line of work upon which the present work builds.

In \citet{Ashton:2025jhn}, we demonstrated that our method outperforms the \maxifar approach on simulated data, achieving superior \ac{ROC} curves and \ac{AUC} scores. 
When applied to the catalogue from the third observing run, O3~\citep{GWTC2-1, GWTC3}, our framework notably assigned a high confidence score to the subthreshold \ac{BNS} candidate GW200311\_103121. 

\begin{figure}
    \centering
    \includegraphics{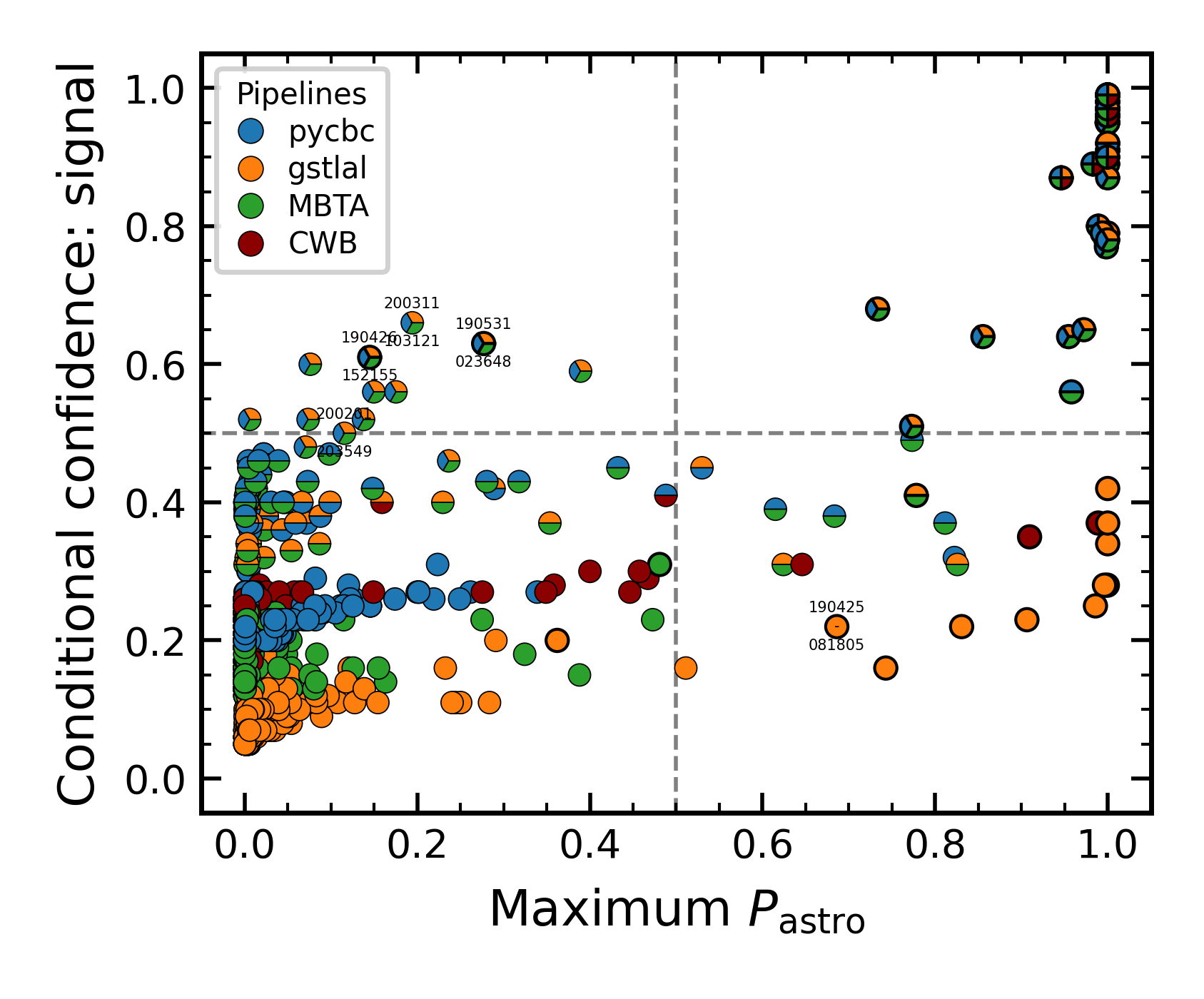}
    \caption{The conditional confidence in the signal label for the \ac{LR} classifier compared to the maximum \pastro for the O3 events in the combined \gwtcTwoPointOne and \gwtcThree catalogues.  
    The classifier is trained and \ac{CP} calibrated on the \ac{MDC} data.
    The dashed lines represent the conditional confidence and \pastro thresholds, respectively. The markers are coloured by the search pipelines which detected each candidate, and high-significance events (\ac{FAR} < 1 per year) are indicated by a bold marker outline.
    We highlight names of candidates discussed in the text.
    The plot is reproduced from Fig. 3 of \citet{Ashton:2025jhn}, but due to a correction in the data assembling now includes several candidates that were previously mistakenly excluded. }
    \label{fig:gwtc3_lr}
\end{figure}

We reproduce the key result of \citet{Ashton:2025jhn} in \cref{fig:gwtc3_lr}, showing conditional confidence scores obtained by combining \ac{LR} with \ac{CP} for \ac{GW} candidates in the third observing run (O3). 
In the bottom-right quadrant, we find candidates with $\pastro \geq 0.5$ but subthreshold (here $0.5$, see \cref{sec:threshold}) conditional confidence.
The majority of these candidates are single-pipeline events, mostly by \gstlal or \cwb, including GW190425\_081805, a single-detector \ac{BNS}~\cite{GWTC2-1, LIGOScientific:2020aai}. 
In the top-left quadrant, we identify several candidates that fall below the \pastro threshold of $0.5$ but with conditional confidence values above $0.5$. 
In addition to the \ac{BNS} candidate GW200311\_103121 with a \ac{FAR} just above threshold ($1.3 \text{ yr}^{-1}$) in two pipelines, we identify the marginally significant \ac{NSBH} candidates GW190426\_152155, GW190531\_023648, and GW200201\_203549~\citep{GWTC3} and several \ac{BBH} events, each found by the \gstlal, \pycbc, and \mbta pipelines. 
A full list of conditional confidence scores and \pastro values for all events in these quadrants is included in \cref{app:gwtc_table}. 

These results are promising, suggesting that several sub-threshold candidates may be of astrophysical origin. 
However, \ac{CP} assumes exchangeability between the data used to calibrate the framework and the data it is applied to~\citep{Shafer2007CP}, which may not hold if there are changes to the search pipelines or detectors, raising the question of whether the conditional confidence scores shown in \cref{fig:gwtc3_lr} are reliable.
Addressing this requires both empirical robustness checks, confirming that results are stable across model and data choices, and physical validation that the classifiers learn meaningful signal properties rather than artefacts of the training data. 

In this work, we extend and validate the framework of \citet{Ashton:2025jhn}. Specifically, we apply our framework to new candidates from the fourth observing run, O4, in \gwtcFour and \gwtcFive, while validating the reliability of these results by comparing different \ac{ML} classifiers, investigating the features driving predictions, and assessing robustness using new mock data.

The paper is structured as follows.
In \cref{sec:method}, we outline the \ac{ML} classifiers and \ac{CP} framework used 
to combine pipeline outputs.
In \cref{sec:exploring_classifiers}, we compare the performance of different classifier architectures and assess the robustness of the framework to the choice of classifier.
In \cref{sec:llpic}, we examine robustness with respect to training and calibration data, by exploring the use of two different mock data sets. 
Furthermore, by examining what events are up-ranked in the mock data and using the new mock data as an unseen test, we investigate the trustworthiness of the catalogue events to which we assign increased confidence. 
In \cref{sec:gwtc4} and \cref{sec:gwtc5}, we apply our framework to the new O4a and O4b candidates in the \gwtcFour and \gwtcFive catalogue, respectively, using parameter estimation and feature investigation to assess how signal-like the candidates with high conditional confidence are. 
Finally, we discuss our results and conclude in \cref{sec:discussion}.

\section{Method and validation}\label{sec:method}
We combine the search pipelines following the same approach as in \citet{Ashton:2025jhn}: first using a \ac{ML} binary classifier to combine pipeline outputs, and second applying \ac{CP}~\cite{vovk2005algorithmic,angelopoulos2021gentle} to quantify the uncertainty and obtain a well-calibrated conditional confidence score for each candidate. 

The classifier takes as input a feature vector assembled from the outputs of the \cwb, \pycbc, \gstlal, and \mbta pipelines, comprising per-pipeline detection statistics and, where available, source parameter estimates from the nearest waveform template.
When applying our method to the new candidates in \gwtcTwoPointOne~\cite{gwtc2p1_data,GWTC2-1}, \gwtcThree~\cite{gwtc3_data,GWTC3}, \gwtcFour~\cite{gwtc4p1_data,GWTC4}, and \gwtcFive~\cite{gwtc5_data,GWTC5}, we restrict the features to the \logIFAR, \ac{SNR}, and chirp mass, setting the chirp mass to zero for \cwb, which does not provide template-based parameter estimates (see \cref{app:method} for details).
Candidates not detected by a given pipeline are assigned feature values of zero, which for the 
\logIFAR is consistent with the absence of a reported \ac{IFAR}.
To train the \ac{ML} classifiers and calibrate \ac{CP} we mainly use the pre-O4 \acf{MDC}~\citep{Chaudhary:2023vec}, while the smaller \ac{LLPIC} dataset~\citep{LLPIC:2026prep} created during O4 is used for additional robustness tests in \cref{sec:llpic}.

In \citet{Ashton:2025jhn}, we focused mainly on the \acf{LR} classifier due to its interpretability. 
In this work, we extend the comparison to \ac{MLP}~\citep{rumelhart1986learning} and XGBoost, a gradient-boosted tree ensemble~\citep{Chen:2016btl}.
The raw output from the \ac{ML} classifiers lacks a statistically rigorous significance measure,  unlike the \ac{FAR} provided by the \maxifar approach, an essential component for assessing the significance of individual candidate events. 
We therefore apply Mondrian (label-conditional) \ac{CP}~\citep{vovk2013conditional}, which provides coverage guarantees separately for signals and noise events, to the \ac{ML} classifier output to produce a \textit{prediction set} of possible labels.
However, \ac{GW} candidate assessment requires a single score quantifying significance of each candidate, for which we use the \textit{conditional confidence}~\citep{Ashton:2024wae}.
The conditional confidence score provides a single well-calibrated significance measure per candidate: values close to one indicate strong support for the signal hypothesis, while values close to zero indicate noise-like behaviour. 

Full details of the data preparation, classifier architectures, and \ac{CP} formalism are given in \cref{app:method}.

\subsection{Robustness to classifier architecture}\label{sec:exploring_classifiers} 
We compare the performance of the \ac{ML} classifiers to assess whether the pipeline combination framework is effective and whether results are robust to the choice of classifier. 
We evaluate each classifier using \ac{ROC} curves and sensitivity and precision metrics under \ac{CP}, and examine consistency on real \ac{GW} candidates from O3. 
Full details and figures are given in \cref{app:classifiers}.

\begin{figure}
    \centering
    \includegraphics[width=0.5\textwidth]{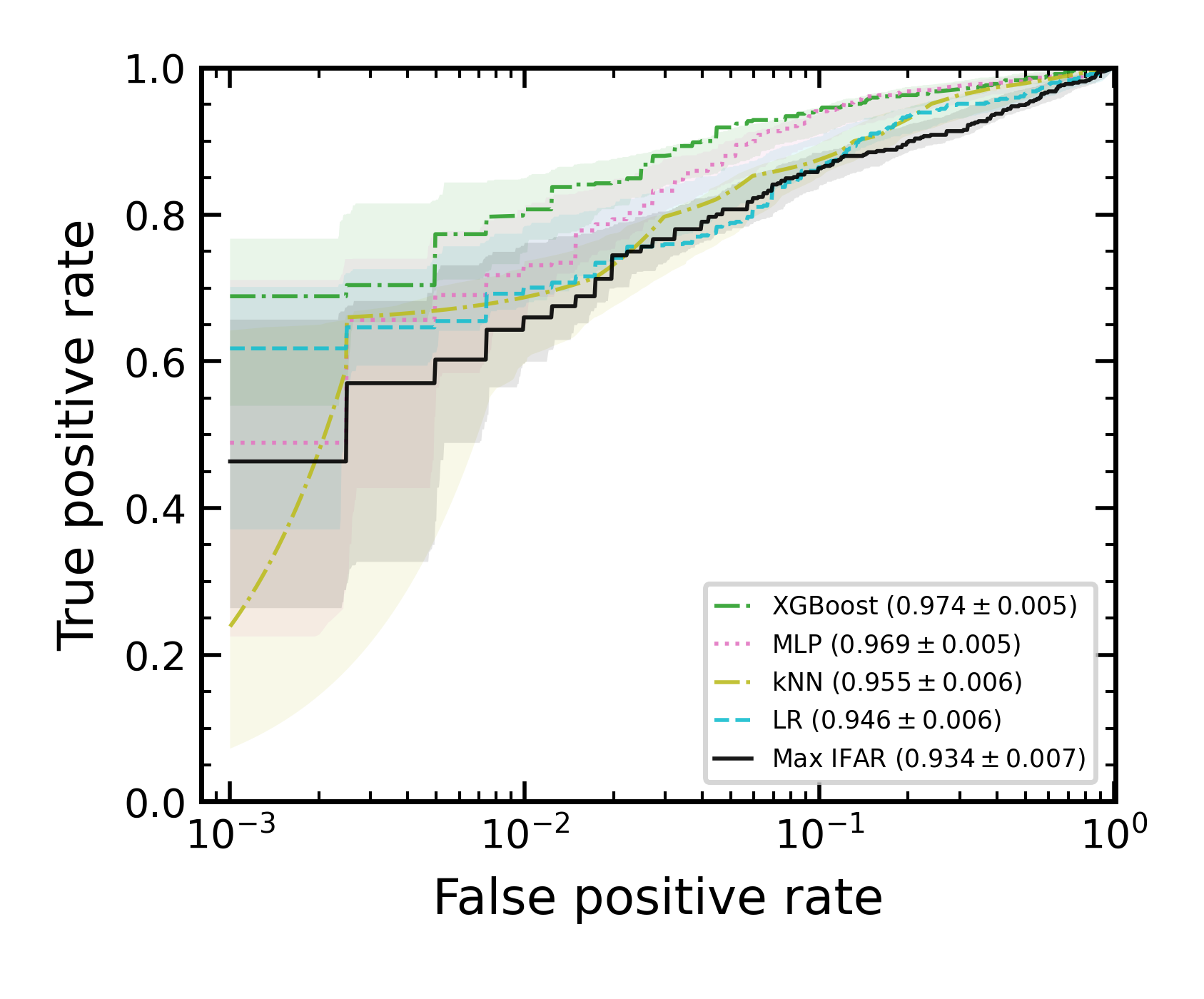}
    \caption{ROC curve comparing the performance of four different \ac{ML} classifiers and the \maxifar method on the \ac{MDC} dataset. 
    The shaded band marks the 90\% interval over 100 random permutations of the training and testing data. 
    The \ac{AUC} is given in the legend for each classifier, including the standard deviation obtained from the train-test data split permutations.} 
    \label{fig:roc_all_classifiers_mdc12}
\end{figure}

We evaluate the \ac{ML} classifiers on the \ac{MDC} dataset, and present the resulting \ac{ROC} curves in \cref{fig:roc_all_classifiers_mdc12}, alongside the \maxifar approach for comparison. 
All \ac{ML} classifiers outperform the \maxifar method, demonstrating the benefit of exploiting correlations between pipeline outputs rather than relying on any single pipeline statistic. 
XGBoost achieves the highest \ac{AUC} on the mock data, approximately one standard deviation above \ac{MLP} and nearly five standard deviations above \maxifar, suggesting that its non-linear, tree-based structure is particularly effective at capturing complex interactions between features, although all \ac{ML} classifiers show a clear and systematic improvement over the \maxifar method. 
XGBoost also leads on the precision and sensitivity metrics, see \cref{sec:sensitivity_precision}. 
At a conditional confidence threshold of $0.5$ (see \cref{sec:threshold} for the threshold justification), \ac{LR} and XGBoost produce zero false positives on the \ac{MDC} data, compared to one for \ac{MLP} and \maxifar.

Applied to the O3 candidates from \gwtcTwoPointOne and \gwtcThree, all three classifiers show broad agreement. 
In particular, the sub-threshold \ac{BNS} candidate GW200311\_103121 listed in the marginal candidate table of \gwtcThree, and the two \gwtcTwoPointOne marginal significance \ac{NSBH} candidates GW190426\_152155 and GW190531\_023648, are consistently assigned above-threshold confidence by all classifiers, lending additional support to their potential astrophysical origin. 
XGBoost is the most conservative classifier, up-ranking only events with strong multi-pipeline support, while \ac{MLP} is the least conservative, also up-ranking single- and two-pipeline candidates.
The conservative behaviour is preferred: false positives, noise events misidentified as astrophysical signals, carry a higher cost than false negatives in \ac{GW} astronomy.

The broad agreement across classifiers demonstrates that the framework is robust to the choice of classifier architecture, with differences arising primarily in the degree of conservatism.
Based on its superior \ac{AUC}, near-zero false positive rate, and conservative behaviour on real candidates, we identify XGBoost as our preferred classifier for the remainder of this analysis.

To understand what drives individual predictions, we perform a \ac{SHAP} analysis~\citep{lundberg2017shap}, which confirms that the classifiers learn physically meaningful patterns: \logIFAR features from the templated pipelines dominate, with higher values pushing toward a signal prediction, while missing detections in sensitive pipelines push toward a noise prediction.
See \cref{sec:shap} for full details and plots.

\subsection{Robustness to training and calibration data}\label{sec:llpic}
Having established our framework on the \ac{MDC} data, we now turn to the more recent \ac{LLPIC} datasets, as described in \cref{sec:data}, to assess its robustness. 
As discussed in \cref{sec:intro}, the exchangeability assumption underpinning \ac{CP} cannot be strictly guaranteed in practice.
Since the \ac{MDC} and \ac{LLPIC} are based on data from different observing runs and therefore reflect differences in detector configuration and pipeline implementations (see \citet{GWTC5method} for details), using the \ac{LLPIC} provides a direct probe of this assumption.

We use this new dataset in two ways: first, as a consistency check to assess whether we obtain results consistent with those from the \ac{MDC} dataset; and second, as independent test data to evaluate whether the conditional confidence can be trusted where it disagrees with the \maxifar classification. 

\subsubsection{Consistency check with \ac{LLPIC}}
We begin by training, calibrating, and testing our framework on the \ac{LLPIC} dataset, and mirror the analysis performed on the \ac{MDC} data, thus checking the consistency and robustness of our method. 

\begin{figure}
    \centering
    \includegraphics[width=0.49\textwidth]{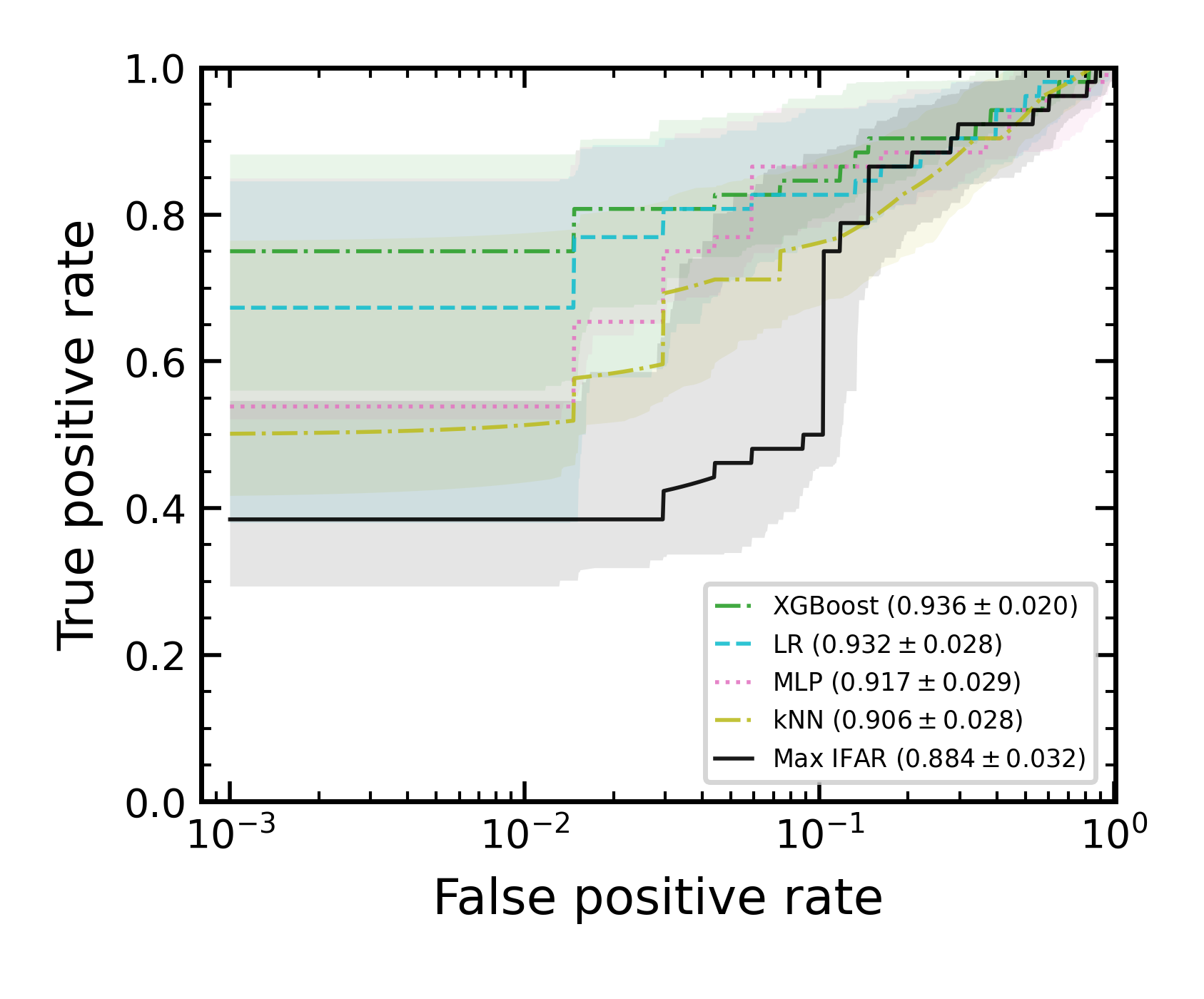}
    \caption{\ac{ROC} curves comparing the performance of four different \ac{ML} classifiers and the \maxifar method as trained, calibrated and tested on the \ac{LLPIC} dataset. 
    The shaded band marks the 90\% interval over the permutations of the training and testing data. 
    The \ac{AUC} is given in the legend for each classifier, including the uncertainty over train-test data split permutations.} 
    \label{fig:roc_all_classifiers_llpic4}
\end{figure}

We assess the performance of the different classifiers using the \ac{ROC} curve shown in \cref{fig:roc_all_classifiers_llpic4}, observing consistent results compared to the \ac{MDC}: all \ac{ML} classifiers outperform the \maxifar combination method, with XGBoost achieving the highest \ac{AUC}. 
At a conditional confidence threshold of $0.5$, no false positives are produced and approximately 50\% of signals are recovered by all classifiers and the \maxifar.
This confirms that the framework generalises to the \ac{LLPIC} dataset and that the result is not dependent on the training dataset used.

\begin{figure*}
    \centering
    \subfigure[\ac{LR}]{\includegraphics[width=0.3\textwidth]{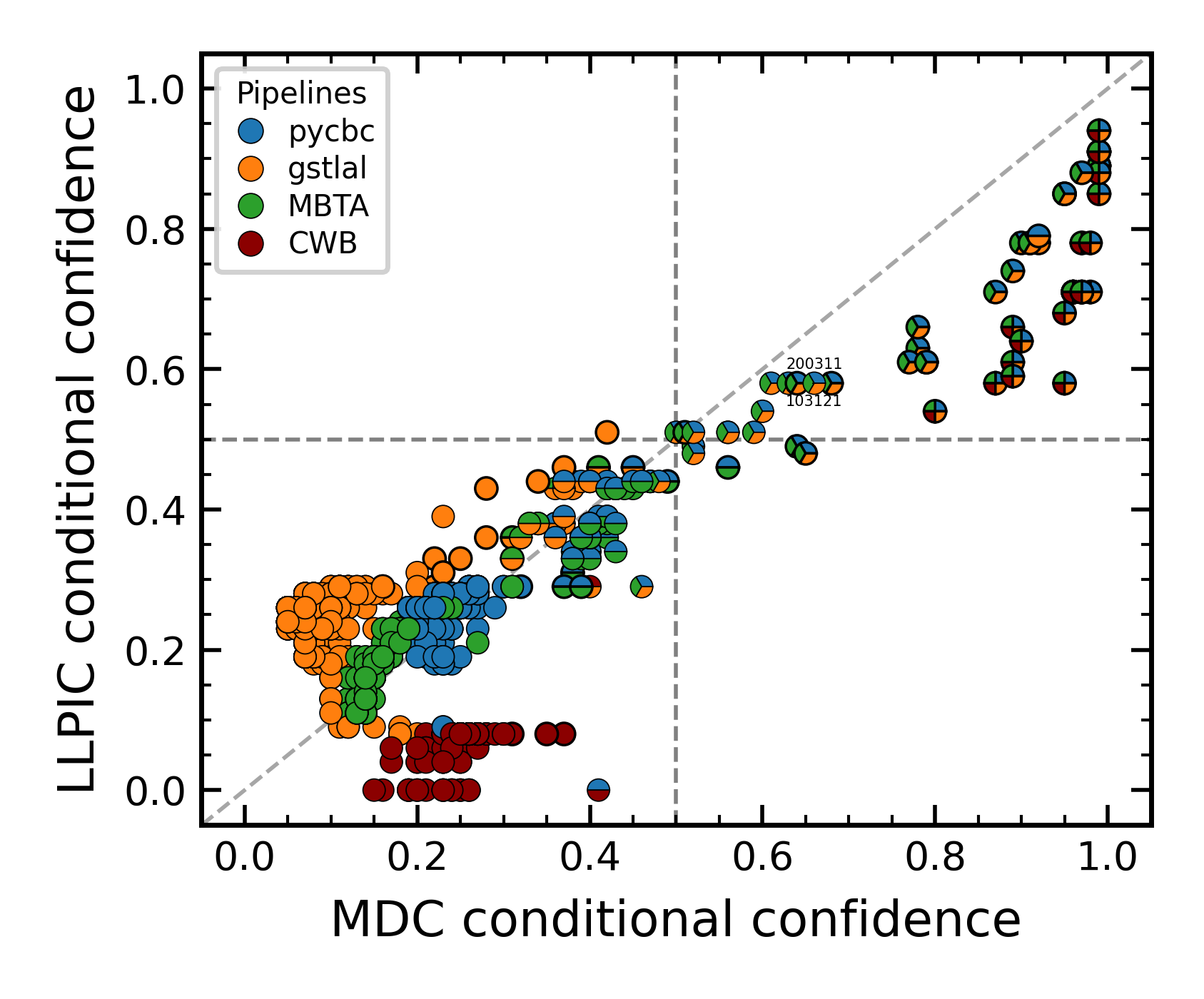} \label{fig:gwtc3_mdc_llpic_logistic}} 
    \subfigure[\ac{MLP}]{\includegraphics[width=0.3\textwidth]{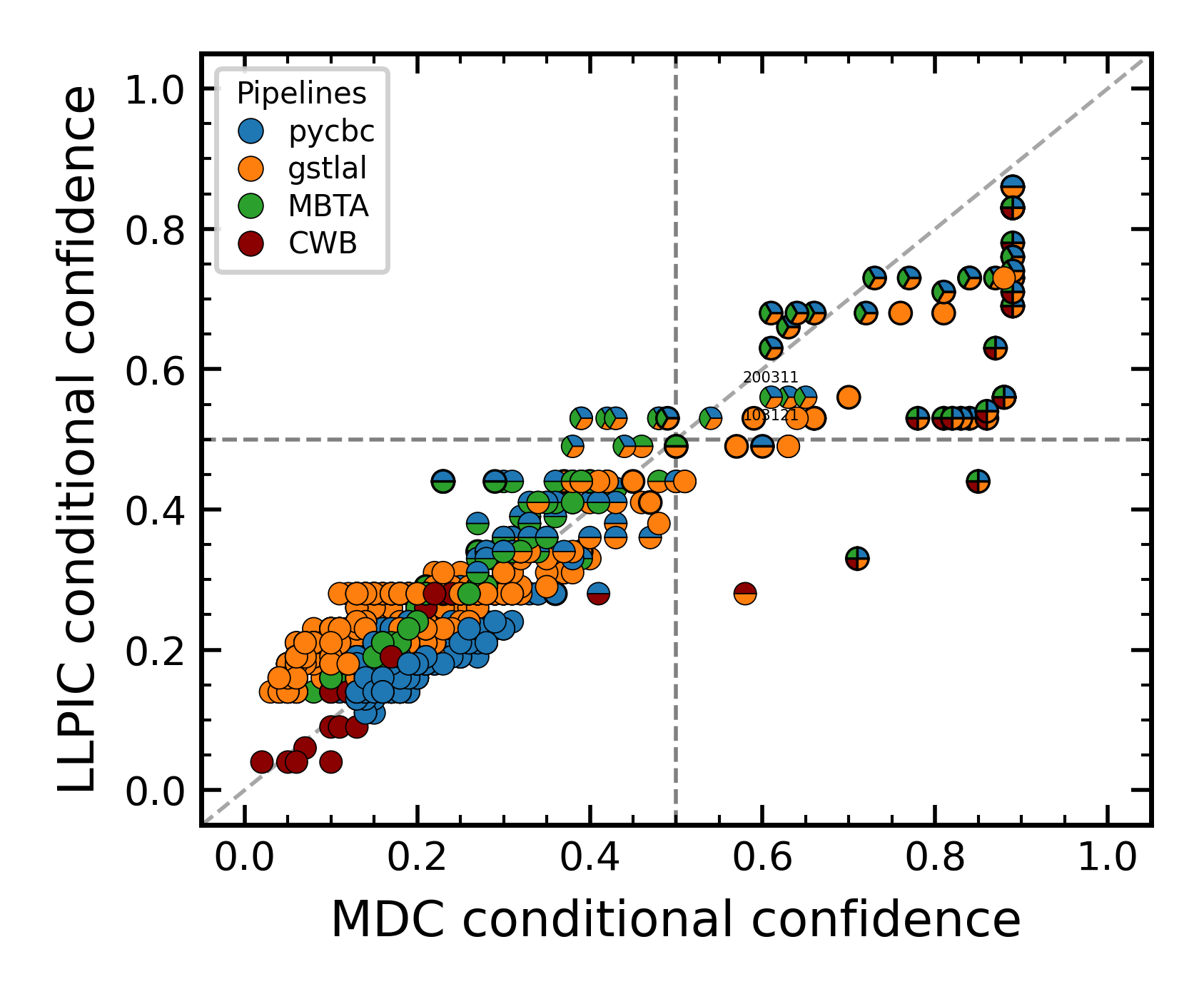}\label{fig:gwtc3_mdc_llpic_mlp}}
    \subfigure[XGBoost]{\includegraphics[width=0.3\textwidth]{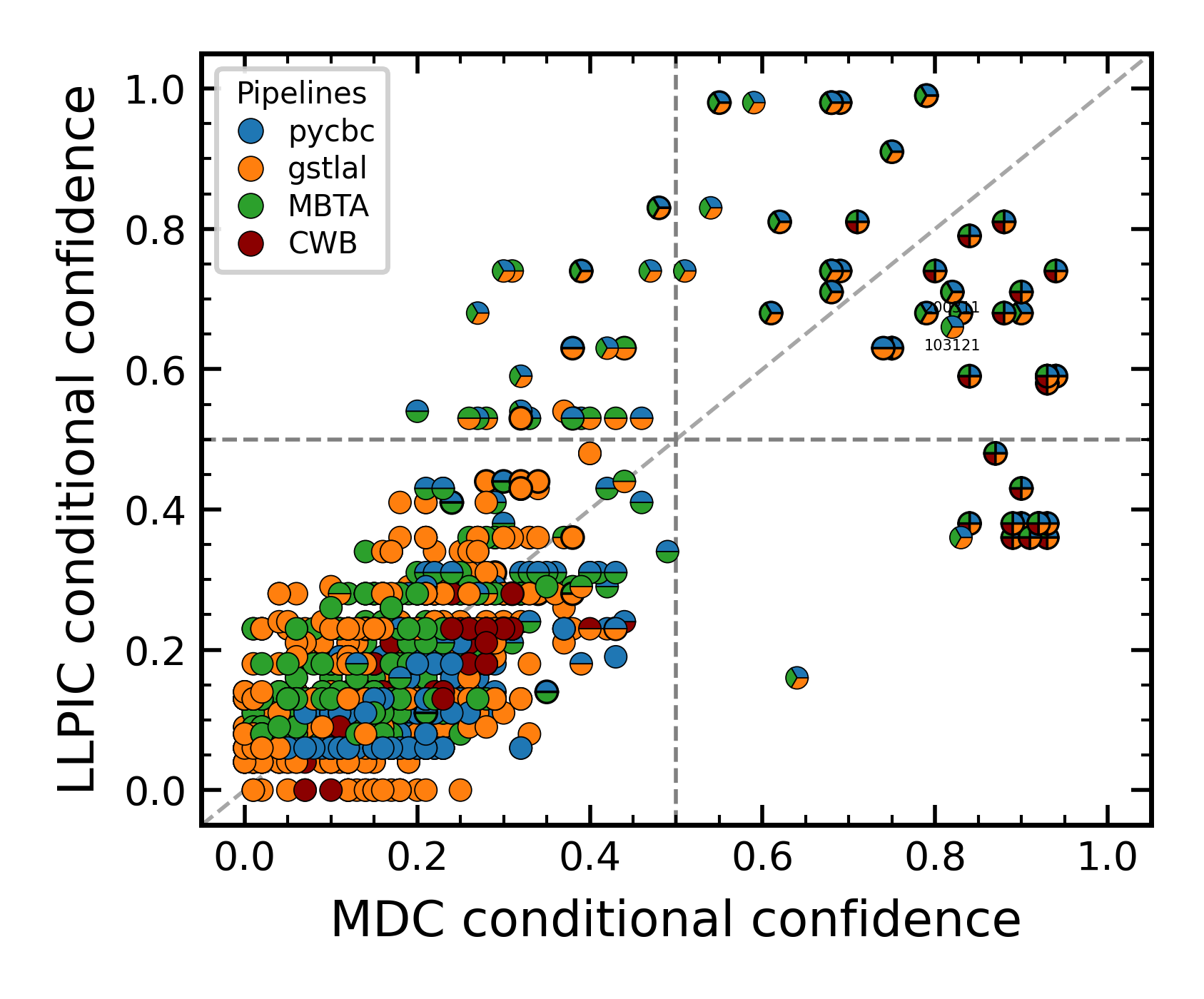}\label{fig:gwtc3_mdc_llpic_xgboost}}
    \caption{The conditional confidence in the signal label for the O3 events in the \gwtcTwoPointOne and \gwtcThree catalogue as obtained from training and calibration on the \ac{MDC} versus \ac{LLPIC} data respectively. 
    The dashed lines represent the conditional confidence threshold at $0.5$. 
    Events with \maxpastro above $0.5$ are indicated by a bold marker outline.} 
    \label{fig:gwtc3_mdc_llpic}
\end{figure*}

Next, we calculate the conditional confidence for the O3 events in the \gwtcTwoPointOne and \gwtcThree catalogues, using \ac{LLPIC} as training and calibration data. 
To highlight the comparison with the \ac{MDC}-trained framework, the confidences as obtained from each of the different training datasets are plotted against each other in \cref{fig:gwtc3_mdc_llpic}. 

The \ac{LLPIC} dataset is significantly smaller than the \ac{MDC}, which affects classifier performance. 
While simple models such as \ac{LR} are generally more robust with limited data due to their lower complexity, more flexible models such as XGBoost may require more training data to generalise reliably~\citep{Chen:2016btl, hastie2009elements}. 

The plots for \ac{LR} and \ac{MLP}, in \cref{fig:gwtc3_mdc_llpic_logistic} and \cref{fig:gwtc3_mdc_llpic_mlp} respectively, show good agreement between confidences for most events. 
For XGBoost (\cref{fig:gwtc3_mdc_llpic_xgboost}), the conditional confidence values show greater variations between the two datasets, suggesting this classifier is more sensitive to the choice of training data.
While \ac{LR} and \ac{MLP} give generally lower confidence when trained and calibrated on the \ac{LLPIC} data, for XGBoost more events are above the confidence threshold with this dataset. 
For all classifiers we observe that the \ac{BNS} candidate GW200311\_103121 and the two \ac{NSBH} candidates GW190426\_152155 and GW190531\_023648 have confidence values above the threshold regardless of training data. 

From this investigation, we conclude that our results remain broadly consistent when switching the training and calibration data despite the smaller size of the \ac{LLPIC}, supporting the robustness of the framework. 
Although exchangeability cannot be guaranteed for the \ac{LLPIC} data either, the consistency across classifiers suggests our approach is reasonably tolerant to the choice of training and calibration dataset. 
The exception is XGBoost, which shows greater sensitivity to this choice, consistent with its higher model variance discussed in \cref{sec:model_comparison}.

\subsubsection{Using \ac{LLPIC} as unseen test data}\label{sec:llpic_unseen_test}
Next, we assess whether the up- and down-rankings produced by our framework can be trusted when exchangeability is broken.
For this purpose, we train and calibrate our framework on the \ac{MDC} data, as before, and use \ac{LLPIC} as independent test data, mimicking the distribution shift between mock and real data. 
This allows us to examine whether events assigned high conditional confidence despite low \maxifar genuinely appear signal-like, and whether events down-ranked despite high \maxifar appear noise-like.

\begin{figure*}
    \centering
    \subfigure[\ac{LR}]{\includegraphics[width=0.49\textwidth]{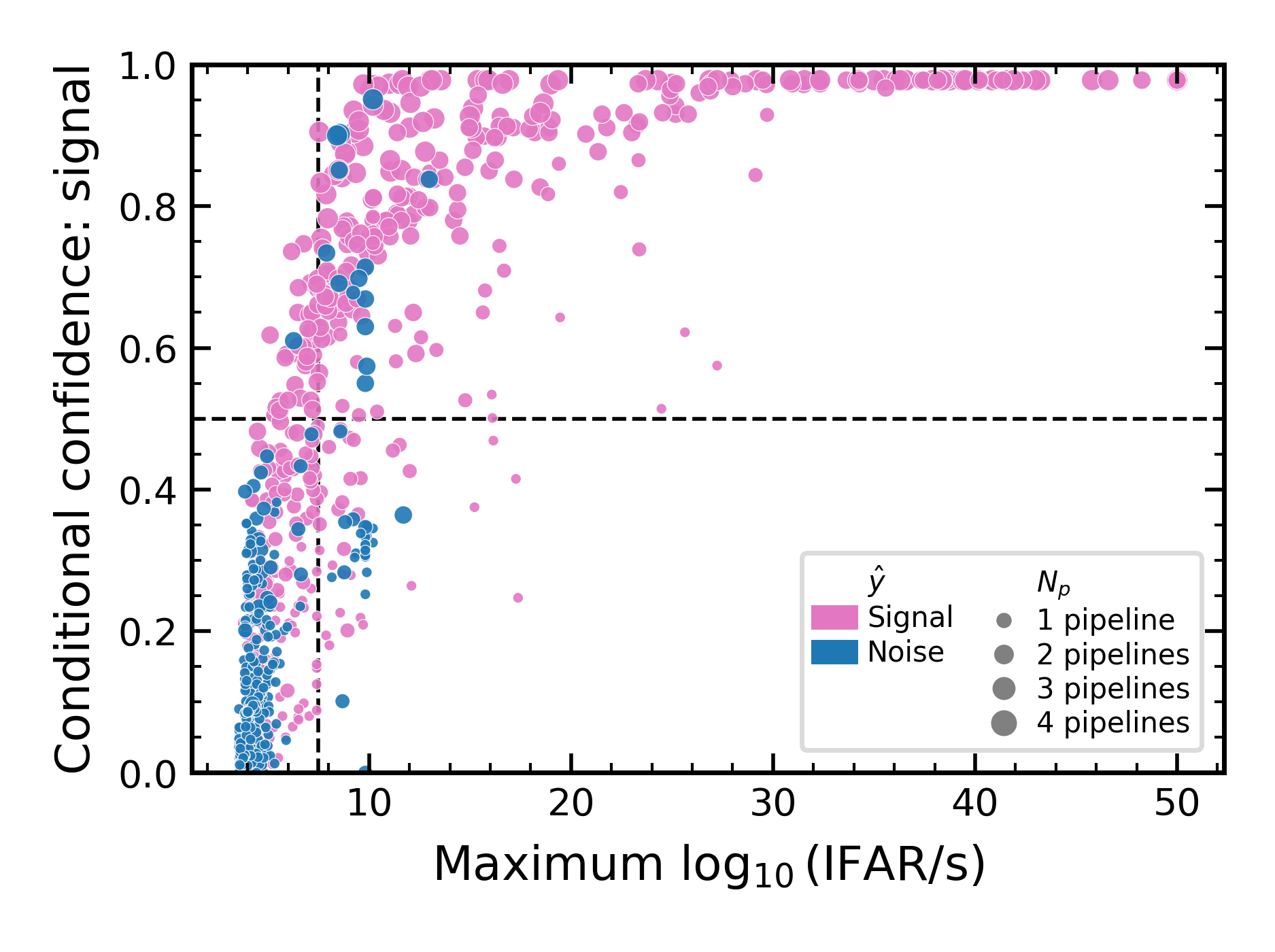} \label{fig:confidence_MDC12-LLPIC4_logistic}} 
    \subfigure[XGBoost]{\includegraphics[width=0.49\textwidth]{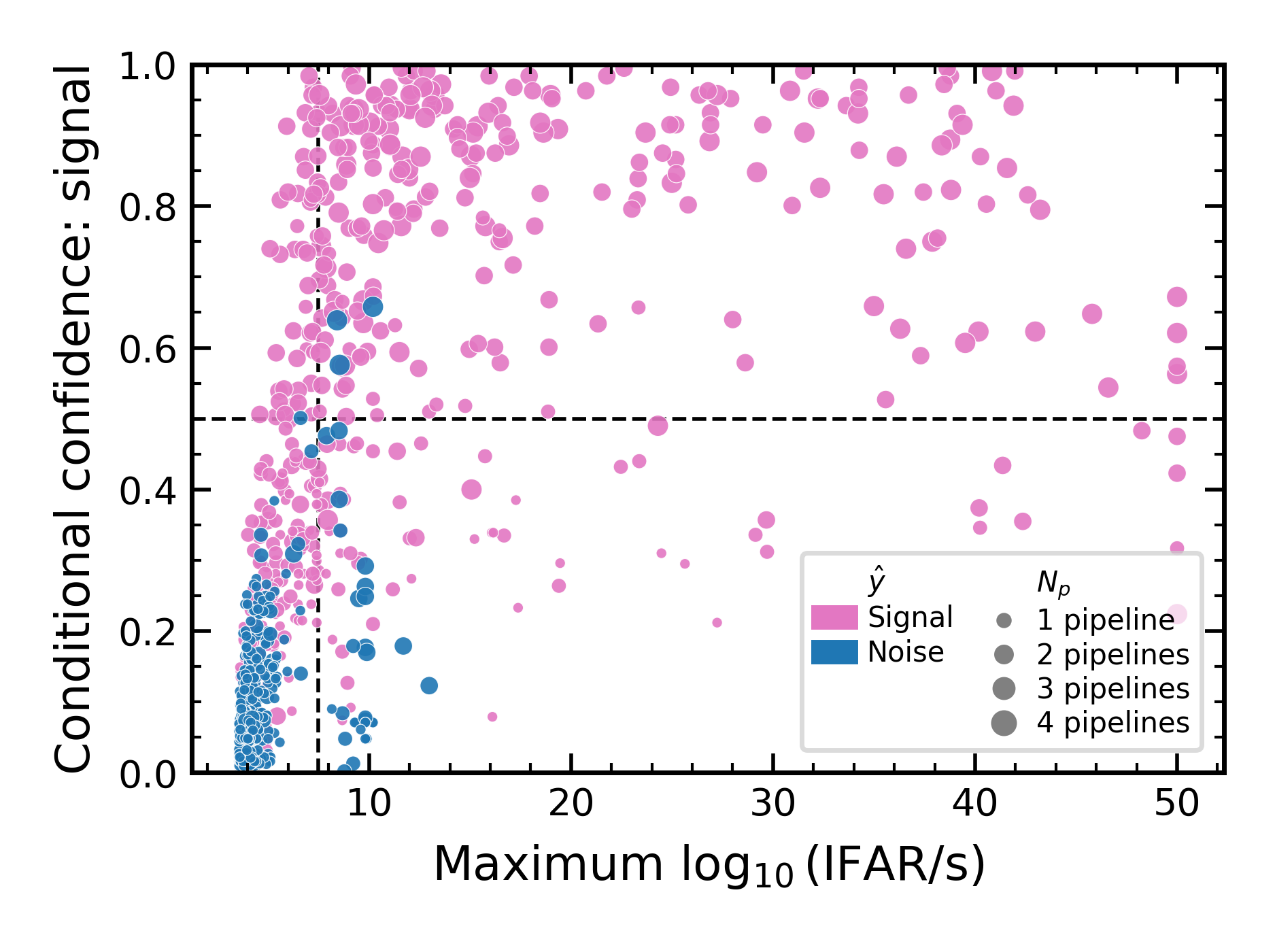}\label{fig:confidence_MDC12-LLPIC4_xgboost}}
    \caption{The conditional confidence in the signal label for the \ac{LR} (left) and XGBoost (right) classifiers, respectively, compared to the maximum \ac{IFAR}, using training and calibration data from the \ac{MDC} and testing on \ac{LLPIC}. The dashed lines show the chosen conditional confidence threshold of $0.5$ and the commonly used \ac{FAR} threshold of 1 per year. 
    The colour of each point indicates the true label $\hat{y}$, while the size indicates the number of pipelines $N_p$ which detected an event. 
    } 
    \label{fig:confidence_MDC12-LLPIC4}
\end{figure*}

\cref{fig:confidence_MDC12-LLPIC4} shows the conditional confidence scores for the \ac{LLPIC} test events (as trained and calibrated on the \ac{MDC} data) plotted against the \maxifar.
For the majority of events, the two methods are in agreement for both the \ac{LR} and XGBoost classifiers, though XGBoost shows greater scatter. 

Investigating the quadrants where the two methods disagree, we find that all candidates up-ranked by our framework correspond to signals, except for one noise candidate located on the threshold for XGBoost and at conditional confidence of $0.6$ for \ac{LR}. 
There are several noise candidates that have been incorrectly classified as signals by the \maxifar method, but which are classified as noise by both classifiers. 
However, there are also several signals that our framework down-ranks. 

While XGBoost returns fewer false positives than \ac{LR}, there are also significantly more high \maxifar signal events with sub-threshold conditional confidence. 
This broader spread in XGBoost confidences was also apparent in the \ac{MDC} results (see \cref{fig:confidence_xgboost}). 
However, using different datasets for training and calibration versus testing, this effect appears to become more pronounced, leading XGBoost to down-rank a substantially larger number of high-\maxifar, multi-pipeline signals. 
Thus, while XGBoost's up-rankings appear reliable, its down-rankings warrant caution.  
\ac{LR} demonstrates more balanced behaviour, down-ranking primarily marginal candidates near the detection threshold and single-pipeline detections.

We quantify the observations from \cref{fig:confidence_MDC12-LLPIC4} in \cref{tab:confusion_matrix_mdc_llpic}, using the same conditional confidence and \maxifar thresholds as before, see \cref{sec:exploring_classifiers}. 
All three \ac{ML} classifiers produce far fewer false positives than the \maxifar method (\ac{LR}: 15, XGBoost: 4, \ac{MLP}: 18, versus 51 for \maxifar), demonstrating a clear advantage in noise rejection. 
XGBoost is the most conservative, achieving the fewest false positives ($0.6\%$) but also the lowest true positive rate ($53.8\%$). 
\ac{MLP} recovers the most signals ($62.7\%$) at the cost of slightly more false positives ($2.7\%$), while \ac{LR} offers an intermediate balance ($59.3\%$ TPR, $2.2\%$ FPR). 
The higher false positive rate of \maxifar ($7.6\%$) likely reflects differences in noise characteristics between the \ac{LLPIC} data and the assumptions underlying the \ac{FAR} threshold.
We further observe that nearly all noise events incorrectly classified as signals by the conditional confidence also have \maxifar values above threshold, suggesting these are borderline cases where both methods struggle. 
Meanwhile, many noise candidates defined as signals by the \maxifar method have very small conditional confidence values. 

\begin{table}
    \centering
    \caption{Performance of the different classifiers + \ac{CP} and the \maxifar method trained and calibrated on the \ac{MDC} and tested on the \ac{LLPIC} dataset. 
    A threshold of $0.5$ on the conditional confidence is applied for the \ac{ML} classifiers and a \ac{FAR} of 1 per year for the \maxifar.}
    \begin{tabular}{l|c|c|c|c}
        \toprule
        Method & TP & FP & FN & TN \\
        \midrule
        LR & 316 & 15 & 217 & 655 \\
        XGBoost & 287 & 4 & 246 & 666 \\
        MLP & 334 & 18 & 199 & 652 \\
        max $\log_{10}(\mathrm{IFAR})$ & 305 & 51 & 228 & 619 \\
        \bottomrule
    \end{tabular}
    \label{tab:confusion_matrix_mdc_llpic}
\end{table}

\begin{figure}
    \centering
    \includegraphics[width=0.49\textwidth]{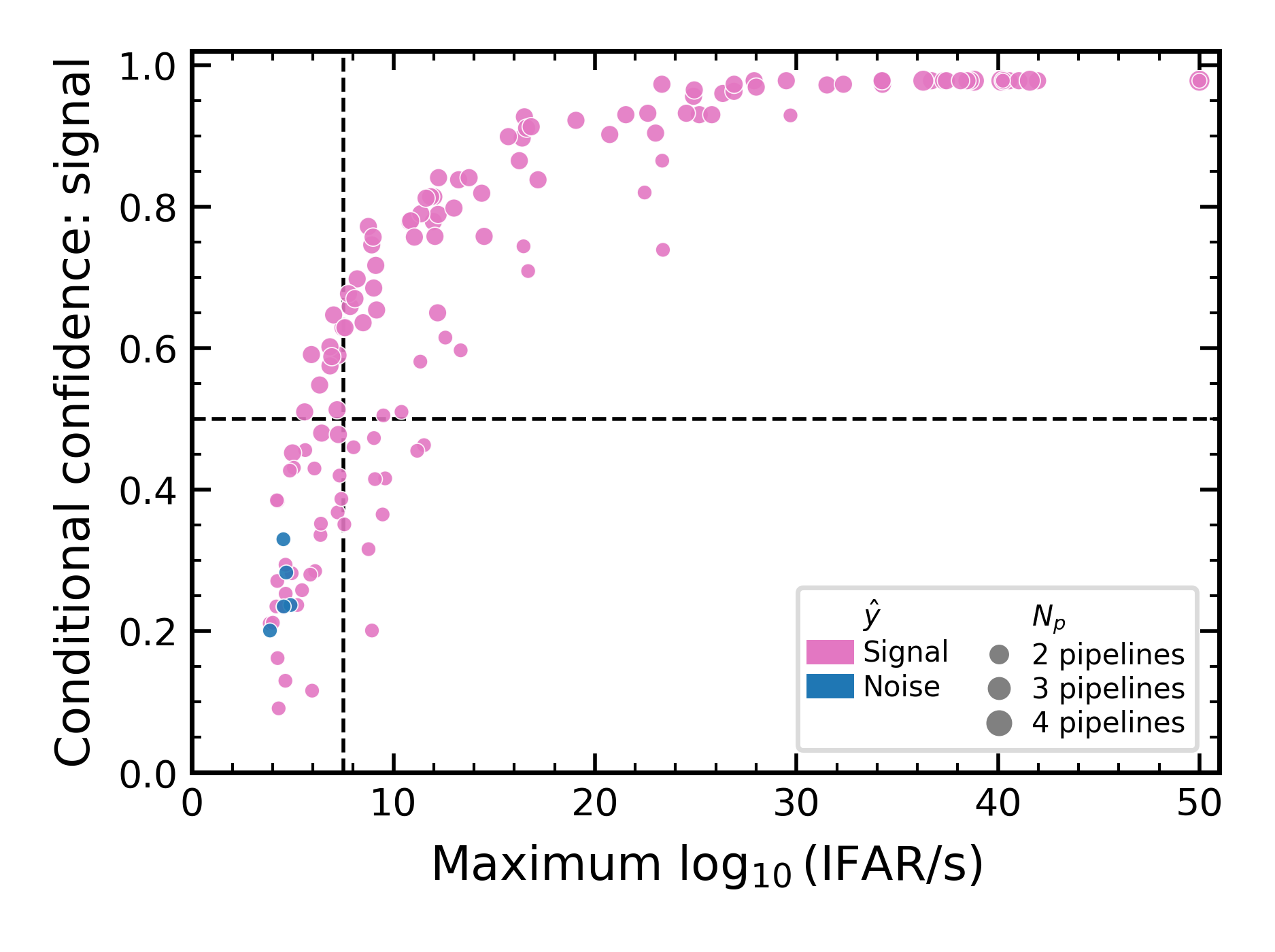}
    \caption{The conditional confidence in the signal label for the \ac{LR} classifier compared to the maximum \ac{IFAR}, using training and calibration data from \ac{MDC} and testing on \ac{LLPIC}, showing only \ac{BNS} injections ($M_1\leq3$) recovered by at least two pipelines for clarity. 
    The dashed lines show the chosen conditional confidence threshold of $0.5$ and the commonly used \ac{FAR} threshold of 1 per year. 
    The colour of each point indicates the true label $\hat{y}$, while the size indicates the number of pipelines $N_p$ which detected an event.
    } 
    \label{fig:confidence_MDC12-LLPIC4_BNS}
\end{figure}

Having validated the framework's general performance, we now focus on whether it correctly identifies \ac{BNS}-like injections, motivated by the up-ranked \ac{BNS} candidate GW200311\_103121 from \gwtcThree.
To obtain an estimate of the performance of our framework on similar events in the \ac{LLPIC} data, we plot the conditional confidence for \ac{BNS} events only (limiting the source frame component masses at $\leq3M_\odot$ to select signals consistent with neutron stars) in \cref{fig:confidence_MDC12-LLPIC4_BNS}. 

Among the three-pipeline \ac{BNS} injections, only three obtain a conditional confidence below threshold, all of which were classified as noise by the \maxifar.
This confirms that sub-threshold \ac{BNS} candidates with multi-pipeline detections, analogous to GW200311\_103121, are correctly identified as signals by our framework even under distribution shift.

To further assess classifier consistency, we construct summary diagrams representing the different regions in the \logIFAR versus conditional confidence plots in \cref{tab:venn_diagrams_mdc_llpic}. 
The diagrams indicate the number of events in each region, and highlight how many events are correctly and incorrectly up and down-ranked. 

We note that the few noise events which are incorrectly up-ranked by one of the classifiers are unique to that classifier. 
Meanwhile, there is broad agreement on the majority of up-ranked signals. 

The classifiers all correctly down-rank some noise events where \logIFAR is above threshold, but also incorrectly down-rank a similar number of signals. 
XGBoost down-ranks significantly more signals than the other classifiers while \ac{MLP} down-ranks the fewest signals.

\begin{table*}
  \caption{Comparison of the three \ac{ML} classifiers trained and calibrated on \ac{MDC} and tested on \ac{LLPIC}. The total number of signals and noise in regions where the conditional confidence and \logIFAR disagree are annotated. We also note how many of these events are up or down-ranked by that classifier only.}
  \centering
  \begin{ruledtabular}
  \begin{tabular}{lll}
    \textbf{LR} & \textbf{XGBoost} & \textbf{MLP} \\
    \midrule
    \includegraphics[width=0.32\textwidth,height=0.4\textheight,keepaspectratio]{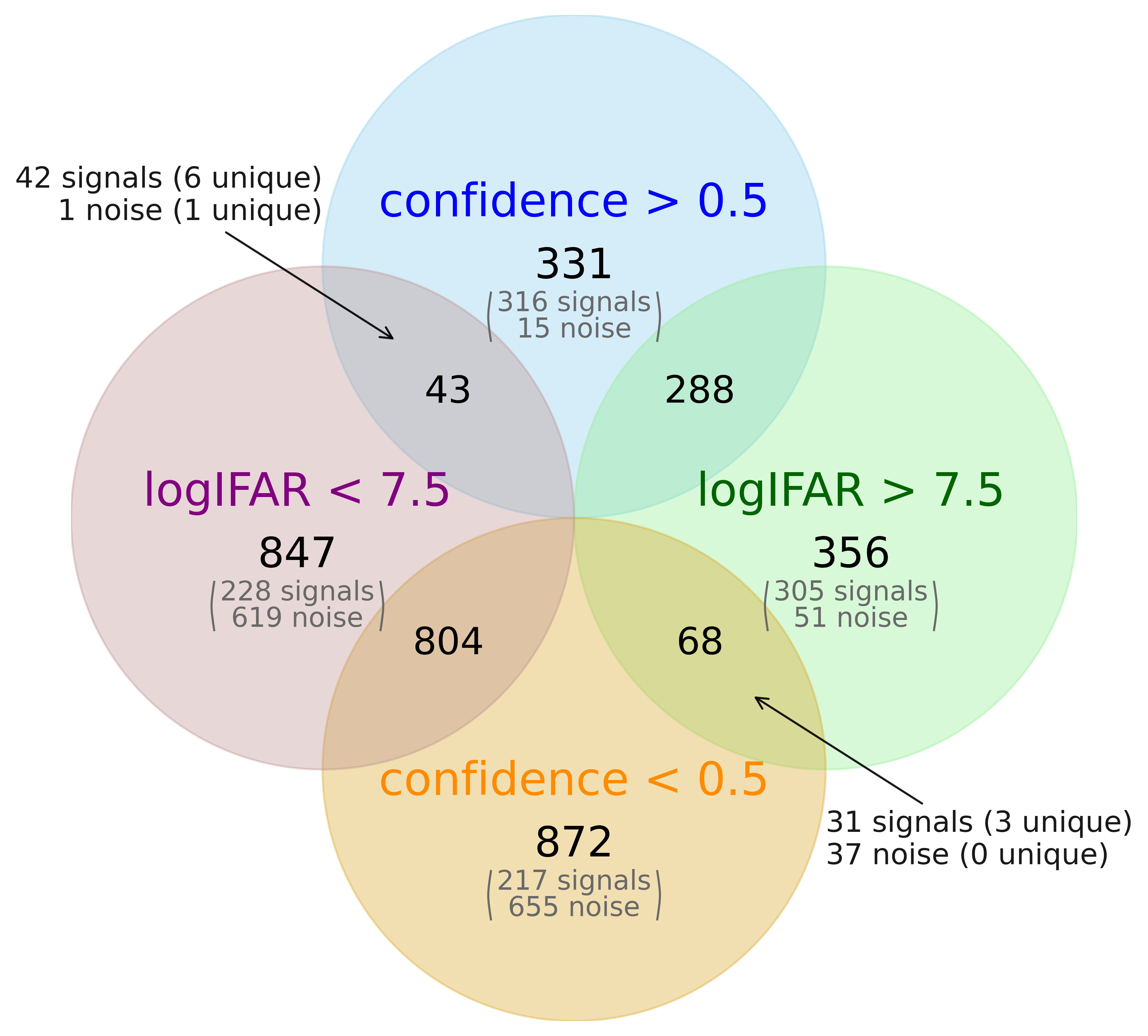} &
    \includegraphics[width=0.32\textwidth,height=0.4\textheight,keepaspectratio]{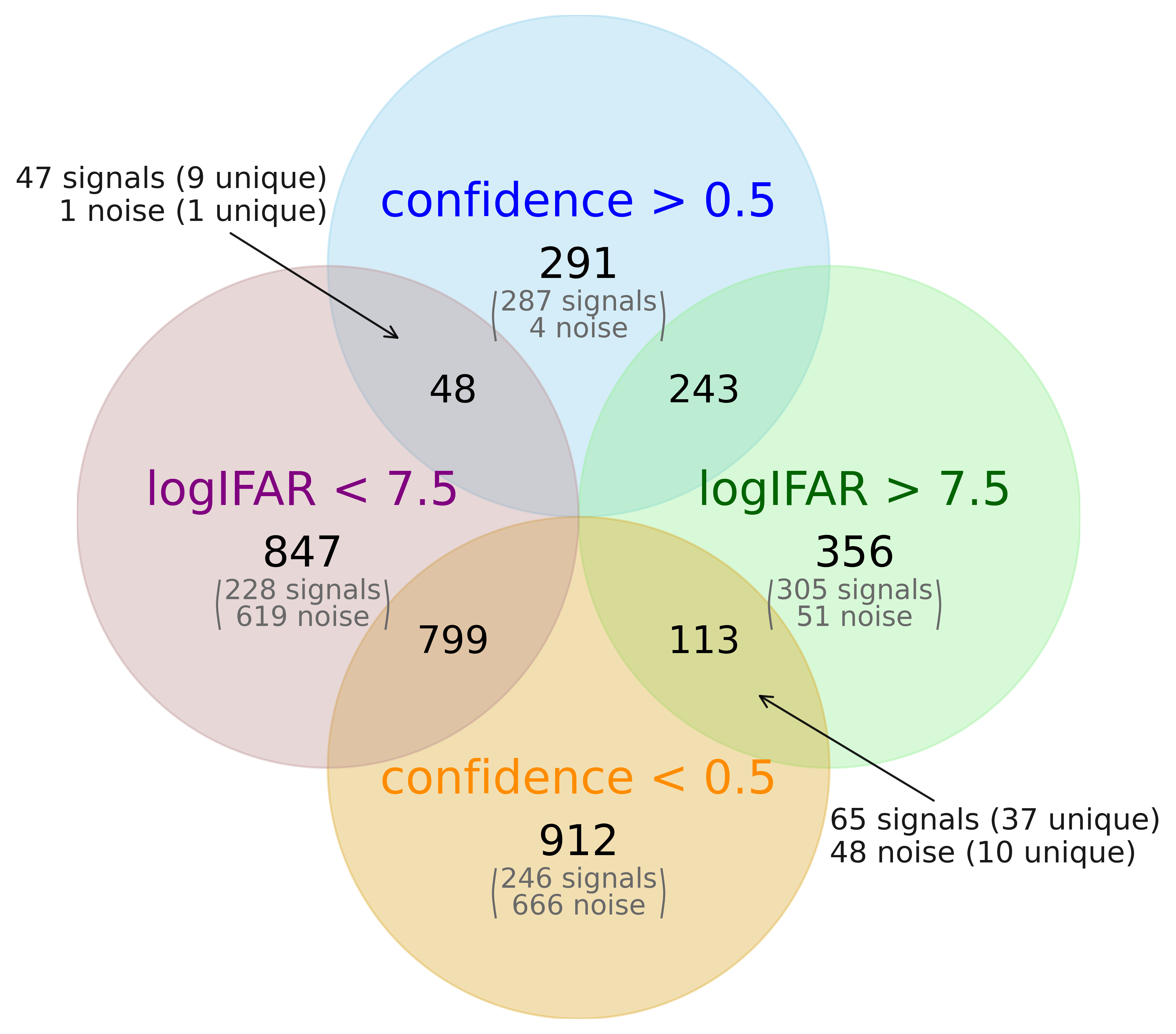} &
    \includegraphics[width=0.32\textwidth,height=0.4\textheight,keepaspectratio]{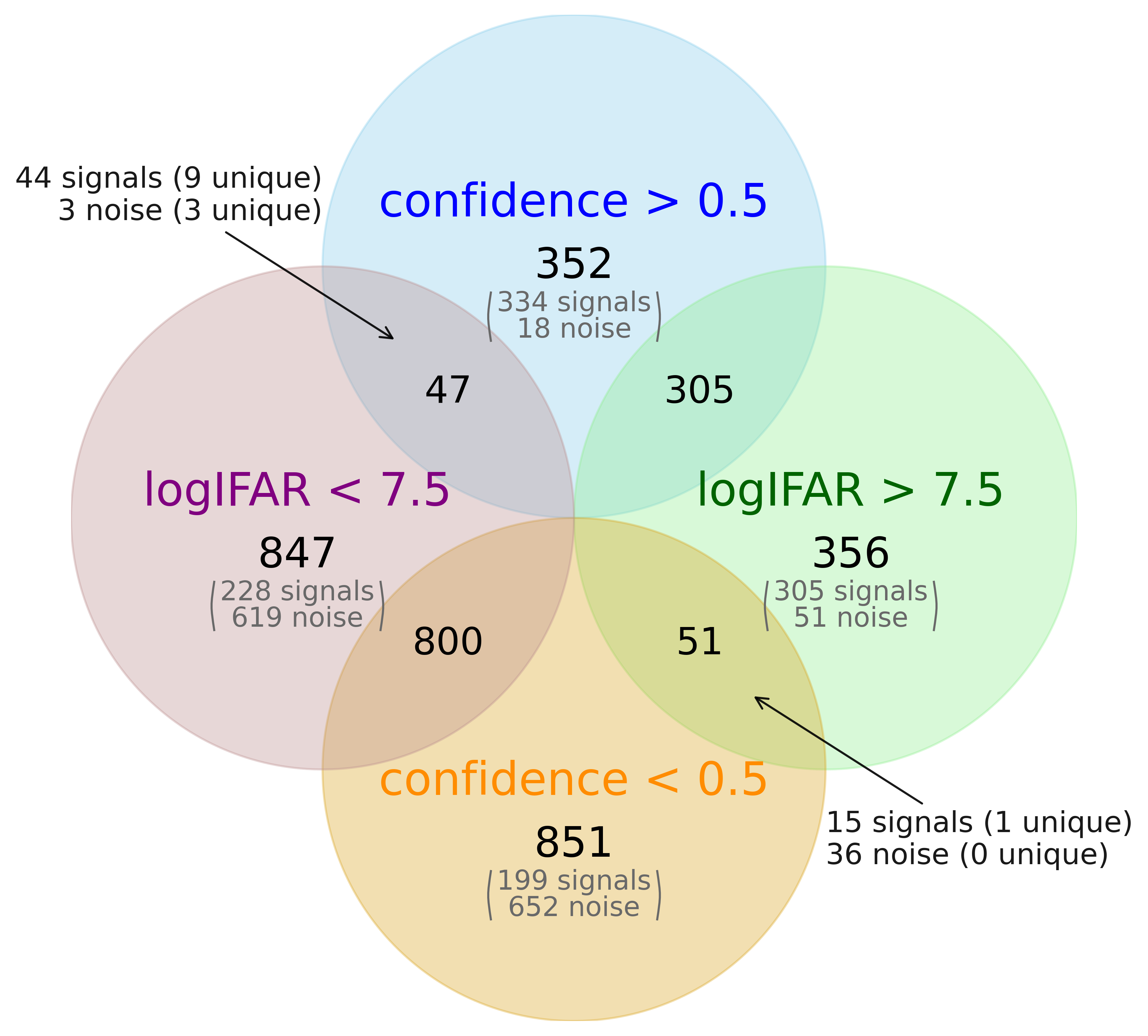} \\ 
  \end{tabular}
  \end{ruledtabular}
  \label{tab:venn_diagrams_mdc_llpic}
\end{table*}

Together, these results demonstrate that our framework remains robust to the choice of training dataset, with broadly consistent conditional confidence scores regardless of whether the \ac{MDC} or \ac{LLPIC} data are used for training and calibration. 
When the exchangeability assumption is deliberately broken (using \ac{MDC} for training and \ac{LLPIC} for testing), the \ac{ML} classifiers continue to outperform \maxifar in noise rejection, though XGBoost's down-rankings become less reliable under this distribution shift.
Importantly, XGBoost's up-rankings remain reliable under this distribution shift, correctly recovering \ac{BNS}-like candidates analogous to GW200311\_103121, and it retains the fewest false positives of all classifiers. 
Its down-rankings warrant some caution, but given its superior \ac{AUC} and correct up-rankings, we continue to use XGBoost as our preferred classifier in the following analysis.

\section{Observational results}\label{sec:obs_results}
\subsection{\gwtcFour}\label{sec:gwtc4}
We now apply our method for the first time to the O4a candidates in the updated \gwtcFour~\cite{GWTC4, GWTC5} catalogue, and investigate the resulting classifications. \gwtcFour is an updated version of the O4a catalogue~\citep{GWTC4}, incorporating a reanalysis of candidates by \pycbc and \gstlal with updated pipeline configurations~\citep{GWTC5}.

\begin{figure}
    \centering
    \includegraphics[width=0.5\textwidth]{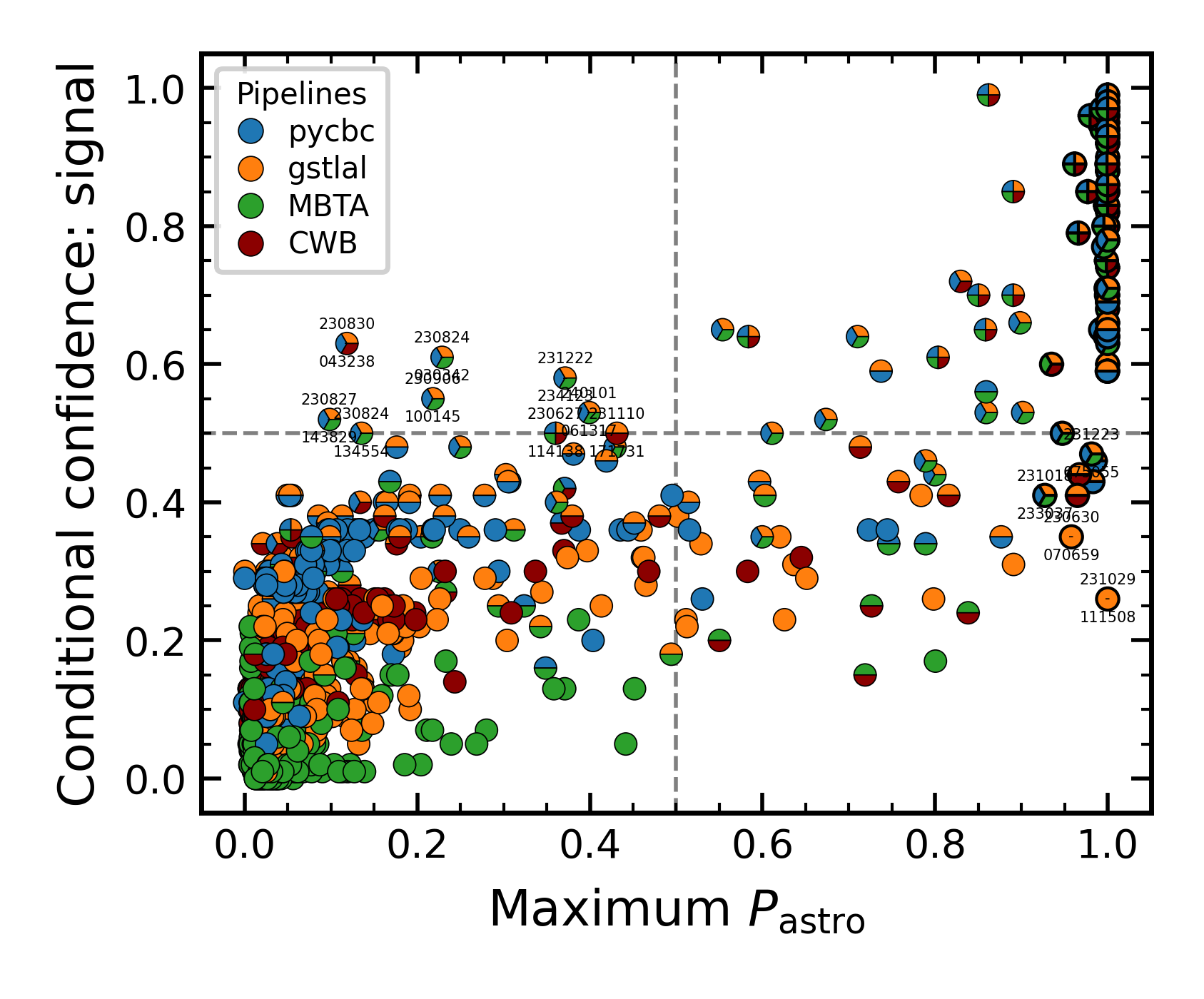}
    \caption{The conditional confidence in the signal label for the XGBoost classifier compared to the maximum \pastro for the O4a events in the \gwtcFour catalogue. The classifier is trained and \ac{CP} calibrated on the \ac{MDC} data.
    The dashed lines represent the conditional confidence and \pastro thresholds, respectively. 
    High-significance events (\ac{FAR} < 1 per year) are indicated by a bold marker outline. 
    We highlight names of candidates discussed in the text.} 
    \label{fig:gwtc4_xgboost}
\end{figure}

We note that only the two LIGO detectors were operating during O4a~\citep{GWTC4}. 
As the larger mock dataset, we use the \ac{MDC} as training and calibration data and restrict the features to the \ac{IFAR}, \ac{SNR}, and chirp mass as discussed in \cref{sec:method}. 
Similarly to O3, we find broad agreement between the conditional confidence and \maxpastro, see \cref{fig:gwtc4_xgboost} for the XGBoost classifier. 

In the bottom-right quadrant of \cref{fig:gwtc4_xgboost}, we find candidates with $\pastro \geq 0.5$ but conditional confidence below $0.5$. 
The majority are low-significance triggers (\ac{FAR} $\geq 1$ per year) identified by only one or two pipelines. 
There are also eight high-significance events found in this quadrant. 
One example is GW230630\_070659, a \gstlal offline-only candidate that was later found to have evidence of instrumental origin~\citep{GWTC4}. 
Follow-up data quality investigations~\citep{LIGO:2024kkz} found excess power consistent with scattered light in both detectors at the time of the event, rendering it inconsistent with a \ac{CBC} signal, and no parameter estimation was performed for the catalogue~\citep{GWTC4}. 
Independent analyses by \citet{Emma:2026zel} find the data consistent with a \ac{BBH} waveform but cannot provide evidence for an astrophysical origin, while \citet{Ashton:2026seh} conclude that a coincident pair of glitches is the likely explanation. 
Another example is GW231029\_111508, a single-detector high-mass event found by \gstlal only. 
Its high detector-frame chirp mass and short duration place it outside the parameter space for single-detector triggers of \mbta and \pycbc, respectively~\citep{LIGOScientific:2025yae, GWTC4}.

The remaining high-significance candidates are low-to-moderate \ac{SNR} events; 
GW230706\_104333 and GW230729\_082317 are low-mass events recovered by \pycbc and \gstlal, while GW231223\_075055 is similarly low-mass but recovered by three pipelines, with high significance from \pycbc only. 
GW231004\_232346 and GW231230\_170116 are intermediate-mass two-pipeline events recovered with high significance in \cwb only. 
GW231018\_233037 is a three-pipeline candidate, recovered with high significance in \mbta only and required glitch mitigation in both detectors~\cite{GWTC4}.
In all cases, the limited pipeline consensus may contribute to their classification as noise by the \ac{ML} combination framework, as correlated high-significance recovery across pipelines is more characteristic of confident signals in the training population.
A full list of all candidates in this quadrant, their \pastro and conditional confidence values, is presented in \cref{tab:gwtc4_down}.

In the top-left quadrant of \cref{fig:gwtc4_xgboost}, we identify several \ac{BBH} candidates that fall below the \pastro threshold but with conditional confidence values above threshold; the full list is given in \cref{tab:gwtc4_up}. 
The majority are three-pipeline events recovered by \gstlal, \mbta, and \pycbc, with low \ac{SNR} values in the range $7$--$8$ and low significance across all pipelines.
GW230627\_114138 is additionally recovered by \cwb, while GW231110\_171731 is found by \cwb and \gstlal only.
Their correlated recovery across multiple pipelines is consistent with the pattern of genuine signals in the training data, resulting in above-threshold conditional confidence.

\begin{figure*}
    \centering
    \subfigure[GW230627\_114138]{\includegraphics[width=0.32\textwidth]{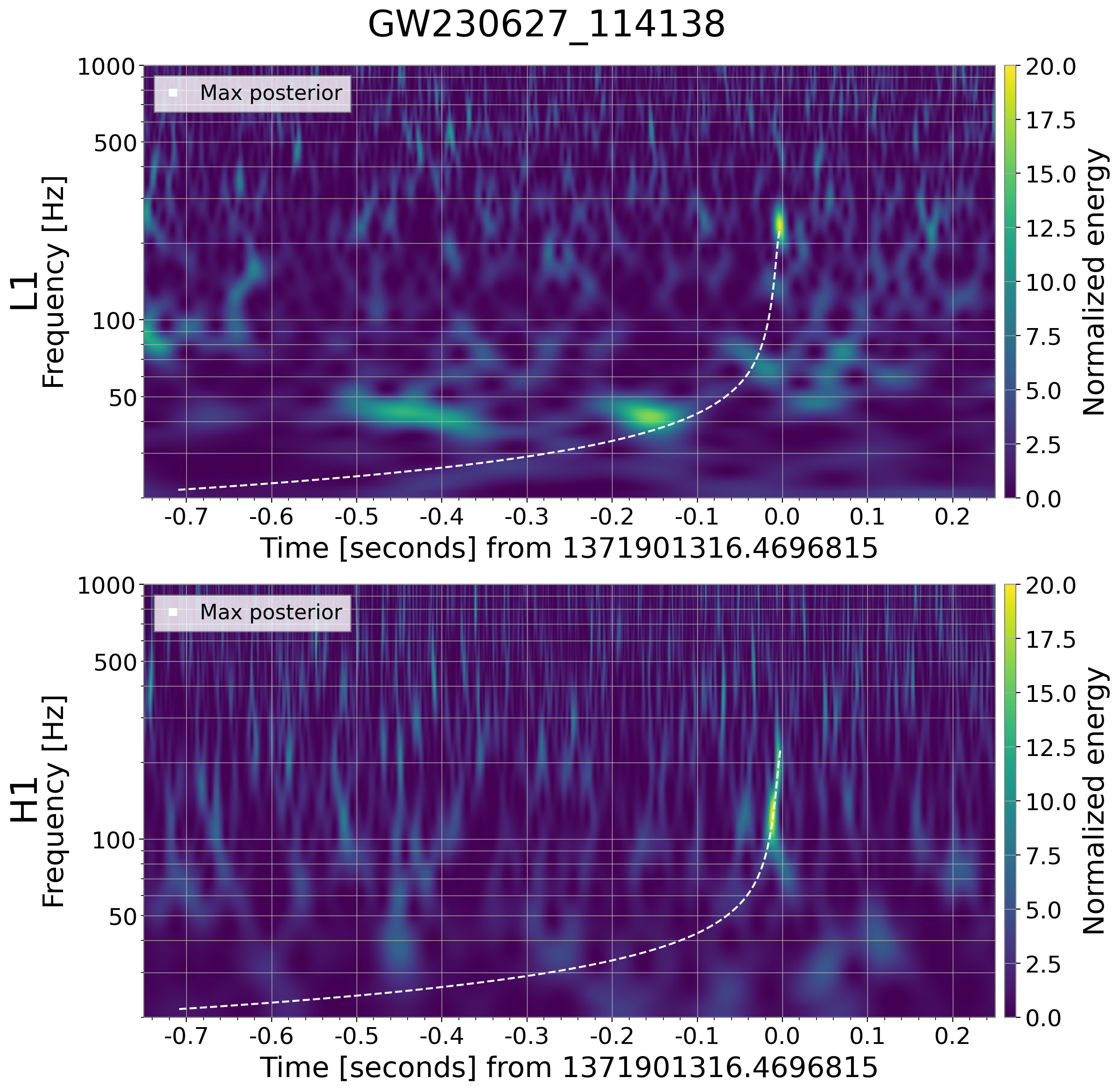} \label{fig:qscan_GW230627_114138}}
    \subfigure[GW230827\_143829]{\includegraphics[width=0.32\textwidth]{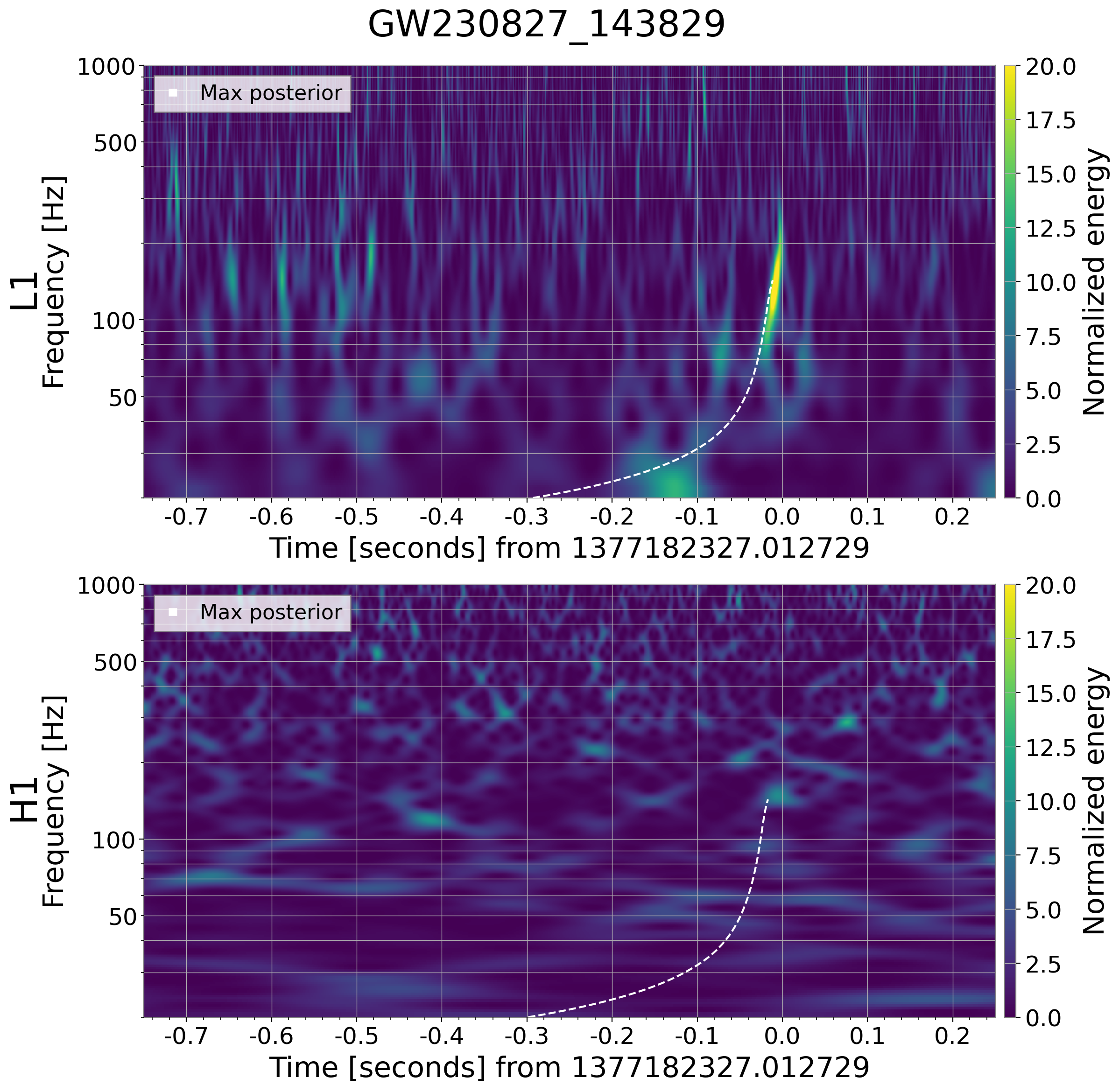}\label{fig:qscan_GW230827_143829}}
    \subfigure[GW231110\_171731]{\includegraphics[width=0.32\textwidth]{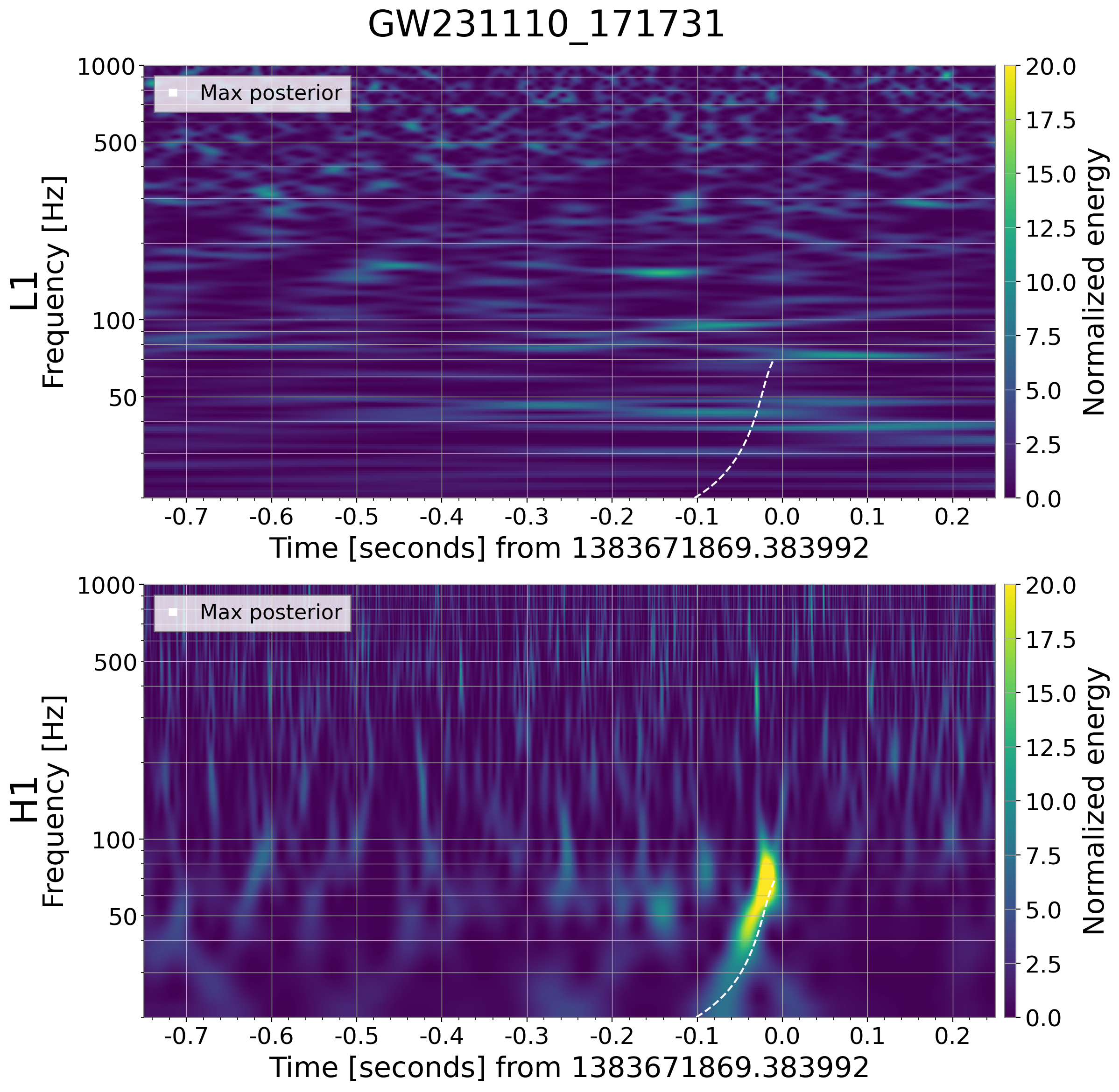}\label{fig:qscan_GW231110_171731}}
    \subfigure[GW230830\_043238]{\includegraphics[width=0.32\textwidth]{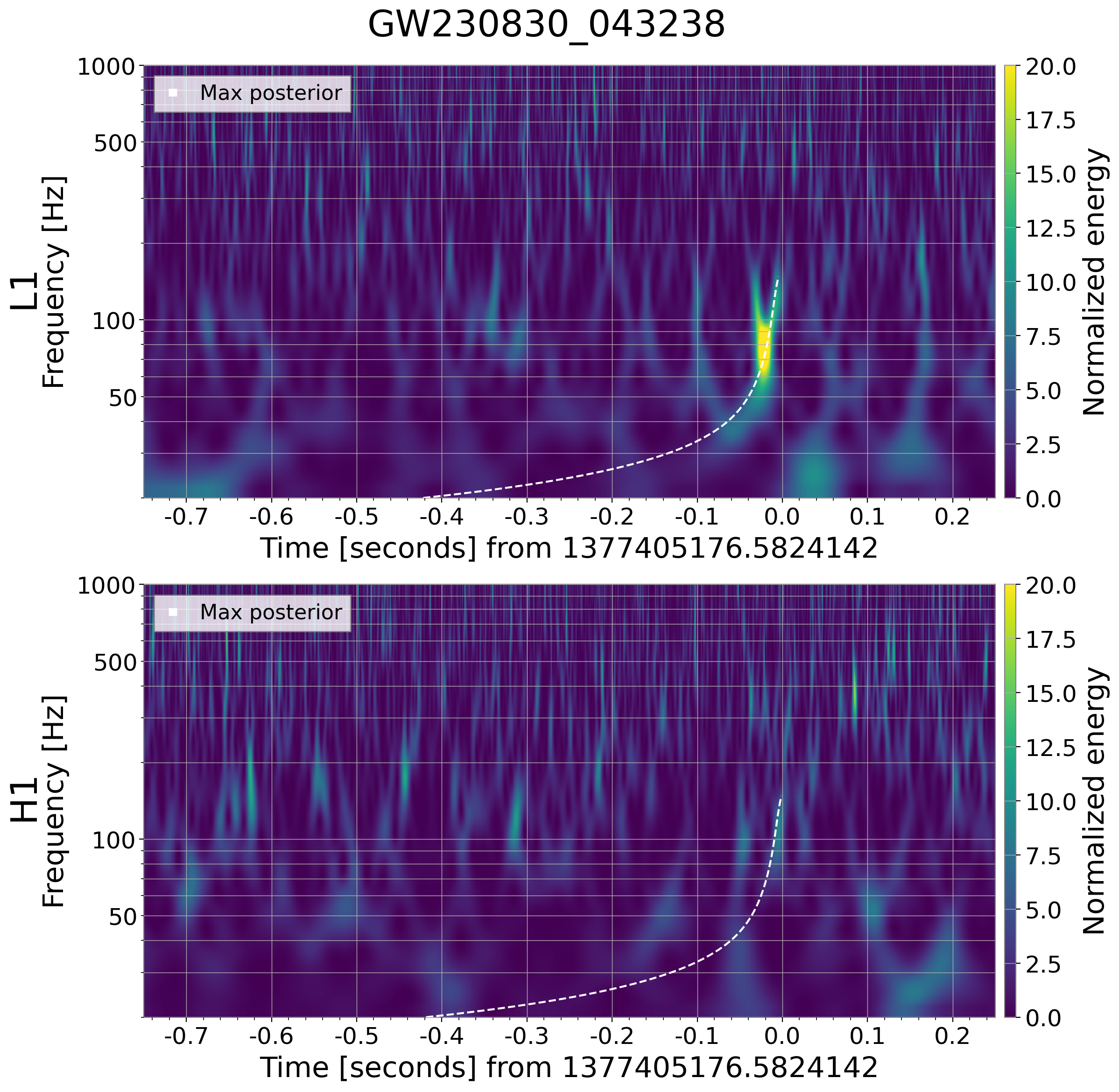}\label{fig:qscan_GW230830_043238}}
    \subfigure[GW230824\_134554]{\includegraphics[width=0.32\textwidth]{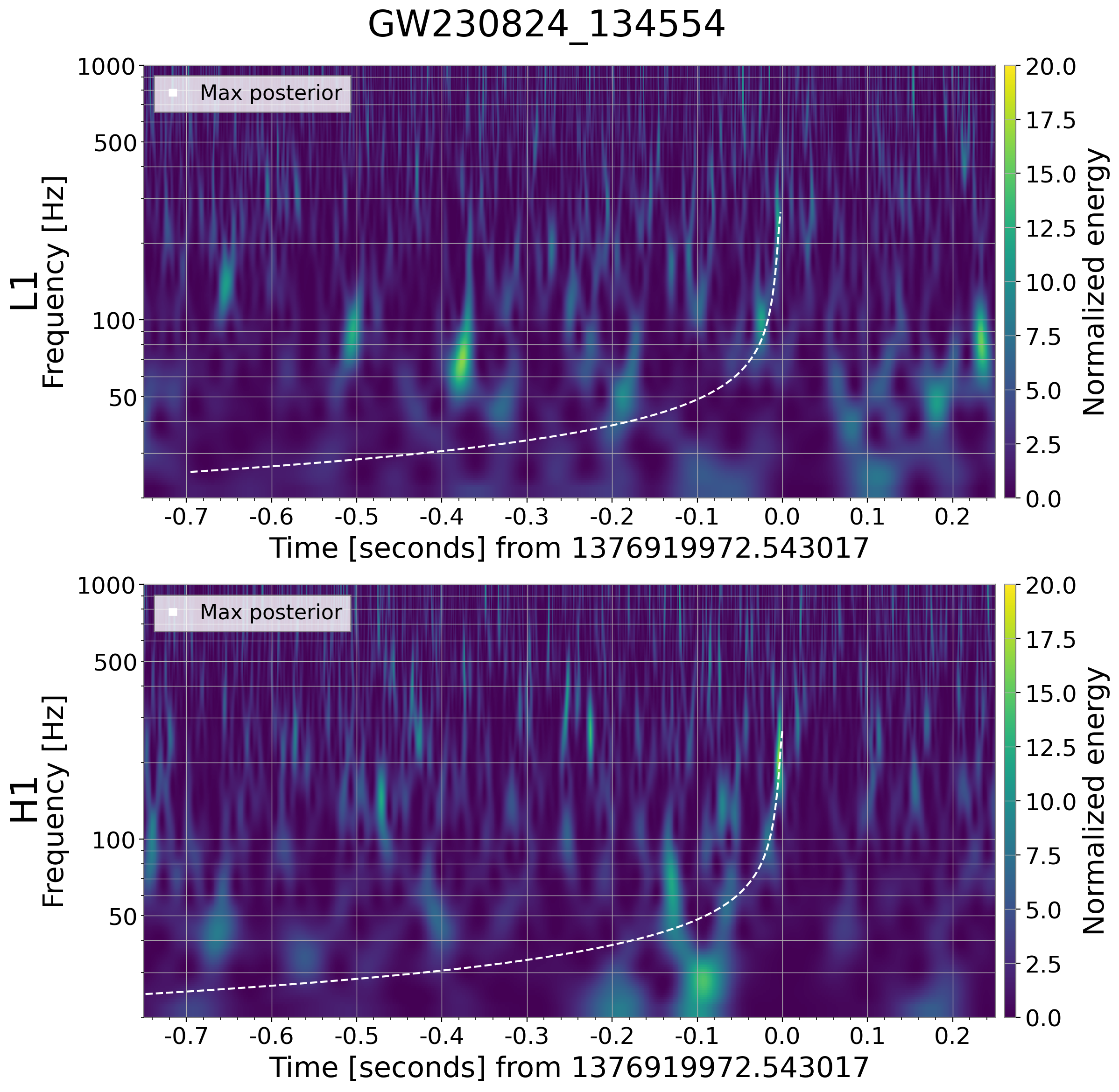}\label{fig:qscan_GW230824_134554}}
    \subfigure[GW230906\_100145]{\includegraphics[width=0.32\textwidth]{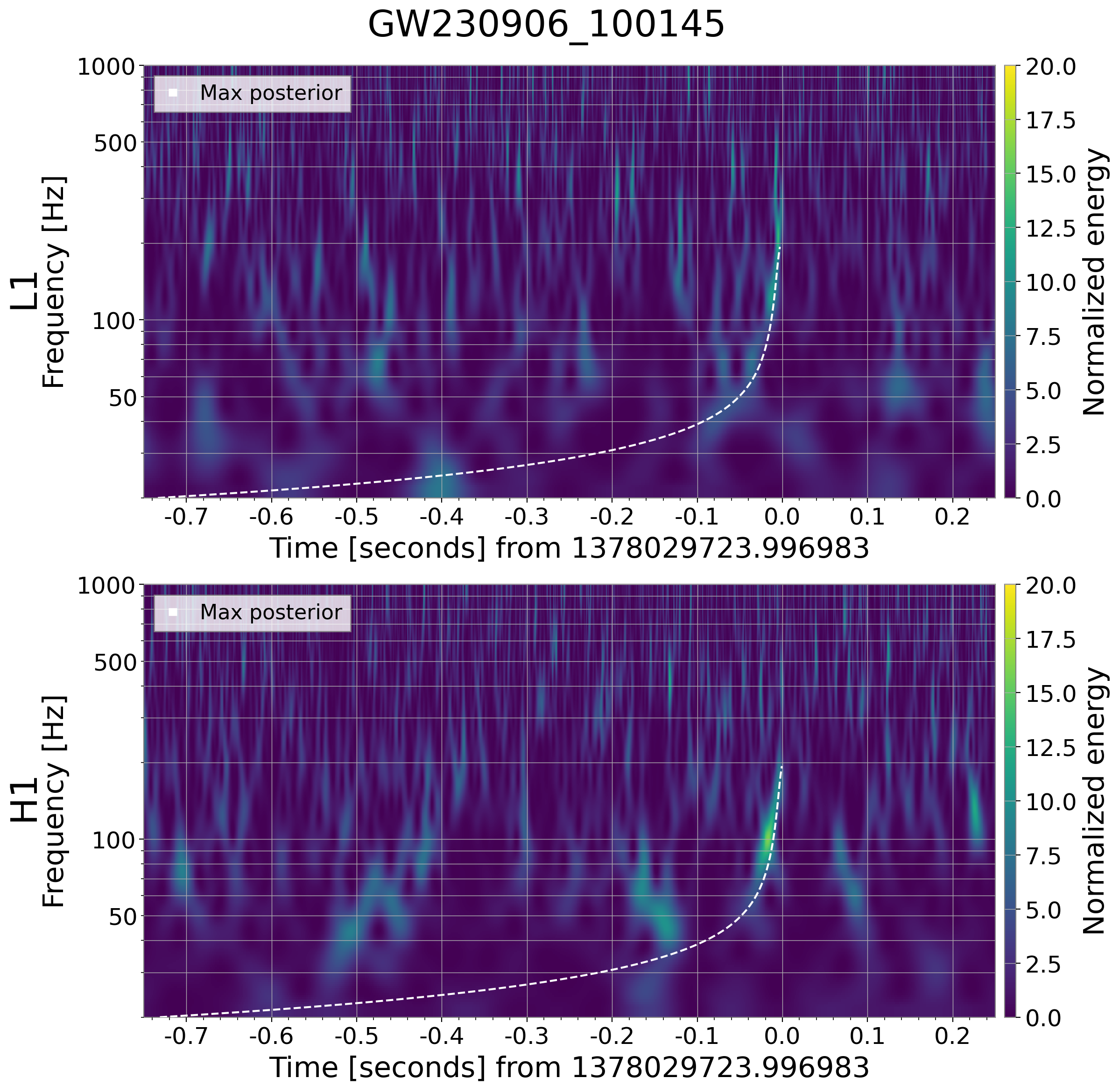}\label{fig:qscan_GW230906_100145}}
    \caption{Time-frequency spectrograms for some of the up-ranked O4a subthreshold candidates in \gwtcFour.
    The dashed white lines show the time-frequency evolution of the inspiral computed from the maximum likelihood parameters of the Bayesian parameter estimation. }
    \label{fig:qscan_gwtc4p1}
\end{figure*}

To further investigate these candidates, we also examine their time-frequency spectrograms in \cref{fig:qscan_gwtc4p1}.
To aid the interpretation, we also overlay the time-frequency evolution of the inspiral computed from the maximum-likelihood parameters of the Bayesian parameter estimation. 
We note that this only represents the peak of the posterior and carries uncertainty, particularly in the absolute merger time. 
At these low \ac{SNR} values, even genuine \ac{CBC} signals would produce only weak or indistinct excess power, so it is unsurprising that most spectrograms show no signal-like features.
Nevertheless, several candidates exhibit some evidence of excess power in the LIGO detectors.
GW230627\_114138, \cref{fig:qscan_GW230627_114138}, shows coincident excess power in both detectors, possibly resembling a chirp in H1, though L1 data quality issues complicate the interpretation.
GW230827\_143829, \cref{fig:qscan_GW230827_143829}, and GW231110\_171731, \cref{fig:qscan_GW231110_171731}, each show a chirp-like feature in L1 and H1, respectively, while GW230830\_043238, \cref{fig:qscan_GW230830_043238}, shows transient power in L1 around the trigger time, resembling a chirp-like feature overlapping transient noise. 
GW230824\_134554 and GW230906\_100145 show faint, possibly chirp-like, coincident excess power in both detectors, see \cref{fig:qscan_GW230824_134554,fig:qscan_GW230906_100145}. 
The remaining candidates up-ranked by the XGBoost classifier show no discernible coincidence excess power. 
All additional time-frequency spectrograms can be found in the accompanying data release \citep{Malz2026:data_release}.

Next, we estimate their source parameters using the Bayesian inference code \bilby~\cite{Ashton:2018jfp}. 
Following the same approach as in \gwtcFive~\citep{GWTC5}, we use the \texttt{IMRPhenomXPHM}~\citep{Pratten:2020ceb, Colleoni:2024knd} waveform model with the \texttt{DYNESTY} nested sampling package~\citep{Speagle:2019ivv}. 
While parameter estimation cannot be used to determine whether a candidate is a real signal, it can provide further support for whether the properties of a candidate appear consistent with a \ac{CBC} signal. 

\begin{figure}
    \centering
    \includegraphics[width=0.5\textwidth]{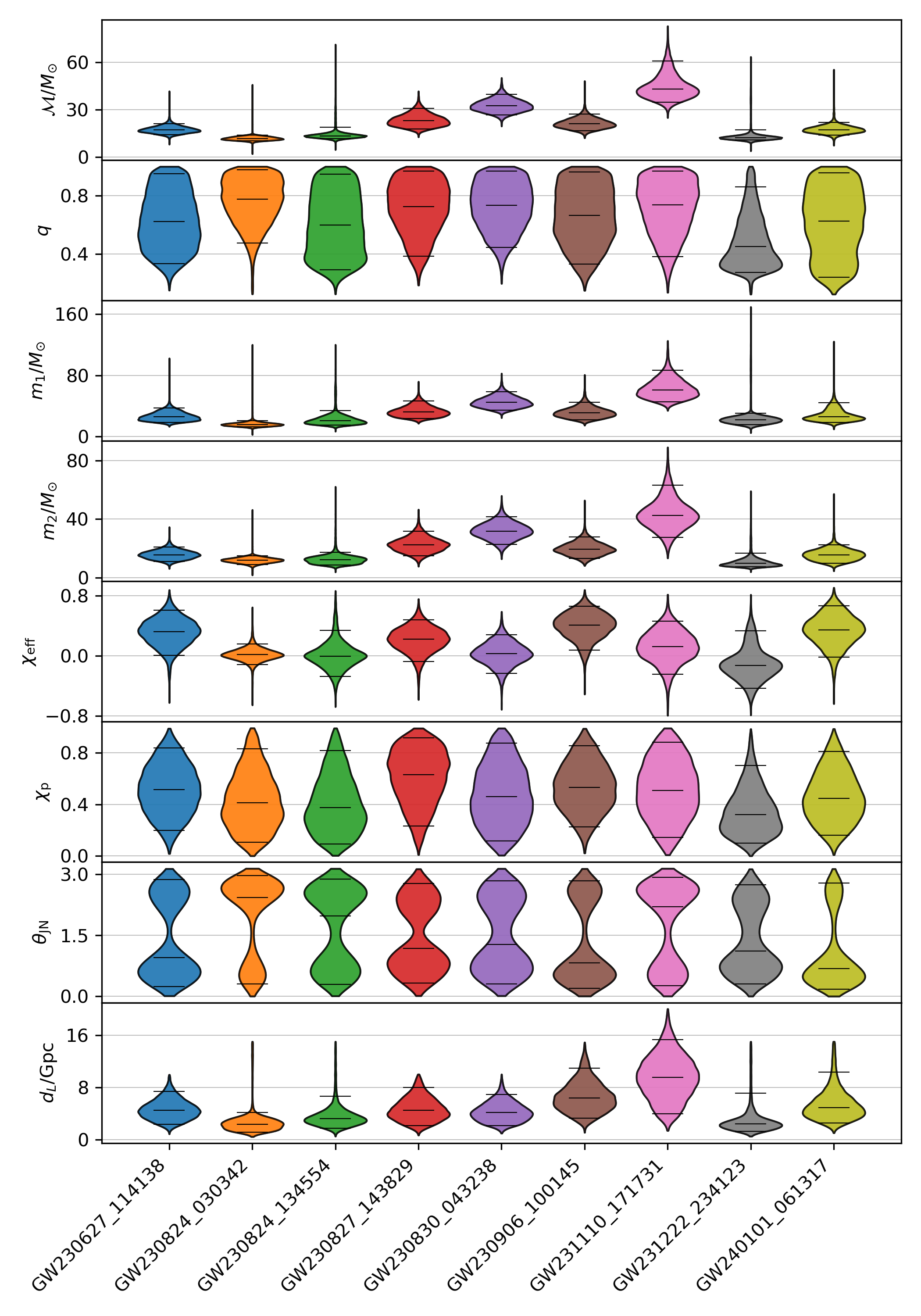}
    \caption{Violin plots of estimated posterior distributions for O4a events in \gwtcFour with subthreshold \pastro but conditional confidence above threshold for XGBoost.
    The parameters shows are the source-frame chirp mass $\mathcal{M}$, mass ratio $q$, the source-frame component masses $m_1$ and $m_2$, effective inspiral spin $\chi_\mathrm{eff}$, effective precession spin $\chi_\mathrm{p}$, source inclination angle $\theta_{JN}$, and luminosity distance $D_L$. 
    Each violin displays the posterior distribution, with the median and the 90\% interval indicated. } 
    \label{fig:gwtc4_violins}
\end{figure}

In \cref{fig:gwtc4_violins}, we show the resulting posteriors for a selection of parameters: source-frame chirp mass $\mathcal{M}$, mass ratio $q$, the source-frame component masses $m_1$ and $m_2$, effective inspiral spin $\chi_\mathrm{eff}$, effective precession spin $\chi_\mathrm{p}$, source inclination angle $\theta_{JN}$, and luminosity distance $D_L$ (see Table 3 in \citet{LIGOScientific:2025hdt} for definitions). 
All of the events return source parameters consistent with stellar-mass \ac{BBH} signals, with GW231110\_171731 having the highest inferred component masses in the set. 
The majority of events are consistent with negligible aligned spins ($\chi_\mathrm{eff} \approx 0$); however, several events, GW230627\_114138, GW230827\_143829, GW230906\_100145, and GW240101\_061317 show support for positive $\chi_\mathrm{eff}$, suggestive of preferentially aligned spins. 
The mass ratios are broadly consistent with approximately equal mass binaries for most candidates; however, GW231222\_234123 shows stronger support for unequal masses. 
The luminosity distances span $\sim2$--$10\,\mathrm{Gpc}$, placing some of these among the most distant \ac{BBH} candidates observed. 
For comparison, the most distant confirmed \ac{GW} event detected to date is GW190403\_051519 at $d_L = 8.28^{+6.72}_{-4.29}\,\mathrm{Gpc}$~\citep{GWTC2-1}, while the most distant candidate here is GW231110\_171731 at $d_L = 9.50^{+5.81}_{-5.54}\,\mathrm{Gpc}$.
Assuming these are actual astrophysical signals, the low significance of these events could at least partly be explained as a consequence of their large distances. 
A table summarising the median and $90\%$ credible intervals for events and parameters shown in \cref{fig:gwtc4_violins} is included in \cref{app:posterior_tabs} in \cref{tab:gwtc4_posteriors}. 

Furthermore, we can investigate the features of each event using SHAP values, and find that all candidates appear signal-like, following the method and discussion in \cref{sec:shap}. 
The waterfall plots for all events are included in the data release \citep{Malz2026:data_release}.

Another candidate of interest from the fourth observing run is GW231109\_235456, a sub-threshold \gstlal-only \ac{BNS} which was later found with high significance by the \gstlal sub-threshold search~\cite{Niu:2025nha}. In our analysis, based on the original pipeline outputs (\ac{SNR} $9.64$, \logIFAR $4.97$, chirp mass $1.3$), the candidate obtains conditional confidence below threshold for all of our classifiers.
The reanalysis finds a \ac{FAR} of 1 per 50 years (which corresponds to a \logIFAR of $9.2$), \ac{SNR} of $9.7$, and detector frame chirp mass of $1.3$~\cite{Niu:2025nha}.
Using these new features, we obtain increased conditional confidence scores still below threshold for the \ac{LR} (from $0.06$ to $0.23$) and XGBoost (from $0.28$ to $0.36$) classifiers. For \ac{MLP}, the confidence increases from $0.3$ to $0.61$, bringing the event comfortably above the threshold. However, as our training data does not include any events from the \gstlal sub-threshold search, this reanalysis violates exchangeability and is purely an illustrative observation. 

In summary, the \ac{ML} framework down-ranks candidates that lack correlated multi-pipeline support, including single-pipeline detections and events with known data quality issues, even where individual pipeline significance is high.
Conversely, it up-ranks low-\ac{SNR} events recovered coherently by multiple pipelines, where no single pipeline reports high significance individually.
\ac{SHAP} analysis confirms that for these up-ranked candidates the classification is dominated by the degree of correlated multi-pipeline recovery, the characteristic  most strongly associated with signals in the training data.
For the up-ranked candidates, their estimated source parameters are broadly consistent with \ac{BBH} mergers observed by the \ac{LVK} at large luminosity distances, lending further plausibility to an astrophysical origin.

\subsection{\gwtcFive}\label{sec:gwtc5}
Next, we apply our method to the O4b candidates in the \gwtcFive~\cite{GWTC5} catalogue, and investigate the resulting classifications. 
We continue using the \ac{MDC} as training and calibration data and restricting the features to the \ac{IFAR}, \ac{SNR}, and chirp mass of each pipeline. 

\begin{figure}
    \centering
    \includegraphics[width=0.5\textwidth]{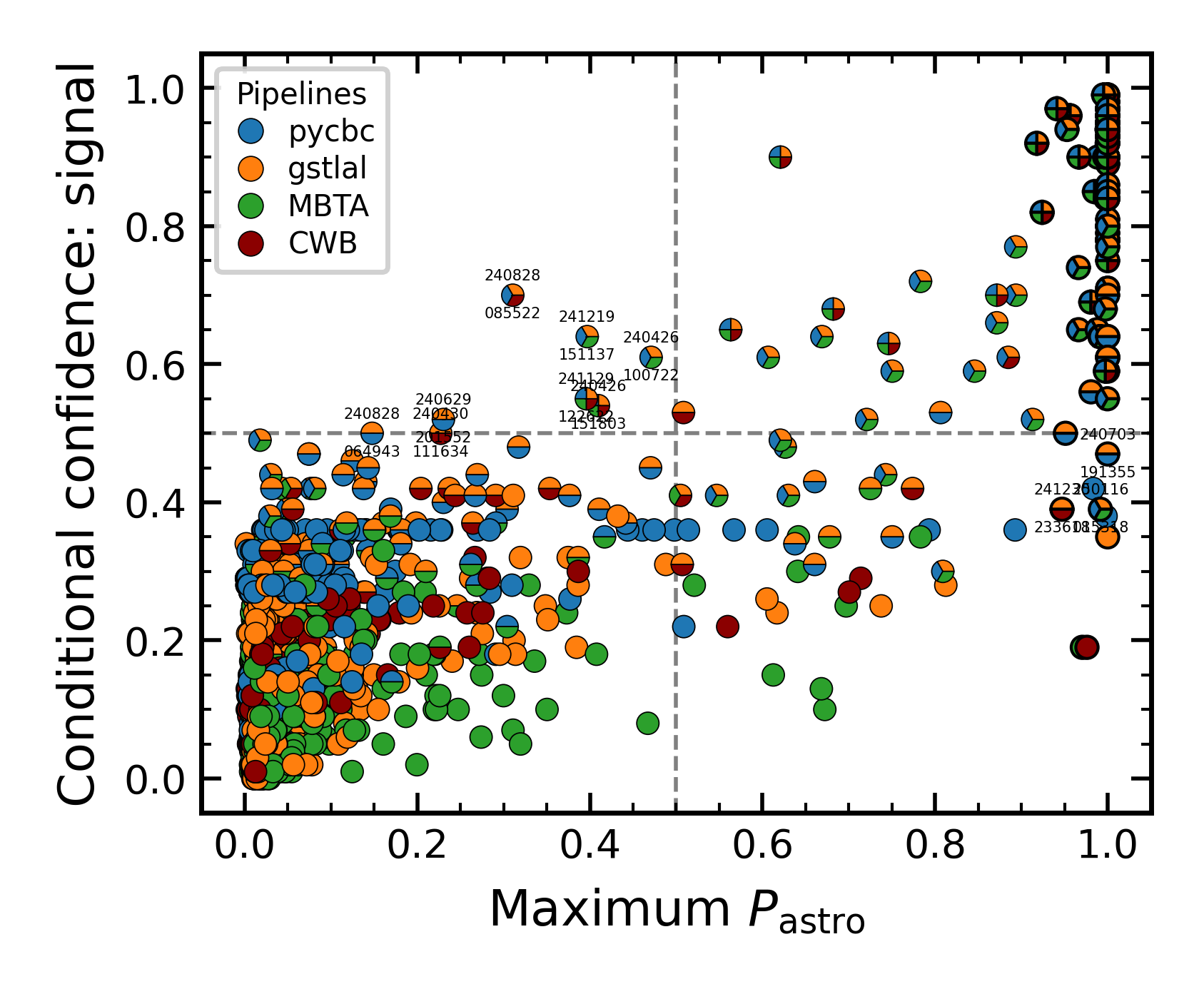}
    \caption{The conditional confidence in the signal label for the XGBoost classifier compared to the maximum \pastro for the O4b events in the \gwtcFive catalogue. 
    The classifier is trained and \ac{CP} calibrated on the \ac{MDC} data.
    The dashed lines represent the conditional confidence and \pastro thresholds, respectively. 
    High-significance events (\ac{FAR} < 1 per year) are indicated by a bold marker outline. 
    We highlight names of candidates discussed in the text.} 
    \label{fig:gwtc5_xgboost}
\end{figure}

Similarly to the previous analyses, we find broad agreement between the conditional confidence and \maxpastro, but a few events where the two methods disagree, see \cref{fig:gwtc5_xgboost} for the XGBoost classifier. 

In the bottom-right quadrant of \cref{fig:gwtc5_xgboost}, we find candidates above the \pastro threshold but with conditional confidence below $0.5$. 
A full list of all candidates in this quadrant, their \pastro and conditional confidence values, is presented in \cref{tab:gwtc5_down}.
The majority are low-significance triggers identified by only one or two pipelines. 

We also find six high-significance \ac{BBH} candidates in this quadrant. 
GW240526\_093944 and GW241230\_084504 are single-pipeline candidates found only by \mbta and \gstlal, respectively.
GW240930\_234614 is recovered exclusively by \cwb, which may reduce confidence in the \ac{BBH} nature of the candidate, though \cwb could potentially identify signals affected by physical effects neglected by matched-filter searches, such as precession and eccentricity~\citep{GWTC5}. 
Additionally, parameter estimation reveals pronounced model-dependent differences across waveform approximants for this event~\citep{GWTC5}.
GW240703\_191355 and GW241230\_233618 are two-pipeline events.
GW240703\_191355 is detected in the Livingston and Virgo detectors only, while GW241230\_233618 has the highest total mass of any O4b candidate in the catalogue ($M \approx 116\,M_\odot$) as well as support for negative effective inspiral spin ($\chi_\mathrm{eff} \approx -0.14$)~\citep{GWTC5}. 
GW250116\_015318 is a three-pipeline event, but the pipeline significances are highly discrepant and only \pycbc recovers it with high significance and above the \pastro threshold; notably only XGBoost down-ranks this candidate, while the other \ac{ML} classifiers assign above threshold conditional confidence. 
In all cases, the limited or inconsistent pipeline consensus results in below-threshold confidence from the \ac{ML} framework. 

In the top-left quadrant of \cref{fig:gwtc5_xgboost}, we identify several \ac{BBH} candidates that fall below the \pastro threshold but with conditional confidence values above threshold. 
The full list is given in \cref{tab:gwtc5_up}. 
The majority of these are three- or four-pipeline candidates with low \ac{SNR}. 
We also note that four of the eight up-ranked candidates occur in pairs on the same day: GW240426\_100722 and GW240426\_151803 are separated by approximately five hours, while GW240828\_064943 and GW240828\_085522 are separated by approximately two hours.
The close temporal separations are notable, and if either pair is confirmed as astrophysical, the origin of the coincidence would warrant further investigation. 

\begin{figure*}
    \centering
    \subfigure[GW240828\_085522]{\includegraphics[width=0.32\textwidth]{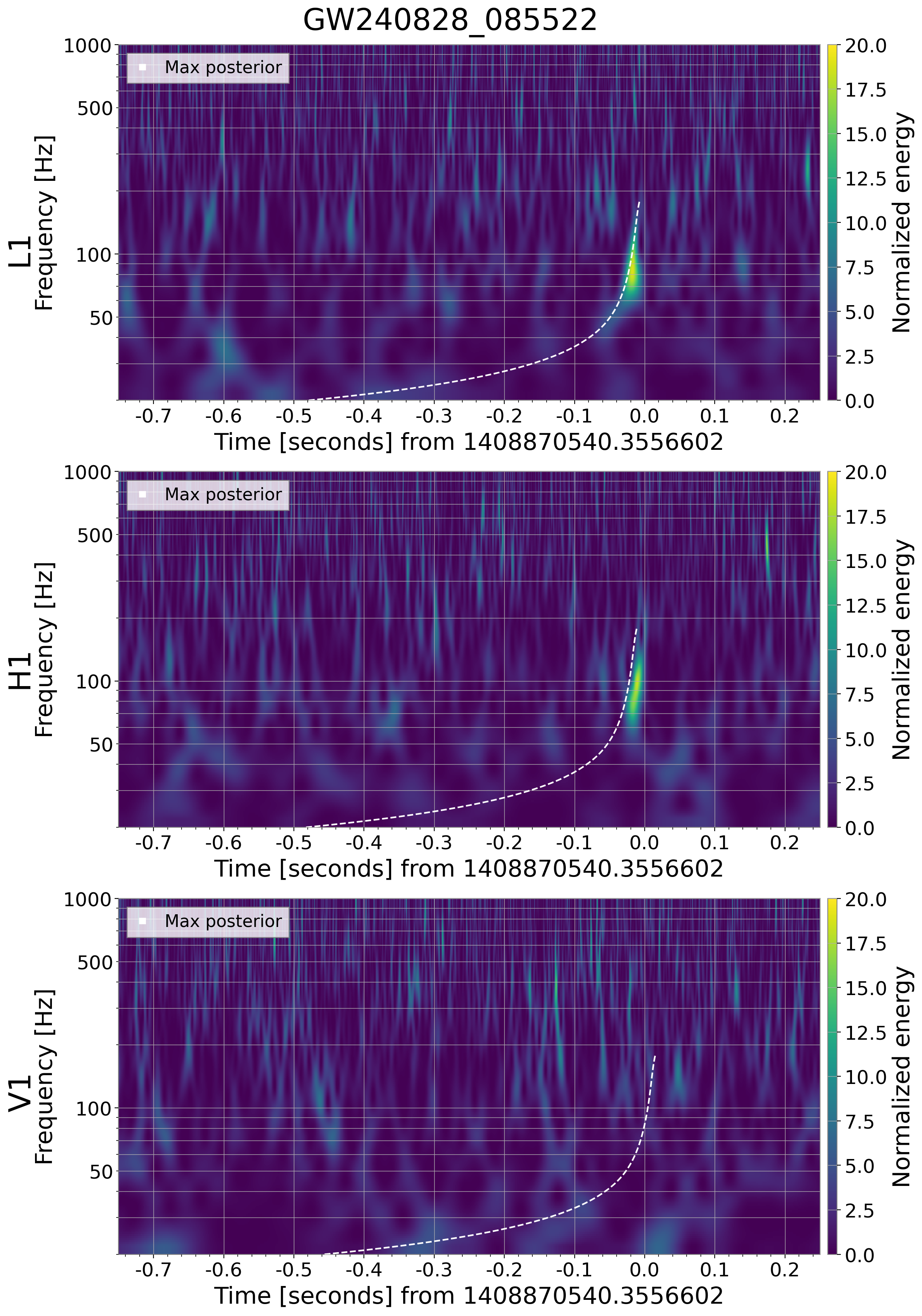}\label{fig:qscan_GW240828_085522}}
    \subfigure[GW240430\_111634]{\includegraphics[width=0.32\textwidth]{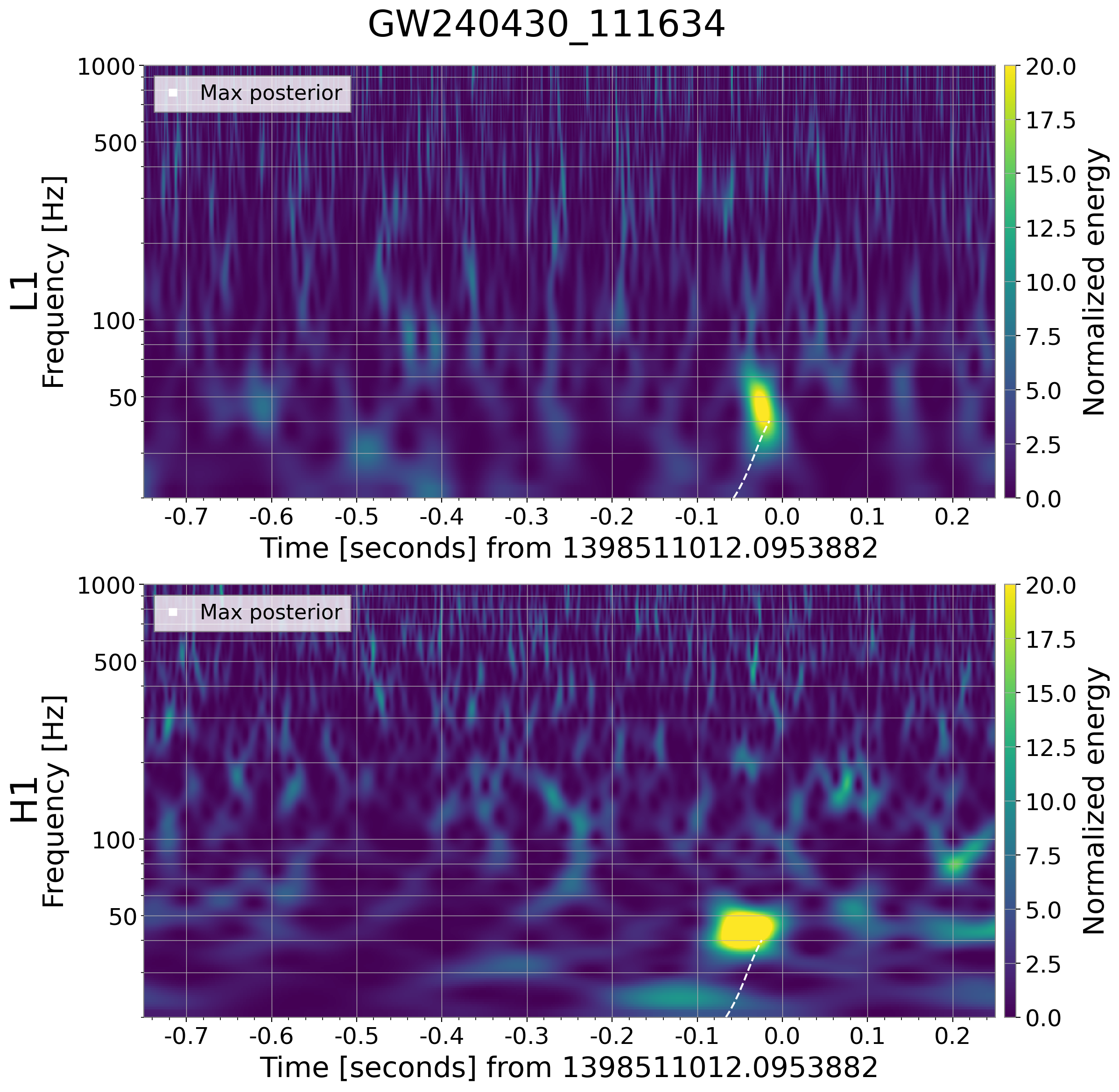}\label{fig:qscan_GW240430_111634}}
    \subfigure[GW241129\_122622]{\includegraphics[width=0.32\textwidth]{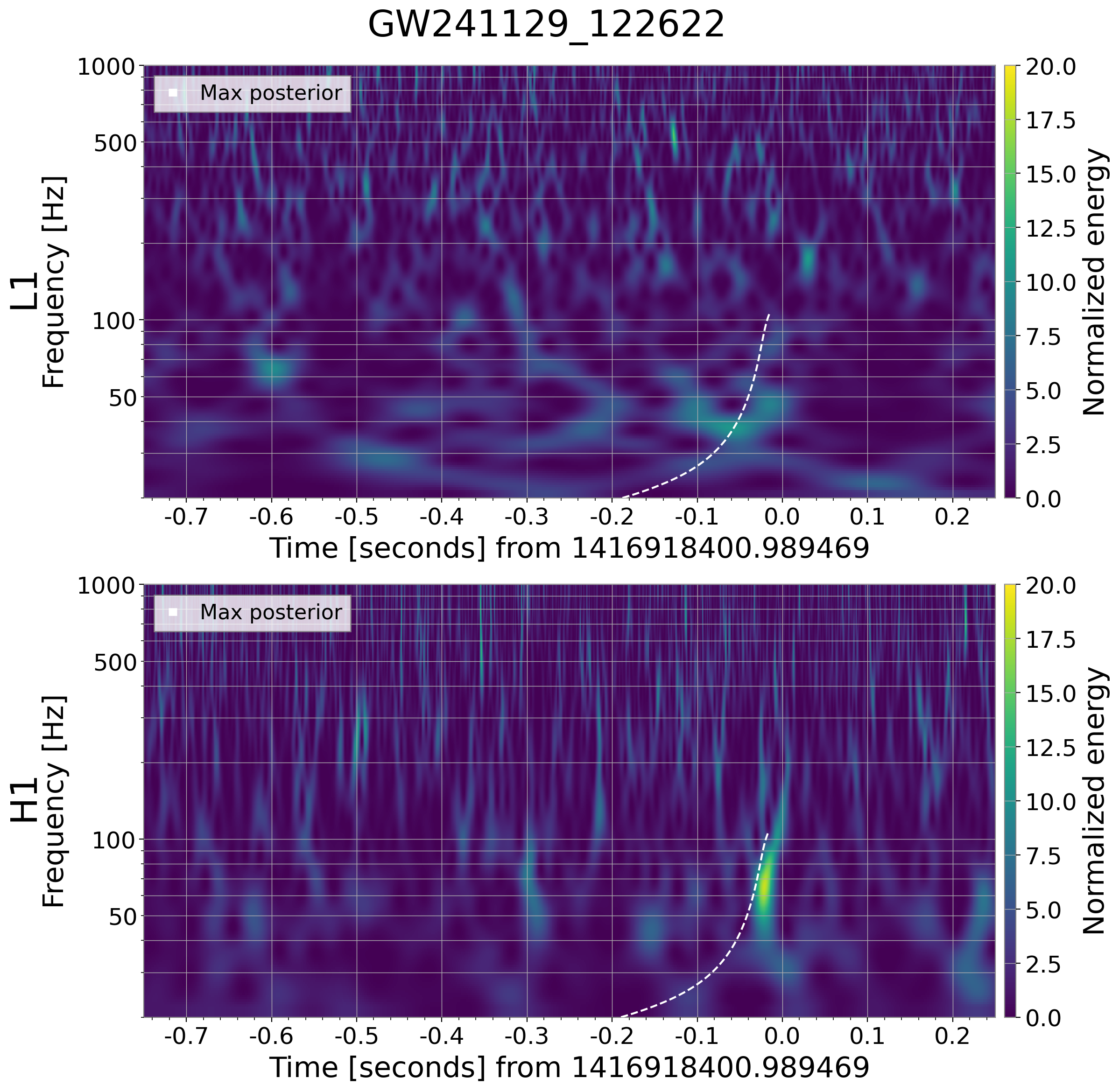}\label{fig:qscan_GW241129_122622}}
    \caption{Time-frequency spectrograms for some of the up-ranked O4b subthreshold candidates in \gwtcFive.
    The dashed white lines show the time-frequency evolution of the inspiral computed from the maximum likelihood parameters of the Bayesian parameter estimation. }
    \label{fig:qscan_gwtc5}
\end{figure*}

Investigating the time-frequency spectrograms, we observe a chirp-like feature coincident in both LIGO detectors for GW240828\_085522, see \cref{fig:qscan_GW240828_085522}. 
GW240430\_111634, \cref{fig:qscan_GW240430_111634}, shows coincident excess power in both LIGO detectors before the merger time. 
GW241129\_122622, \cref{fig:qscan_GW241129_122622} shows excess power around the time of merger in H1.

The other up-ranked events show no signal-like features, but this is not unexpected as they are all low \ac{SNR} ($7$--$8$) candidates and even genuine \ac{CBC} signals would produce only weak or indistinct excess power at these \ac{SNR} values.

\begin{figure}
    \centering
    \includegraphics[width=0.5\textwidth]{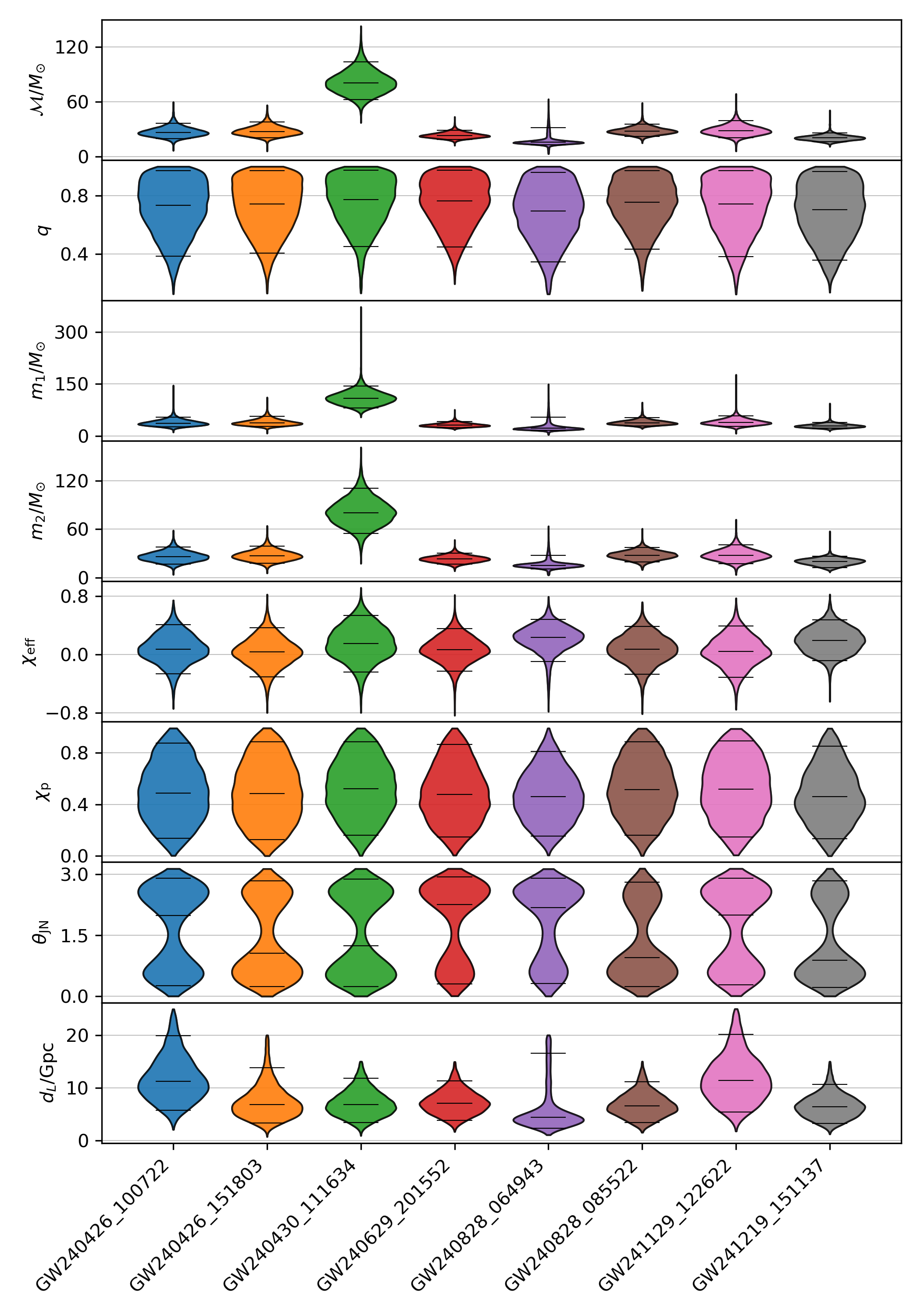}
    \caption{Violin plots of estimated posterior distributions for O4b events in \gwtcFive with subthreshold \pastro but conditional confidence above threshold for XGBoost.
    The parameters shows are the source-frame chirp mass $\mathcal{M}$, mass ratio $q$, the source-frame component masses $m_1$ and $m_2$, effective inspiral spin $\chi_\mathrm{eff}$, effective precession spin $\chi_\mathrm{p}$, source inclination angle $\theta_{JN}$, and luminosity distance $D_L$. 
    Each violin displays the posterior distribution, with the median and the 90\% interval indicated. } 
    \label{fig:gwtc5_violins}
\end{figure}

Similarly to the previous section, we estimate the source parameters for all candidates up-ranked by the XGBoost classifier, as shown in \cref{fig:gwtc5_violins}. 
All candidates return posteriors consistent with \ac{BBH} signals, with mass ratios broadly consistent with near-equal mass binaries. 
Chirp masses are mostly in the stellar-mass range, with the exception of GW240430\_111634 ($\mathcal{M} \approx 80\,M_\odot$), which lies in the high-mass regime. 
With a total mass of $M = 190.15^{+52.67}_{-42.57}\,M_\odot$\footnote{This value had been corrected from a previous version.}, GW240430\_111634 would, if confirmed astrophysical, be one of the most massive \ac{BBH} candidates detected to date (though not as massive as GW231123\_135430, with $M = 236^{+29}_{-48}\,M_\odot$ \citep{GWTC4, LIGOScientific:2025rsn}).
Notably, the primary component mass of GW240430\_111634 ($m_1 \approx 109\,M_\odot$) falls in the intermediate-mass black hole regime. 

Effective inspiral spins are broadly consistent with zero for all candidates, though GW240828\_064943 and GW241219\_151137 show mild support for positive $\chi_\mathrm{eff}$, suggestive of preferentially aligned spins.
Luminosity distances span $4$--$11 \,\mathrm{Gpc}$, suggesting that, if of astrophysical origin, some of these would be among the most distant \ac{BBH} candidates observed, with the most distant being GW241129\_122622 at $d_L = 11.36^{+8.83}_{-5.99}\,\mathrm{Gpc}$, followed by GW240426\_100722 at $d_L = 11.26^{+8.61}_{-5.54}\,\mathrm{Gpc}$.
Both of these exceed the most distant up-ranked O4a candidate GW231110\_171731 at $d_L = 9.50^{+5.81}_{-5.54}\,\mathrm{Gpc}$.
The median parameter values and $90\%$ credible intervals are tabulated in \cref{tab:gwtc5_posteriors}.

Lastly, we investigate the features of each event using SHAP values, and find that all up-ranked candidates appear signal-like, following the method and discussion in \cref{sec:shap}. The plots are included in the data release \citep{Malz2026:data_release}. 

As in the O4a analysis, the up-ranked candidates in O4b share a common profile of low-\ac{SNR} events recovered coherently by multiple pipelines, with source parameters consistent with \ac{BBH} mergers at large luminosity distances. 
Their correlated multi-pipeline recovery is the pattern most strongly associated with genuine signals in the training data, and provides the basis for the above-threshold conditional confidence assigned by the \ac{ML} framework.

\section{Discussion and conclusion}\label{sec:discussion}
In this work, we have presented a systematic evaluation of our framework \citep{Ashton:2025jhn} for combining the outputs of multiple \ac{GW} search pipelines using supervised \ac{ML} together with \ac{CP}.
Rather than relying on simple combination rules such as the maximum \ac{IFAR} or \pastro, our approach learns an optimal mapping from multi-pipeline information to a calibrated confidence measure, trained on data where the ground truth is known.
This enables the model to exploit correlations between pipelines and to adapt naturally to parameter-space-dependent sensitivities.

Applying our framework to real data from O3 and O4, we find several events with subthreshold \pastro values but \ac{CP} conditional confidence above threshold, most notably the marginal \ac{BNS} candidate GW200311\_103121.
If real, it is only the third \ac{BNS} signal detected.
If our framework were applied in real time, an up-ranking of a subthreshold candidate such as this could lead to electromagnetic follow-up and the possibility of a multi-messenger detection, enabling the rich multi-messenger science that would follow.
We also consistently up-rank the two marginal \ac{NSBH} candidates GW190426\_152155 and GW190531\_023648 from \gwtcTwoPointOne, whose low \pastro values reflect prior uncertainty on the \ac{NSBH} merger rate rather than weak pipeline significance.

When applied to O4a and O4b candidates, our framework identifies several subthreshold \ac{BBH} candidates with elevated confidence.
The up-ranked candidates share a common profile: low-\ac{SNR} events in the range $7$--$8$, recovered coherently by multiple pipelines, with no single pipeline reporting high individual significance.
If of astrophysical origin, their source parameters are consistent with \ac{BBH} mergers spanning stellar to intermediate masses, at luminosity distances of $\sim2$--$11\,\mathrm{Gpc}$, which would place some among the most distant \ac{GW} events observed to date.
Notably, the O4b candidate GW240430\_111634 has an estimated total mass of $\sim190\,M_\odot$, which would be among the most massive \ac{BBH} candidates detected to date~\citep{GWTC4, LIGOScientific:2025rsn}.
Several candidates also show chirp-like features in their time-frequency spectrograms, lending further support to a potential astrophysical origin.
Their sub-threshold individual significance is plausibly explained by their large distances, and their correlated multi-pipeline recovery is what distinguishes them from noise in the eyes of the classifier.

We validated the reliability of these results through a series of robustness checks.
The framework produces broadly consistent results across different \ac{ML} classifier architectures, indicating that the findings are driven by the multi-pipeline information rather than model-specific choices.
\ac{CP} is classifier-agnostic, so the framework is readily compatible with new classifiers as they become available.
Results are also largely consistent when training and calibrating on two different mock datasets, the \ac{MDC} and the smaller \ac{LLPIC}, despite differences in dataset size and observing run configuration.
A \ac{SHAP} analysis confirms that the classifiers learn physically meaningful patterns, with detection significance features dominating and results consistent between the mock and real datasets.
Based on its superior false positive control under distribution shift, we identify XGBoost as our preferred classifier.

A key consideration is whether the up-ranked events can be trusted.
The validity of \ac{CP} rests on the exchangeability of the calibration and test data, which cannot be guaranteed when training on mock data and testing on real events.
We also note that the \ac{MDC} has known limitations as a proxy for real data: the injected event rate is far higher than astrophysically expected, the pipelines were still under development, and the minimum \ac{IFAR} threshold restricts the available noise statistics.
A more realistic \ac{MDC} with lower \ac{IFAR} thresholds would strengthen the exchangeability assumption and improve confidence in the results.
Nevertheless, while exchangeability cannot be guaranteed, we have empirically assessed the reliability of the framework through a series of consistency checks.

First, when training and calibrating on the \ac{MDC} and evaluating on the \ac{LLPIC}, mimicking the distribution shift between mock and real data, all \ac{ML} classifiers achieve substantially lower false positive rates than \maxifar, with XGBoost performing best.
Moreover, most noise events misclassified by \maxifar obtain sub-threshold confidence from all \ac{ML} classifiers, suggesting the framework is conservative where it matters most.

Second, \ac{SHAP} analysis of the up-ranked events confirms they exhibit signal-like characteristics, and parameter estimation returns source parameters consistent with compact binary mergers, lending further support to their astrophysical interpretation.

Taken together, these consistency checks provide evidence that the up-ranked events correspond to genuine signal-like candidates, supporting their astrophysical interpretation, including that of GW200311\_103121 as a \ac{BNS} signal.

The framework also assigns sub-threshold conditional confidence to some candidates above the \pastro threshold, an effect that is most pronounced for XGBoost, which prioritises a low false positive rate over sensitivity. 
This reflects what the framework has learned from the training data: lack of agreement between pipelines is strongly associated with noise, so candidates without corroborated multi-pipeline recovery receive lower confidence, even when one pipeline reports high significance. 
Where individual pipelines have missed a real signal for legitimate reasons, such as gaps in template bank coverage or the absence of coincident triggers, genuine signals may thus be incorrectly down-ranked, as the apparent lack of pipeline agreement does not reflect their astrophysical nature.
Conversely, in some cases the lower confidence correctly identifies noise, as illustrated by GW230630\_070659, a high-significance single-pipeline candidate subsequently identified as having an instrumental origin~\citep{GWTC4}.

More broadly, this framework is not intended to replace detailed pipeline-specific expertise, but to complement it by distilling diverse information into a unified measure that is readily usable by downstream analyses and astronomers responding to \ac{GW} alerts.
The approach is flexible: it can incorporate additional input features such as astrophysical probabilities or detector-state information, extends naturally to multi-class classification and real-time applications, and is readily extensible to additional pipelines provided mock data is available.

We therefore view this work as a step towards a more systematic and interpretable combination of \ac{GW} search pipelines.
By combining \ac{ML} models with \ac{CP}, this framework offers a robust path to producing catalogues and alerts with well-calibrated uncertainty estimates, supporting both confident detections and careful interpretation near threshold.

\begin{acknowledgments}
We want to thank Thomas Sainrat for valuable suggestions and feedback, Tom Dent and Achal Kumar for their helpful discussions and assistance in diagnosing the confusion regarding the \cwb pipeline outputs, and Sebastian Khan for discussions and feedback on the \ac{ML} architecture and feature-importance methodology.
We also thank Michael Coughlin, Deep Chatterjee, Tito Dal Canton, Reed Essick, Shaon Ghosh, Sushant Sharma-Chaudhary, Max Trevor, and Andrew Toivonen for the development of the \ac{MDC} results used in this work. 
We use data from \citet{LLPIC:2026prep} and thank the authors for preparing this dataset, and Sushant Sharma-Chaudhary and Thomas Sainrat for their help and advice in utilising it. 
Implementation of our \ac{ML} models was done using \sklearn \cite{Buitinck:2013fcp} and \texttt{XGBoost}~\citep{Chen:2016btl}, while we utilise \texttt{NumPy} \cite{harris_2020}, \texttt{Pandas} \cite{reback2020pandas}, and \texttt{Matplotlib} \cite{Hunter:2007} for data handling and visualisation. 
This material is based upon work supported by NSF’s LIGO Laboratory, which is a major facility fully funded by the National Science Foundation. 
The authors are grateful for computational resources provided by the LIGO Laboratory and supported by National Science Foundation Grants PHY-0757058 and PHY-0823459.
This work is supported by the Science and Technology Facilities Council (STFC) grant UKRI2488.

\end{acknowledgments}

\bibliography{bibliography}

%\onecolumngrid

\appendix
\crefalias{section}{appendix}
\crefalias{subsection}{appendix}
\crefalias{subsubsection}{appendix}

\section{Method details}\label{app:method}
\subsection{Data preparation}\label{sec:data}
The pipeline outputs include both significance statistics and source parameter estimates. 
From the template-based pipelines, we use the reported \ac{FAR}, \ac{SNR}, and $\chi^2$ (a matched-filter signal-consistency test~\cite{Allen:2004gu,Messick:2016aqy}). 
These pipelines also provide estimates of source parameters from the nearest waveform template. We record the detector-frame chirp mass, component masses, and aligned spin.
As the \cwb pipeline does not rely on an explicit waveform model, it does not provide template-based parameter estimates, and we thus only record the \ac{FAR} and \ac{SNR}. 
Although \cwb\ internally derives a detector-frame chirp mass estimate from the time-frequency morphology of the reconstructed signal for use in its vetoes and classifier~\citep{Tiwari:2015bda, GWTC5method}, we do not include it here since this estimate relies on a different methodology than the template-based pipelines and may be subject to systematics that are not yet well understood; incorporating it is left for future work.
We note that \cwb records \ac{SNR}$^2$ as the parameter named \ac{SNR}, and correct for this when assembling our dataset.
The \ac{SNR} reported by \cwb is a coherent, energy-based statistic~\cite{GWTC5method} rather than a matched-filter \ac{SNR} as used by the template-based pipelines, although both quantify the overall detection significance of a candidate.
In all cases, we use the logarithm of the \ac{IFAR} in units of seconds instead of the \ac{FAR}, since variations in statistical significance are meaningful on a logarithmic rather than a linear scale. 
This choice also ensures that non-detections, for which no \ac{IFAR} is reported, can be consistently assigned a value of zero.
Both modelled and unmodelled searches produce additional diagnostic features which we do not include in the present analysis and whose utility would need to be assessed for a production-level implementation.

To train the \ac{ML} classifier and calibrate \ac{CP}, a labelled dataset that is exchangeable with the real data is required. 
In \citet{Ashton:2025jhn} and for the majority of this work, we use the pre-O4 \ac{MDC} study \citep{Chaudhary:2023vec}, which applied the \gstlal, \pycbc, \spiir, \mbta, and \cwb pipelines to 40 days of replayed data from the third observing run with injected signals. Secondly, we also consider the \ac{LLPIC} dataset~\citep{LLPIC:2026prep}, produced during the maintenance break towards the end of the fourth observing (O4) run by injecting signals into replayed O4 background data. 
The \ac{LLPIC} consisted of four one-week replay cycles; we use only the last 7 days, as this cycle is covered by the most pipelines.
Candidates in both datasets with \ac{IFAR} > 1~hr are time-clustered, retaining only the maximum \ac{SNR} candidate per pipeline. 

This produces feature datasets containing per-pipeline detection statistics and template source parameter estimates, with feature values for non-detections set to zero. 
Ground-truth labels are assigned by matching candidates to known injections and astrophysical signals, yielding 9948 candidates, of which 5910 are signals for the \ac{MDC} dataset and 1204 candidates, including 534 signals, for the \ac{LLPIC} data.  
We note that the signal rate is far higher than astrophysically expected in both mock datasets. 
This inflated signal rate means that \pastro values and precision estimates computed from the mock data are not physically representative~\citep{Chaudhary:2023vec}, and should be interpreted only as relative performance comparisons within the mock-data context.
For this reason, we do not use \pastro as a comparison metric for the mock data studies, and instead compare against the \maxifar method.
For most experiments in this work, each dataset is split into training (80\%), calibration (10\%), and testing (10\%) subsets. However, we also use the entire \ac{LLPIC} dataset for testing only in \cref{sec:llpic_unseen_test}. 

In this work, we only use the \cwb, \pycbc, \gstlal, and \mbta pipelines, as they have been applied to all datasets considered in this study. 
However, this framework can be applied to other pipelines as well, provided sufficient labelled training and calibration data exist. 
For mock data studies, we use all 21 features per pipeline, where available, as discussed in \citet{Ashton:2025jhn}: the network \logIFAR and \ac{SNR}, and the template chirp mass, plus the per-detector \ac{SNR}, $\chi^2$, component masses, and aligned spins for each of the H1, L1, and V1 detectors.
This establishes the classifier performance achievable when all pipeline information is available.
When applying our method to the new candidates in \gwtcTwoPointOne~\cite{gwtc2p1_data,GWTC2-1}, \gwtcThree~\cite{gwtc3_data,GWTC3}, \gwtcFour~\cite{gwtc4p1_data,GWTC4}, and \gwtcFive~\cite{gwtc5_data,GWTC5}, we restrict the features to the measured \ac{IFAR}, \ac{SNR}, and chirp mass (setting the chirp mass to zero for \cwb). 
These are chosen to ensure better exchangeability across datasets: the \ac{IFAR} and \ac{SNR} are fundamental detection statistics that are consistently calibrated and validated across pipeline versions, while the chirp mass, though template-derived, is the most robustly measured source parameter and provides source classification information. 
This choice is further supported by the finding in \citet{Ashton:2025jhn} that including additional features does not improve classification performance significantly, though further feature selection analysis could refine this choice.

However, differences in detector noise characteristics, pipeline configurations, and astrophysical signal rates between the mock data and the different catalogues may introduce systematic differences in the pipeline behaviour. 
Furthermore, in \gwtcThree the \pycbc pipeline performed a second, \ac{BBH}-only search \cite{GWTC3}, which we exclude to ensure exchangeability. 
\subsection{Machine learning classifiers}\label{sec:ml}
In our previous work \citep{Ashton:2025jhn}, we focused primarily on \ac{LR} due to its interpretability. 
Here, we explore additional classifiers and compare the performance of \ac{LR}, \ac{MLP}, \ac{KNN}, and XGBoost, a tree-based method. 
We implement the first three using \sklearn~\citep{Buitinck:2013fcp}, and use the \texttt{XGBoost} package~\citep{Chen:2016btl} for the latter. 

For each classifier, the \ac{ML} model receives a feature vector $\vec{X}$ (the pipeline outputs for a candidate) and outputs a probability $P_y(\vec{X}; \lambda)$ for each label $y$, where $\lambda$ denotes the model parameters. 

The model parameters $\lambda$ (weights and biases) are learned from the training data by minimising the cross-entropy loss, or equivalently maximising the log-likelihood. 
For a labelled training dataset of size $N$, the log-likelihood is written as
\begin{equation}
     \sum_{n=1}^N \left(Y_n\log(P_{y=1}(\vec{X}_n;\lambda))+(1-Y_n)\log(1-P_{y=1}(\vec{X}_n;\lambda))\right)\,,
    \label{eq:log-likelihood}
\end{equation}
where $Y_n\in\{0,1\}$ is the true label of the $n$th sample (one for signal, zero for noise), $\vec{X}_n$ is its feature vector, and $P_{y=1}(\vec{X}_n;\lambda)$ is the predicted probability for the signal label. 

During training, the model parameters that maximise the log-likelihood in \cref{eq:log-likelihood},  $\hat{\lambda}$, are estimated. 
The specific maximisation algorithm varies by classifier, as described below.
After training, the learned model can evaluate previously unseen data, $\vec{X}'$ by computing $P_y(\vec{X}';\hat{\lambda})$.

\subsubsection{Logistic Regression (\ac{LR})} 
\ac{LR}~\citep{cox1958regression} is a simple \ac{ML} model, with $\lambda = (\vec{w}, b)$, that maps a linear combination of input features to a probability through the sigmoid activation function $\sigma(z)$,
\begin{equation}
    P_{y=1}(\vec{X}; \lambda) = \sigma(z) = \frac{1}{1 + \exp(-z)}\,,
    \label{eq:sigmoid}
\end{equation}
where 
\begin{equation}
   z = \vec{w}\cdot\vec{X} + b\,.
\end{equation}
The model is trained by maximising the log-likelihood in \cref{eq:log-likelihood} using the \texttt{L-BFGS-B} optimisation algorithm~\cite{Zhu:1997plu}. 

The magnitude of the \ac{LR} coefficients $w_i$ indicates the strength of feature $i$, while the sign indicates the direction of influence. 
\ac{LR} is simple, interpretable, and computationally efficient, but its predictions are constrained by the linear relationship with the input features, which may limit flexibility on more complex datasets.

\subsubsection{Multi-layer Perceptron (\ac{MLP})} 
\ac{MLP}~\citep{rumelhart1986learning} is a feedforward artificial neural network in which the input feature vector $\vec{X}$ is sequentially transformed through layers of neurons to produce a probability for each class. 
In our implementation, the neural network consists of a single hidden layer of $K=100$ neurons (we find only minor improvements when investigating deeper models, but at the cost of decreased computational efficiency). 
Each neuron in the hidden layer computes a weighted sum of the input features plus a bias and applies the ReLU (rectified linear unit) activation function~\cite{nair2010rectified, Agarap:2018uiz} (which is defined as $\mathrm{ReLU}(x) = \max(0, x)$). 
The outputs of the hidden layer are then combined in the output layer using the sigmoid activation function to produce the final probability $P_{y=1}(\vec{X};\lambda)=\sigma(z)$, with 
\begin{equation}
    z=b^o+\sum_{k=1}^K w^o_k \, \mathrm{ReLU}\left(\vec{w}^{h}_k \cdot \vec{X}+b^h_k\right) \,.
\end{equation}
The weights and biases $\vec{w}^h$, $b^h$ correspond to the $K$ neurons in the hidden layer, while $\vec{w}^o$, $b^o$ are associated with the output neuron.

Similarly to \ac{LR}, the model parameters are optimised by maximising the log-likelihood in \cref{eq:log-likelihood}, here using \texttt{Adam}, a stochastic gradient-based optimiser~\cite{Kingma:2014vow}. 
Unlike \ac{LR}, \ac{MLP}s can capture complex, non-linear relationships between features through their hidden layers, but require careful tuning and interpretation is less straightforward.

\subsubsection{k-Nearest Neighbours (\ac{KNN})} 
\ac{KNN}~\citep{cover1967nearest} is a non-parametric, instance-based model. For a given test feature vector $\vec{X}'$, the classifier identifies the $k$ closest training points according to a chosen metric (we use the Minkowski distance with $p=2$, equivalent to Euclidean distance). 
The output probability $P_y(\vec{X}')$ is then calculated as the fraction of neighbours belonging to each class:
\begin{equation}
    P_y(\vec{X}') = \frac{1}{k} \sum_{i \in \mathcal{N}_k(\vec{X}')} \mathbf{1}\{Y_i=y\}\,,
\end{equation}
where $\mathcal{N}_k(\vec{X}')$ denotes the indices of the $k$ nearest neighbours of $\vec{X}'$ in the training set, $Y_i$ is the true label of neighbour $i$, and $\mathbf{1}\{\cdot\}$ is the indicator function. 

Unlike \ac{LR} and \ac{MLP}, \ac{KNN} does not require explicit training and has no weights to optimise. 
The predictions are obtained directly by computing the distances from the test sample to the stored training data. 
\ac{KNN} naturally adapts to non-linear decision boundaries, but its performance depends on $k$, feature scaling, and distance metric. 
We use cross-validation to obtain an optimal $k=19$ for our data. 

We note that \ac{KNN} is excluded from the \ac{CP} analysis in \cref{sec:model_comparison}: its discrete output probabilities (fractions of $k$ neighbours) hinder the ranking required by the \ac{CP} algorithm, producing plateaued confidence values. 
It is therefore only compared via the \ac{ROC} curve.

\subsubsection{XGBoost} 
XGBoost (eXtreme Gradient Boosting)~\cite{Chen:2016btl} is a gradient-boosted decision tree ensemble, where the input feature vector $\vec{X}$ is passed through a series of $M$ decision trees $f_m(\vec{X}; \lambda_m)$, each trained sequentially to correct the residual errors of the previous trees. 
The final model output is a sum over all trees, converted to a probability via the sigmoid function $P_{y=1}(\vec{X};\lambda)=\sigma(z)$, with
\begin{equation}
    z=\sum_{m=1}^{M} f_m(\vec{X}; \lambda_m)\,, 
\end{equation}
where $\lambda_m$ represents the parameters of tree $m$, including the split features, split thresholds, and leaf weights. 

Each tree $f_m$ is constructed by recursively splitting the training data to minimise a regularised objective function, which for binary classification consists of the logistic (log-loss) term (analogous to \cref{eq:log-likelihood}) plus a penalty on the complexity of the tree, using the sum of squared leaf weights. 
The tree is constructed to fit the negative gradient of the loss (the residuals) from the current ensemble of previous trees, effectively performing a step of functional gradient descent. 
This sequential process allows each new tree to focus on the errors made by the current model, gradually improving predictive performance. 

Regularisation parameters, including the learning rate and maximum tree depth, are used to prevent overfitting and control model complexity. 
XGBoost can capture complex, non-linear interactions between features and generally achieves high predictive performance~\citep{Chen:2016btl}. 

The nature of decision trees means that missing (NaN) values can be handled, and it is thus not necessary to replace non-detections by zero for the XGBoost classifier. 
However, for comparability between classifiers, we replace non-detections with zero for XGBoost as well. 
We implement XGBoost using the \texttt{XGBoost}\xspace package and, using cross-validated grid-search, obtain the optimal classifier with 96 trees and a maximum depth of 6.

\subsection{Conformal prediction}
\ac{CP}~\cite{vovk2005algorithmic, angelopoulos2021gentle} is a framework developed to provide statistically rigorous and well-calibrated uncertainties for any point prediction. 
\ac{CP} does not modify the underlying algorithm but instead uses its predictions on a labelled calibration dataset to learn the uncertainty. 
Thus, \ac{CP} requires no assumptions about the model or data distributions. 
The only requirement is that the data are exchangeable.

First, a nonconformity measure $A(\vec{X}, y)$ is defined to quantify how unusual a sample-label pair $(\vec{X}, y)$ is, with smaller values indicating better agreement with the model. 
The choice of nonconformity measure depends on the problem (see \citet{Malz:2024zjd} for discussion), and here we define it as the complement of the classifier prediction probability,
\begin{equation}
    A(\vec{X},y)=1-P_y(\vec{X}; \lambda)\,,
    \label{eq:nonconf_measure_simple}
\end{equation}
where $P_y(\vec{X}; \lambda)\in[0,1]$ is the classification score for label $y$.

In this work, we employ Mondrian (label-conditional) \ac{CP}~\cite{vovk2013conditional, ding2023CP} to guarantee conditional coverage for each class. 
During calibration, for each label, a nonconformity score $s_n^y = A(\vec{X}_n, y)$ is computed for each feature vector $\vec{X}_n$ in the calibration dataset. 
The scores are then sorted in ascending order, and the per-label $(1-\alpha)$ quantile is obtained as
\begin{equation}
    \hat{q}_y = s^y_{\lceil (N_y+1)(1-\alpha) \rceil}\,,
    \label{eq:qhat}
\end{equation}
where $\alpha \in [0,1]$ is the user-chosen error rate and $N_y$ is the number of data points in the calibration data of label $y$.

For a new test sample $\vec{X}'$, the nonconformity scores $A(\vec{X}', y)$ are computed for all possible labels $y$. 
A prediction set $\Gamma^\alpha$ is then defined as containing all labels (signal and noise, in our binary case) whose nonconformity score falls at or below the $(1-\alpha)$ quantile of the calibration scores for that label:
\begin{equation}
    \Gamma^\alpha = \{ y : A(\vec{X}', y) \le \hat{q}_y \}\,.
\end{equation}

\ac{CP} guarantees that, in the limit of a large calibration set, the true label $Y'$ is included in the prediction set $\Gamma^\alpha$ with probability approximately $1-\alpha$. 
More formally, for a finite calibration set of size $N_y$ per label:
\begin{equation}
    1-\alpha\leq\textrm{Pr}(Y'\in\Gamma^\alpha(\vec{X'})|Y'=y)\leq 1-\alpha+\frac{1}{N_y+1}\,.
    \label{eq:cp_validity}
\end{equation}
By using Mondrian \ac{CP}, we ensure conditional coverage for each class individually, in addition to marginal coverage across the entire dataset.

To quantify the significance of individual events, we use the conditional confidence in the signal label, defined as the largest value of $\alpha$ for which the signal label is included in the prediction set $\Gamma^{\alpha}$ \citep{Ashton:2024wae}. 
Formally, since $\Gamma^{\alpha=0}(\vec{X})$ contains all labels by construction, this maximum is always well-defined:
\begin{equation}
    \texttt{conf}_{y=1}(\vec{X}) = \max \left\{ \alpha \in [0,1] : y=1 \in \Gamma^{\alpha}(\vec{X}) \right\} \,.
\end{equation}
In practice, we evaluate $\Gamma^\alpha$ over a discrete grid of decreasing $\alpha$ values, taking the largest $\alpha$ for which $y=1 \in \Gamma^\alpha(\vec{X})$ as the conditional confidence.
By placing a threshold on the confidence (e.g. 0.5), a catalogue with well-defined purity can be constructed.

\section{Classifier evaluation: full results}\label{app:classifiers}
In this Section, we compare the performance of the \ac{ML} classifiers introduced in \cref{sec:ml} to assess whether the pipeline combination framework is effective and whether results are robust to the choice of classifier. 
We evaluate each classifier using \ac{ROC} curves and sensitivity and precision metrics under \ac{CP}, and examine whether our framework produces consistent results.

\subsection{Model comparison}\label{sec:model_comparison}
The raw \ac{ML} classifier output lacks a statistically rigorous significance measure,  unlike the \ac{FAR} provided by the \maxifar approach, an essential component for assessing the significance of individual candidate events and motivating the use of \ac{CP}. 
We therefore apply \ac{CP} to the \ac{ML} classifier predictions and use the conditional confidence as our significance measure for each candidate. 

We apply \ac{CP} to each classifier in turn, with the exception of \ac{KNN}: its discrete classification probabilities hinder ranking in the \ac{CP} algorithm, leading to plateaued confidence values. 
Increasing $k$ removes the plateau but reduces the \ac{AUC} significantly and introduces a large number of false positives. 
In figures where showing all classifiers would be redundant, we use XGBoost as a representative example, chosen due to its superior \ac{AUC}.
As a tree-based ensemble, XGBoost exhibits higher model variance than \ac{LR} or \ac{MLP}, with small changes in training data potentially affecting split points and tree structure; this becomes apparent when comparing results across datasets.

\begin{figure}
    \centering
    \includegraphics[width=0.5\textwidth]{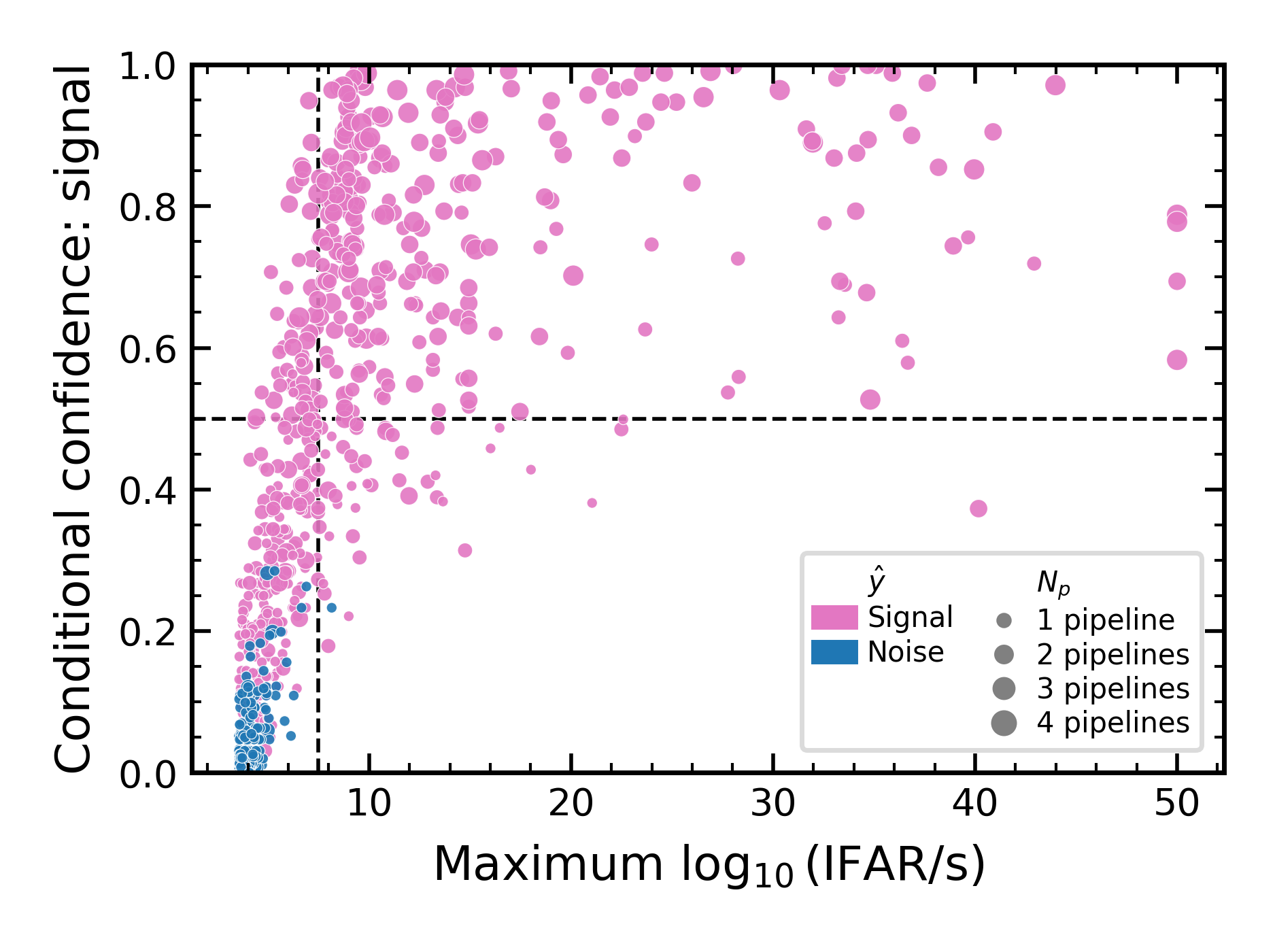}
    \caption{The conditional confidence in the signal label for the XGBoost classifier compared to the maximum \ac{IFAR}. 
    The dashed lines show the chosen confidence threshold of $0.5$ and the commonly used \ac{FAR} threshold of 1 per year. 
    The colour of each point indicates the true label $\hat{y}$, while the size indicates the number of pipelines $N_p$ which detected an event.} 
    \label{fig:confidence_xgboost}
\end{figure}

Using the \ac{MDC} data, we calculate the conditional confidence and compare the resulting predictions to those from the \maxifar method, as shown in \cref{fig:confidence_xgboost}. 
The confidence scores are broadly consistent with the \maxifar ranking.  
At a confidence threshold of $0.5$ (see \cref{sec:threshold} for discussion), the \ac{CP} method produces no false positives, compared to one for the \maxifar method at the standard \ac{FAR} threshold of 1 per year.
The hard cut-off at the lower end of the \ac{IFAR} axis reflects the 1-hour \ac{IFAR} noise threshold (see \cref{sec:data}), below which no candidates are retained.
The plot presents results for XGBoost, but we note that all classifiers exhibit similar agreement with the \maxifar method. 

To provide a quantitative summary of classifier performance, the number of \ac{TP}, \ac{FP}, \ac{FN} and \ac{TN} for each classifier are shown in \cref{tab:confusion_matrix_mdc}, along with that for the \maxifar method. 
\ac{LR} and XGBoost achieve zero \acp{FP}, while \ac{MLP} and the \maxifar method each produce one. 
The number of \ac{TP} is marginally higher for the \ac{ML} classifiers ($\sim 52-53\%$) than for \maxifar ($51\%$), though the differences are small. 
This confirms that all methods perform comparably on the mock data, with minimal false positives.

\begin{table}
    \centering
    \caption{Performance of the different classifiers + \ac{CP} and the \maxifar method on the \ac{MDC} dataset. 
    A threshold of $0.5$ on the conditional confidence is applied for the \ac{ML} classifiers and a \ac{FAR} of 1 per year for the \maxifar.}
    \begin{tabular}{l|c|c|c|c}
        \toprule
        Method & TP & FP & FN & TN \\
        \midrule
        LR & 312 & 0 & 279 & 403 \\
        XGBoost & 310 & 0 & 281 & 403 \\
        MLP & 312 & 1 & 279 & 402 \\
        max $\log_{10}(\mathrm{IFAR})$ & 302 & 1 & 289 & 402 \\
        \bottomrule
    \end{tabular}
    \label{tab:confusion_matrix_mdc}
\end{table}

\begin{figure}
    \centering
    \includegraphics[width=0.5\textwidth]{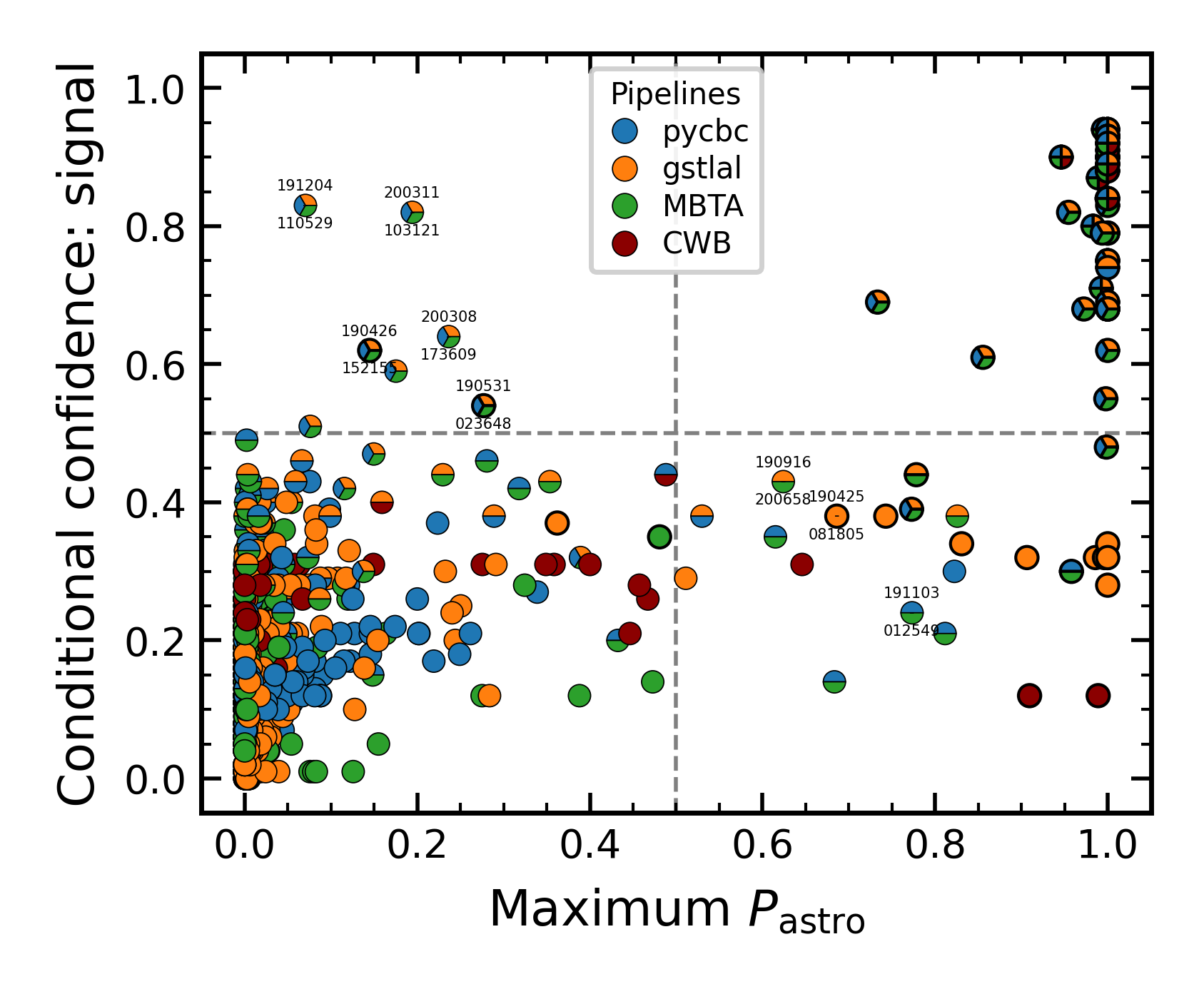}
    \caption{The conditional confidence in the signal label for the XGBoost classifier compared to the maximum \pastro for the O3 events in the combined \gwtcTwoPointOne and \gwtcThree catalogue. 
    The classifier is trained and \ac{CP} calibrated on the \ac{MDC} data. 
    The dashed lines represent the confidence and \pastro thresholds, respectively. 
    High-significance events (\ac{FAR} < 1 per year) are indicated by a bold marker outline. } 
    \label{fig:gwtc3_xgboost}
\end{figure}

Having established that the conditional confidence behaves consistently on the mock data, we apply our combination approach to real events from the third observing run~\cite{GWTC2-1, GWTC3}. 
In \cref{fig:gwtc3_xgboost}, we compare the resulting confidence scores with the corresponding \maxpastro, analogous to the comparison shown in \cref{fig:gwtc3_lr} for the \ac{LR} classifier. 
Most candidates are found in the top-right and bottom-left quadrants, indicating agreement between the confidence and \maxpastro estimates for signals and noise, respectively. 

In the bottom-right quadrant, we identify candidates classified as signals according to their \pastro but assigned sub-threshold confidence values. 
As in \cref{fig:gwtc3_lr}, the majority of these candidates are identified by only a single pipeline, most often \gstlal or \cwb, as indicated by the colours of the markers. 
This reflects the empirical information in the training data: events seen by multiple pipelines are more likely to be astrophysical, while single-pipeline triggers are more often associated with noise or pipeline-specific artefacts. 
For instance, candidates identified only by \cwb may be less likely to correspond to \ac{CBC} signals, as the pipeline relies on coherence rather than a templated search. 
As the only pipeline configured to identify single-detector signals in O3 \citep{GWTC3}, \gstlal naturally reports a larger number of candidates than pipelines requiring multi-detector coincidence. 
We also note that many of the \gstlal single-pipeline candidates pass the \ac{FAR} threshold. 
Although differences in search methods, tuning, and configurations mean that pipelines are not expected to always identify the same candidates, our framework has learned from the training data that these single-pipeline events are statistically less likely to be astrophysical signals.

A notable example among the down-ranked single-pipeline events is GW190425\_081805, a single-detector \ac{BNS} identified only by \gstlal with a \pastro of $0.69$ in O3~\citep{GWTC2-1, LIGOScientific:2020aai}. 
Its total mass of $\sim 3.4 \, M_\odot$ exceeds that of all known galactic \ac{BNS} systems, and \ac{GW} data cannot rule out the possibility that one or both components of the system are black holes \citep{LIGOScientific:2020aai}. 

There are also two three-pipeline and nine two-pipeline candidates in this quadrant. 
Notably, both three-pipeline candidates are high-significance events (\ac{FAR} < 1 per year), suggesting that the XGBoost classifier is penalising some aspect of their feature distribution beyond raw significance. 

In the top-left quadrant, we identify several events with sub-threshold \pastro values but conditional confidence scores exceeding the threshold, all detected by three pipelines each. 
We find that XGBoost independently recovers several candidates previously up-ranked by \ac{LR} in \cref{fig:gwtc3_lr}, lending additional weight to their potential astrophysical origin.
Most notably, this includes GW200311\_103121, with conditional confidence $0.82$, the sub-threshold \ac{BNS} candidate listed in the marginal candidate table of \gwtcThree. 
\citet{Ashton:2021cub} performed parameter estimation on GW200311\_103121, finding source masses of $1.4M_\odot$ and $1.2M_\odot$ and concluded that ``analysing the event under the assumption that it is an astrophysical signal, its properties are highly consistent with that of a binary neutron star merger''.

We also find GW190426\_152155 and GW190531\_023648, two marginal significance \ac{NSBH} candidates in the \gwtcTwoPointOne catalogue~\citep{GWTC2-1}, which were also up-ranked by \ac{LR} in \cref{fig:gwtc3_lr}. 
Both candidates pass the \ac{FAR} threshold in the \gstlal pipeline (with $0.91 \text{ yr}^{-1}$ and $0.41 \text{ yr}^{-1}$, respectively), but obtain low \pastro values of $0.14$ and $0.28$.

Furthermore, XGBoost also up-ranks GW200308\_173609 and GW191204\_110529, two \ac{BBH} candidates with \maxpastro of $0.24$ in \mbta, and $0.07$ in \gstlal, respectively, but with conditional confidences of $0.64$ and $0.83$. 
Notably, both of these events are also detected by the \pycbc-\ac{BBH} search with a \pastro of $0.86$ and $0.74$, respectively, but these results are excluded from our test data set for exchangeability reasons previously discussed. 
These cases illustrate an important consistency check: our framework independently up-ranks events that a separate, excluded pipeline search also considers significant, providing an external validation of the up-rankings.

To assess the consistency of results across classifiers, we construct summary plots representing the different regions in the \pastro versus conditional confidence plots. 
\Cref{tab:venn_diagrams_gwtc3} shows these diagrams for the \ac{LR}, \ac{MLP}, and XGBoost classifiers. 
The diagrams indicate the number of events in each region, and highlight multi-pipeline events where the confidence and \pastro statistics disagree. 
Events annotated in black appear in the same quadrant for two or more classifiers, indicating consistent classification.

\begin{table*}
  \caption{Comparison of the three \ac{ML} classifiers on \gwtcThree data. 
  Events in regions where the confidence and \pastro disagree are annotated if detected by more than one pipeline. 
  Events where two or more classifiers agree are annotated in black, while events unique to that classifier are shown in grey. }
  \centering
  \begin{ruledtabular}
  \begin{tabular}{lll}
    \textbf{LR} & \textbf{XGBoost} & \textbf{MLP} \\
    \midrule
    \includegraphics[width=0.32\textwidth,height=0.4\textheight,keepaspectratio]{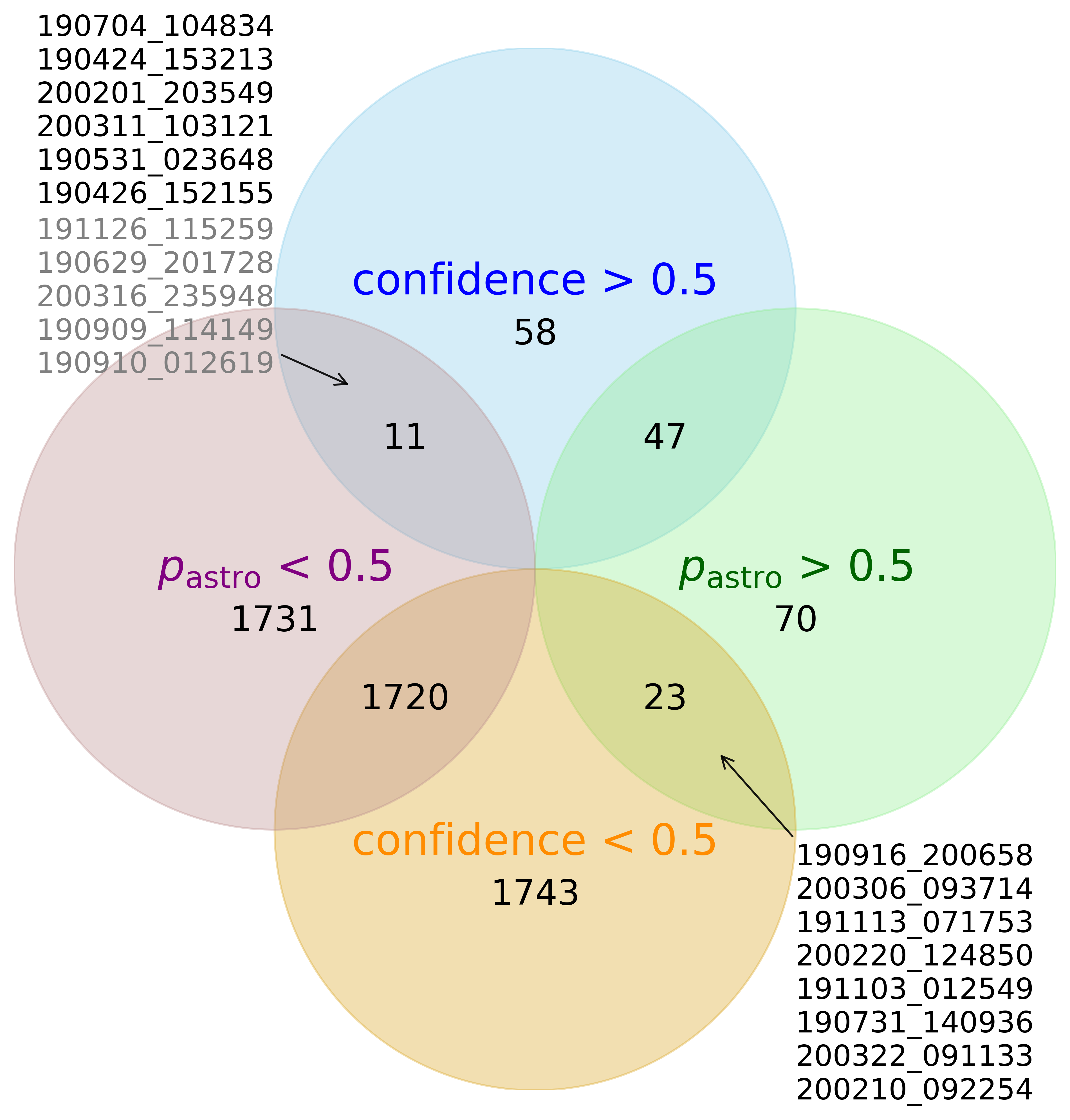} &
    \includegraphics[width=0.32\textwidth,height=0.4\textheight,keepaspectratio]{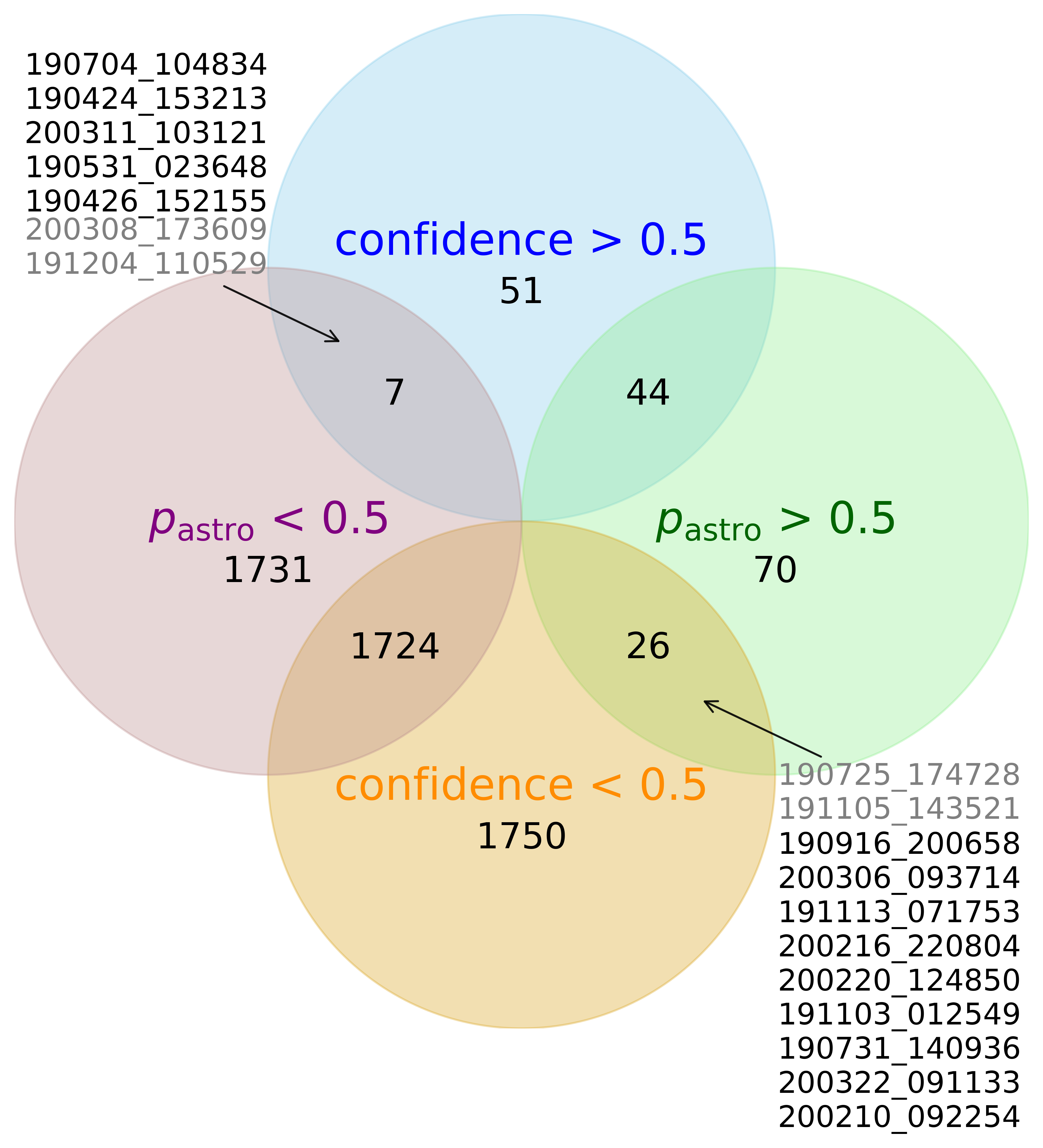} &
    \includegraphics[width=0.32\textwidth,height=0.4\textheight,keepaspectratio]{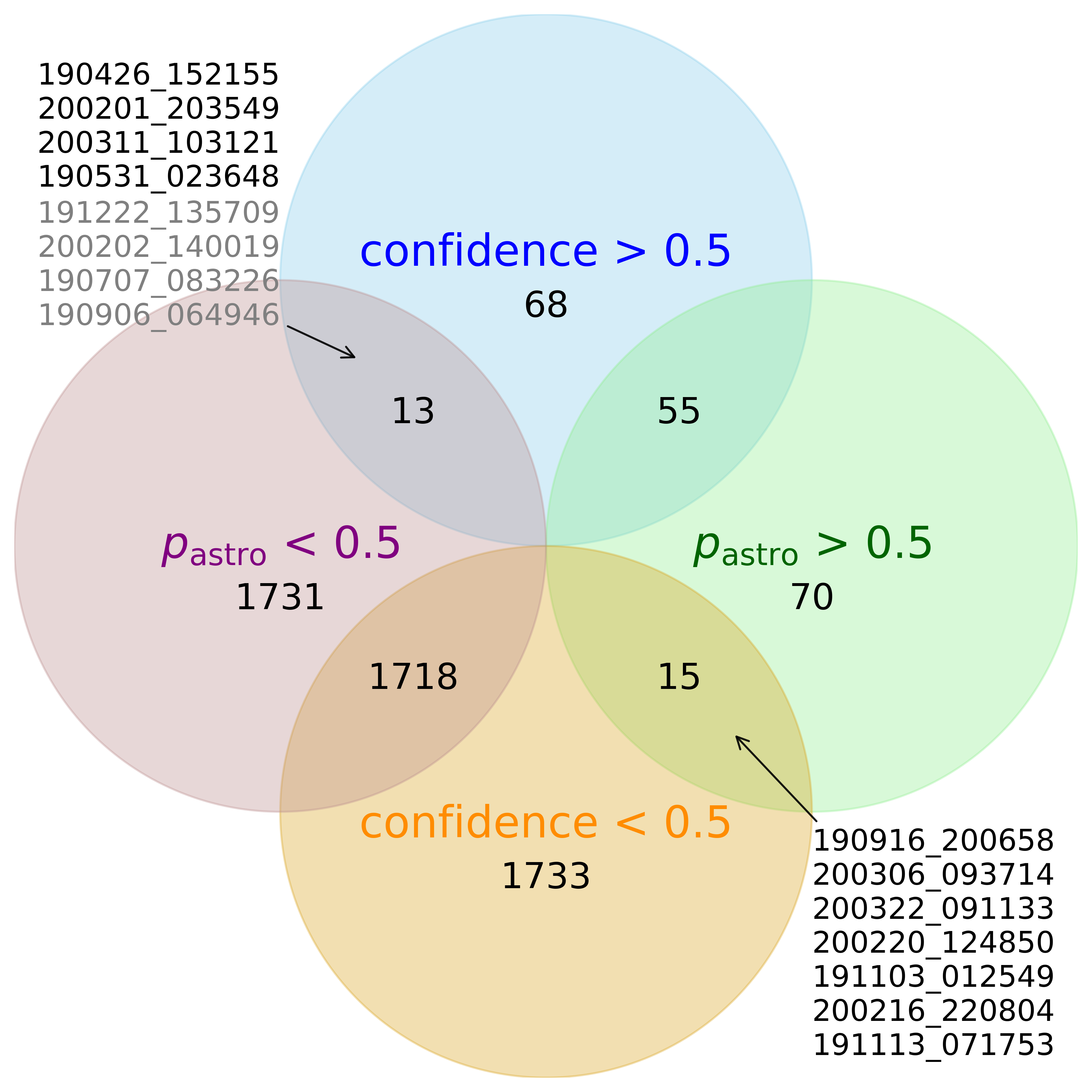} \\ 
  \end{tabular}
  \end{ruledtabular}
  \label{tab:venn_diagrams_gwtc3}
\end{table*}

From \cref{tab:venn_diagrams_gwtc3}, we find that GW200311\_103121, GW190426\_152155 and GW190531\_023648 consistently have confidence values above threshold for all classifiers. 
This demonstrates robustness and consistency; multiple classifiers independently assigning high confidence to the same sub-threshold candidate increases the reliability of the prediction. 
Notably, these candidates correspond, if real, to a \ac{BNS} and two \ac{NSBH} systems. 

The low \pastro values of these candidates reflect the known population uncertainties discussed in \citet{GWTC3}. 
While the \ac{BBH} mass distribution is now relatively well constrained, the \ac{BNS} and \ac{NSBH} populations remain poorly characterised, leading to larger uncertainties in the astrophysical prior used to compute \pastro. 
For such systems, these uncertainties can exceed $0.1$ in \pastro for candidates near the detection threshold, meaning that \pastro values may be particularly sensitive to assumptions about the underlying population~\cite{GWTC3}.
Additionally, the higher detectable \ac{BBH} merger rate means \ac{BBH} candidates receive higher \pastro values than low-mass systems at comparable \ac{FAR}, causing \pastro-based thresholds to preferentially exclude \ac{BNS} and \ac{NSBH} candidates. 

This behaviour is illustrated by GW190426\_152155, which was included as a confident candidate in \gwtcTwo under the \ac{FAR} < 2 yr$^{-1}$ detection criterion \cite{GWTC2}, but was reclassified as a marginal candidate in \gwtcTwoPointOne which uses \pastro $\geq 0.5$ as the detection threshold \cite{GWTC2-1}. 
Notably, \gwtcTwo did not assign a \pastro to this event, citing strong prior dependence on assumed \ac{NSBH} signal rates. 
\gwtcTwoPointOne subsequently estimated a low \pastro of $p_{\rm NSBH}=0.14$~\citep{GWTC2-1}, reflecting the lower inferred rate of detectable \ac{NSBH} mergers rather than weak search significance (\ac{FAR} of $0.91 \text{ yr}^{-1}$). 
In contrast, our framework relies primarily on empirical multi-pipeline information rather than population modelling. 
A dedicated reanalysis of GW190426\_152155 found it to be consistent with either an \ac{NSBH} merger or a low-mass \ac{BBH} \citep{Li:2020pzv}.
The consistent up-ranking of these \ac{BNS} and \ac{NSBH} candidates across all classifiers therefore suggests coherent support from multiple search pipelines that is not fully captured by current population-informed \pastro estimates. 

While XGBoost up-ranks only a few events, \ac{LR} and \ac{MLP} up-rank considerably more candidates. 
For \ac{LR}, these are all three-pipeline events, consistent with XGBoost. 
\ac{MLP}, however, also up-ranks several single- and two-pipeline candidates, suggesting it is less conservative than the other classifiers, which we discuss further in later sections.

Similarly to GW200308\_173609 and GW191204\_110529 (up-ranked by XGBoost), GW191126\_115259 (up-ranked by \ac{LR}) was also detected by the \pycbc-\ac{BBH} search with a \pastro of $0.7$. 
GW200201\_203549, up-ranked by both \ac{LR} and \ac{MLP}, is (if real) another \ac{NSBH} low-\pastro candidate that appears in the marginal candidate table of \gwtcThree~\citep{GWTC3}. 

Although these events are not yet confirmable as astrophysical signals, these results illustrate the potential of pipeline combination to identify sub-threshold candidates that other methods may overlook; such identifications may become more reliable with improved control over systematic differences. 

There are six multi-pipeline events with \pastro $\geq 0.5$ but confidence $<0.5$ for all classifiers, all of which are two-pipeline candidates with low significance. 
GW191103\_012549 and GW190916\_200658 are \ac{BBH} candidates identified with a \maxpastro in \pycbc and \mbta, respectively, but the highest \pastro was obtained by \pycbc-\ac{BBH}, which is excluded from our test data (as discussed in \cref{sec:data}). 
The other four events, which are consistently down-ranked, have a \pastro $\geq 0.5$ in the \mbta pipeline only. 
In addition, all of them were also seen by the \pycbc-\ac{BBH} search (albeit with small \pastro). 
Furthermore, there are four multi-pipeline high significance events which are down-ranked; all are down-ranked by XGBoost, while \ac{LR} and \ac{MLP} each independently down-rank one candidate from this group. 

Overall, the three classifiers show broad agreement on which events to up-rank and down-rank, with differences arising primarily from XGBoost being more conservative. 
XGBoost achieves the highest \ac{AUC} and is the most conservative classifier, up-ranking only events with strong multi-pipeline support, while also down-ranking more events than the other classifiers. 
\ac{LR} performs comparably and up-ranks a similar set of three-pipeline candidates but down-ranks fewer events than XGBoost. 
\ac{MLP} is the least conservative, up-ranking the most candidates, including single- and two-pipeline events, and down-ranking the fewest events.

A full list of confidence scores for all \ac{ML} classifiers and \pastro values for all events in the \gwtcThree catalogue where the \maxifar method disagrees with our framework is included in \cref{app:gwtc_table}.
\subsection{Sensitivity and precision}\label{sec:sensitivity_precision}
In the \ac{GW} literature, the sensitive volume~\cite{Usman:2015kfa, Schafer:2021cml} is a key metric used to compare pipeline performance on mock data. 
It is estimated from injections as the fraction of simulated signals recovered above a given \ac{FAR} threshold (equivalently, the sensitivity at that threshold) scaled by the maximum survey volume~\cite{Usman:2015kfa, Schafer:2021cml}.
Since sensitive volume is directly proportional to sensitivity at a fixed threshold, we consider the latter first. 
Sensitivity, also referred to as recall or true positive rate, is the fraction of true events correctly identified as signals and is formally defined as
\begin{equation}
    \texttt{sensitivity} = \frac{\text{TP}}{\text{TP}+\text{FN}}\,,
\end{equation}
where \ac{TP} is the number of events correctly predicted as signals and \ac{FN} the number incorrectly predicted as noise. 
By this definition, a high sensitivity can be achieved even if many noise triggers are misclassified as signals, since \acp{FP} are not accounted for. 
Precision complements sensitivity by quantifying how reliable the detections are, and is defined as the fraction of positive predictions that are true signals,
\begin{equation}
    \texttt{precision} = \frac{\text{TP}}{\text{TP}+\text{FP}}\,.
\end{equation}
Together, these metrics provide a more complete picture of classifier performance than the \ac{ROC} curve alone, which is insensitive to class imbalance and does not directly account for precision~\citep{davis2006relationship, saito2015precision}.
In real \ac{GW} searches, noise candidates vastly outnumber signals, making precision particularly informative. 
However, the \ac{MDC} dataset has an artificially high signal rate, so the precision values reported here should be interpreted accordingly.

\begin{figure}
    \centering
    \includegraphics[width=0.49\textwidth]{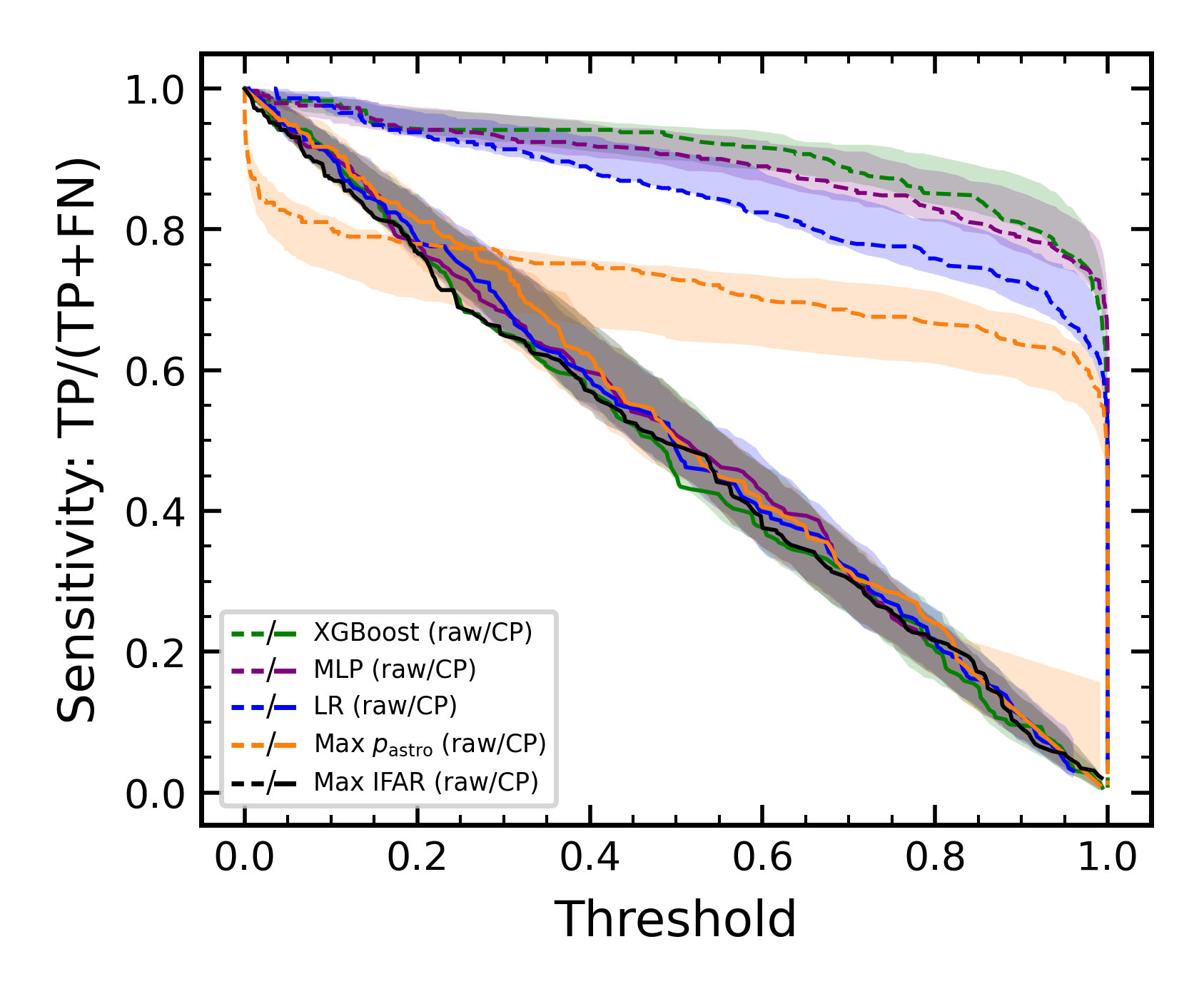} 
    \caption{Sensitivity for the \ac{LR} and XGBoost classifiers trained and evaluated on the \ac{MDC} data. 
    Dashed lines represent methods with \ac{CP} applied. 
    The \maxifar and \maxpastro methods with \ac{CP} applied are also shown for comparison. 
    The shaded bands show the 90\% confidence interval over 100 permutations of data splits, while the lines illustrate one of the realisations.} 
    \label{fig:recall}
\end{figure}

For our pipeline combination approach, however, the sensitivity curve computed by varying the detection threshold is necessarily diagonal. 
This arises from the validity guarantee of \ac{CP} and the conditional confidence definition chosen. 
\ac{CP} guarantees that the true label is included in the prediction set $\Gamma^\alpha$ with a probability of approximately $1-\alpha$ (see \cref{eq:cp_validity}). 
Using Mondrian (label-conditional) \ac{CP}, we can make this guarantee for each class (signal or noise) individually. 
Thus, by varying the threshold $\alpha$, \ac{CP} guarantees an approximately diagonal coverage (sensitivity) curve when the calibration set is sufficiently large. 
Because the conditional confidence used to assess the significance of each candidate corresponds to the value of $\alpha$ at which the signal label is included in the prediction set, the resulting plot of sensitivity versus confidence is also necessarily diagonal. 
This is illustrated in \cref{fig:recall}; all methods that include \ac{CP} yield a diagonal sensitivity curve. 

A direct comparison with the \maxifar curve is not straightforward, as the threshold scales non-linearly. 
Instead, we include the \maxpastro for a rough comparison, while noting that the \pastro values in the \ac{MDC} dataset are not physically realistic since the enhanced signal rate of the dataset means the \pastro is not well calibrated~\citep{Chaudhary:2023vec}. 

Therefore, to make a fair comparison, we apply \ac{CP} to the \maxifar and \maxpastro as well, ensuring all approaches are evaluated under the same framework. 
Applying \ac{CP} does not affect the underlying ranking of events, but converts the raw significance measure into a calibrated confidence value on the same $[0,1]$ scale as the \ac{ML} classifiers, enabling a direct comparison at a common threshold. 

This diagonal behaviour also implies that sensitive volume would be identical across all \ac{CP}-calibrated classifiers, rendering it uninformative for comparison.

\begin{figure}
    \centering
    \includegraphics[width=0.49\textwidth]{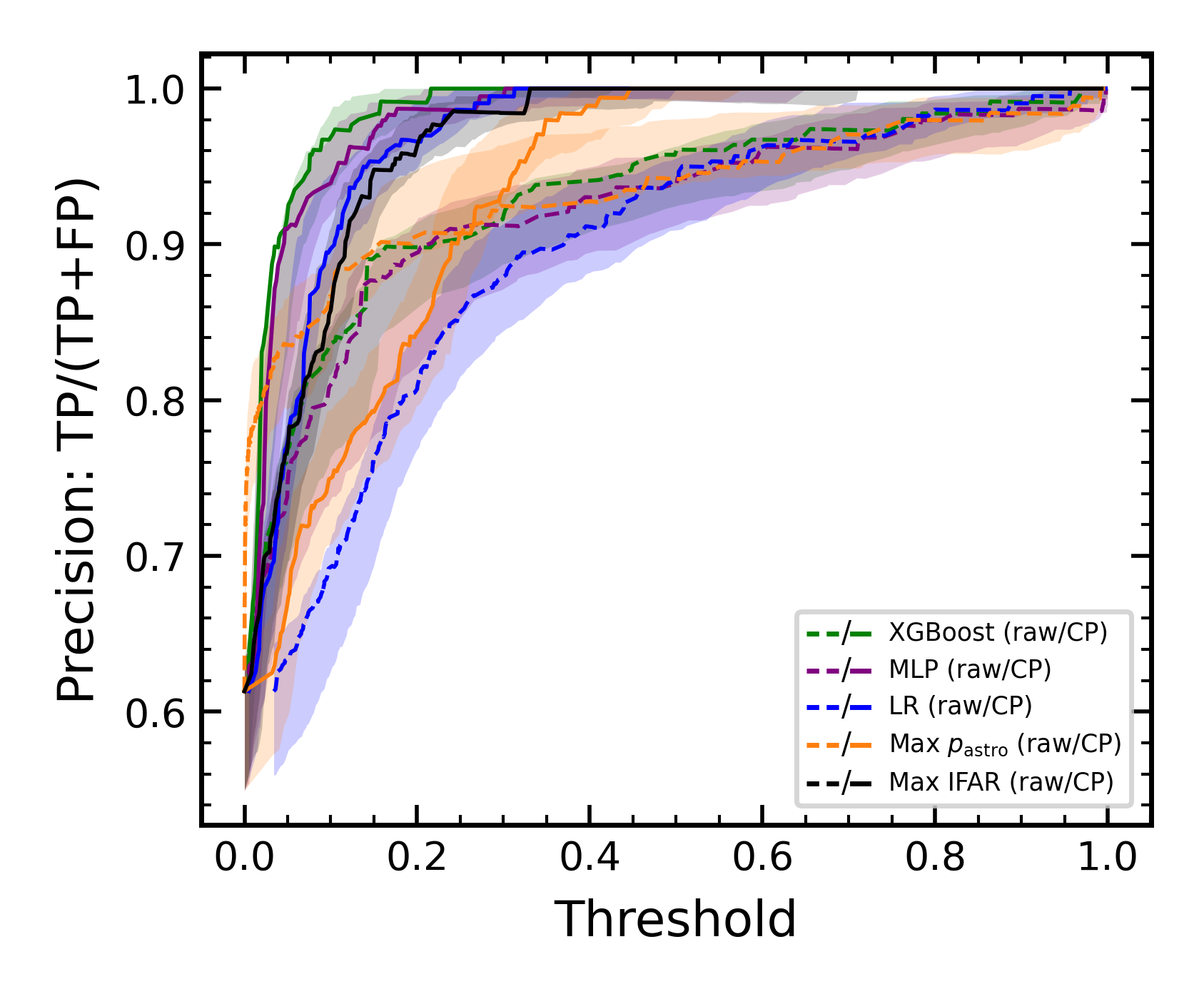} 
    \caption{Precision curves for the \ac{LR} and XGBoost classifiers trained and evaluated on the \ac{MDC} data. 
    The dashed lines represent methods with \ac{CP} applied. 
    The \maxifar and \maxpastro methods with \ac{CP} applied are also shown for comparison. The shaded bands show the 90\% confidence interval.} 
    \label{fig:precision}
\end{figure}

While the sensitivity plot in \cref{fig:recall} is not informative for comparing the performance of different methods, the precision plot in \cref{fig:precision} provides a more meaningful assessment of how reliably each method identifies true signals. 
The figure shows the precision curves for the \ac{ML} classifiers, both with and without the application of \ac{CP}, along with the \maxifar and \pastro methods, each with \ac{CP} applied. 

We observe that for the \ac{ML} classifiers shown, applying \ac{CP} increases the precision significantly. 
All \ac{ML} classifiers also perform better than the \maxifar method. 
For the \maxpastro, applying \ac{CP} appears to increase the precision only for threshold values above $\sim 0.3$. 
However, this might be an artefact of the physically unrealistic \pastro. 

By construction, \ac{CP} precision approaches one at high confidence thresholds, since events assigned high confidence are very likely to carry the signal label as their true class.

\begin{figure}
    \centering
    \includegraphics[width=0.49\textwidth]{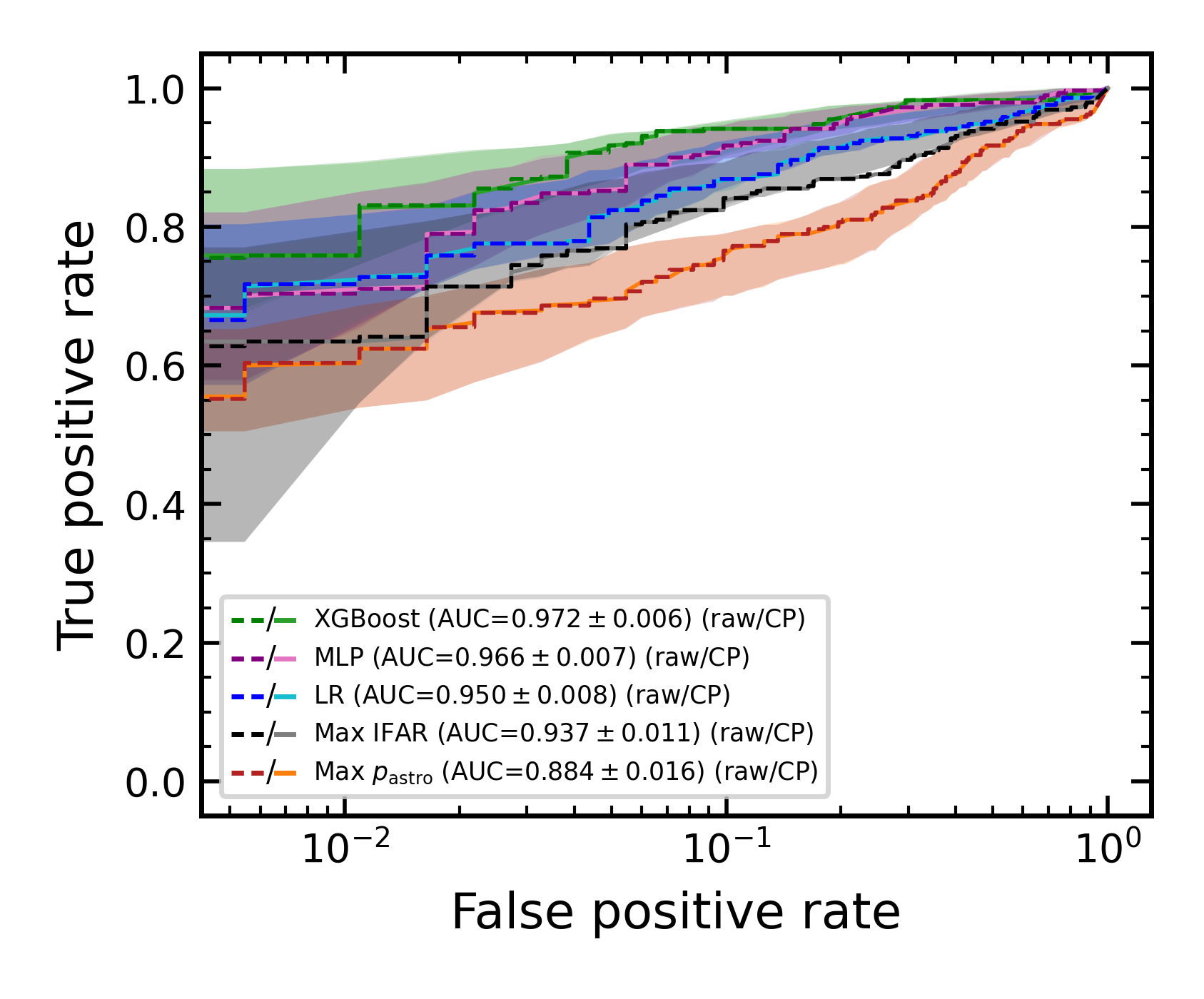} 
    \caption{\ac{ROC} curves using the \ac{CP} confidence as threshold. The \ac{ROC} curves for the classifiers directly and the \maxifar and \maxpastro methods are also included for comparison. 
    The shaded bands show the 90\% confidence interval over 100 permutations of data splits, and the \ac{AUC} is given as the mean and standard deviation of the same.} 
    \label{fig:roc_cp_conf}
\end{figure}

\begin{figure}
    \centering
    \includegraphics[width=0.49\textwidth]{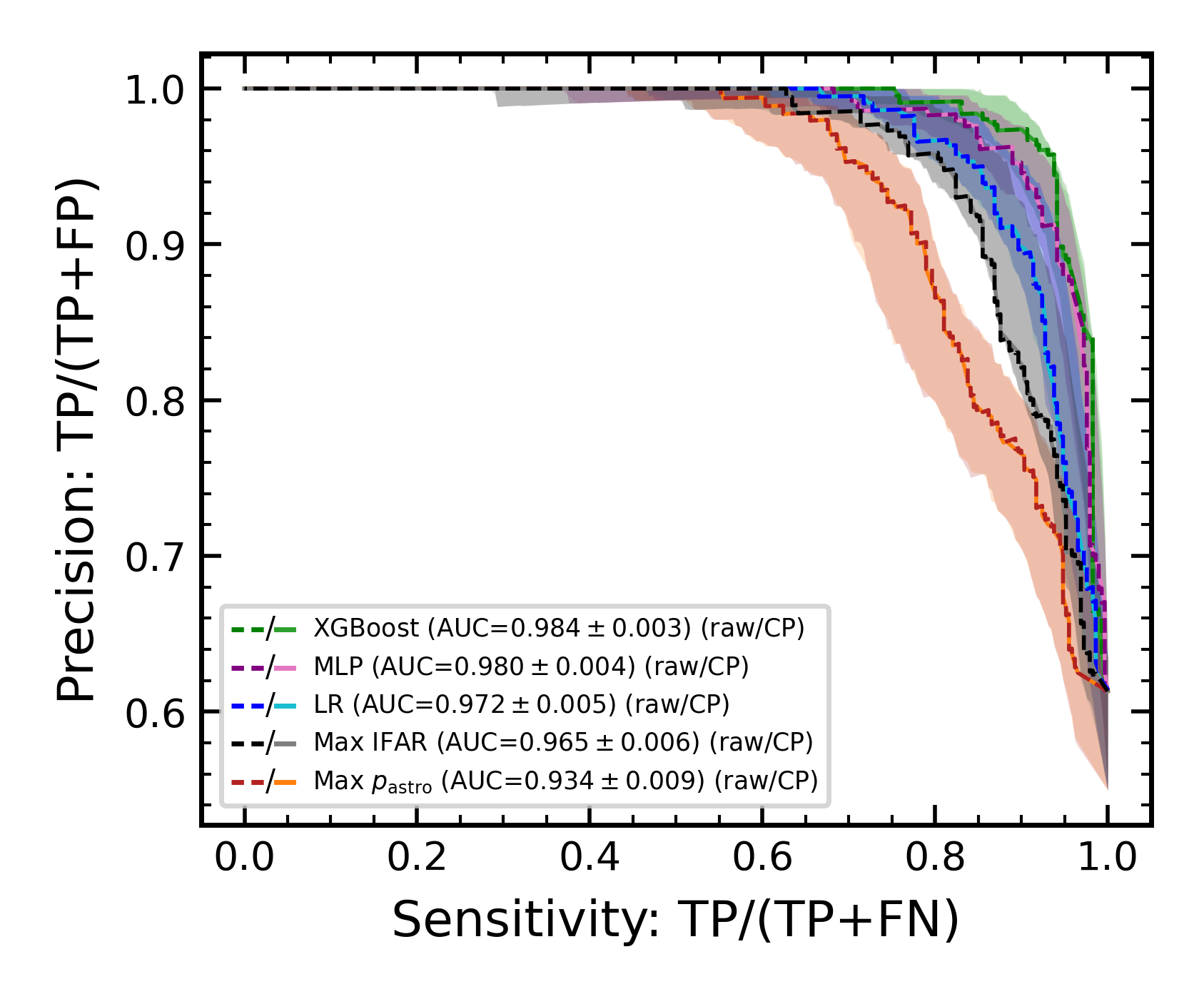} 
    \caption{Precision-sensitivity curve obtained from varying the respective thresholds for the different methods. 
    The shaded bands show the 90\% confidence interval over 100 permutations of data splits, and the \ac{AUC} is given as the mean and standard deviation of the same permutation set.} 
    \label{fig:precision-recall}
\end{figure}

We can also plot the \ac{ROC} and the precision-sensitivity curves, using the \ac{CP} confidence as threshold, and calculate the \ac{AUC} for both, see \cref{fig:roc_cp_conf} and \cref{fig:precision-recall} respectively. 
For both plots we note that adding \ac{CP} on top of the underlying classification method does not change the curves or \ac{AUC}. 
This arises because both curves only depend on the ranking of scores and since \ac{CP} preserves rank, its application calibrates the output but does not change the classifier performance. 

In summary, the \ac{ROC} curves (\cref{fig:roc_all_classifiers_mdc12,fig:roc_cp_conf}) and precision-sensitivity curve (\cref{fig:precision-recall}) characterise overall classifier performance and are unaffected by the application of \ac{CP}, which preserves rank order.
The sensitivity plot (\cref{fig:recall}) is necessarily diagonal for all \ac{CP}-calibrated methods and serves as a validity check rather than a discriminating metric. 
The precision plot (\cref{fig:precision}) is the most informative for comparing methods, showing that \ac{CP} substantially improves precision for all classifiers and that all \ac{ML} classifiers outperform maximum approaches.

Across all metrics, the XGBoost classifier achieves the best performance, both with and without \ac{CP} applied.
Combined with its conservative behaviour when assigning confidence to both mock data and real \ac{GW} candidates discussed in \cref{sec:model_comparison}, we choose XGBoost as our preferred classifier for the remainder of this analysis.
\subsection{Choosing the conditional confidence threshold}\label{sec:threshold}
For our analysis so far, we have used the arbitrary confidence threshold of $0.5$ to distinguish between signal and noise; we now briefly explore alternative thresholds. 

If the main goal is catalogue purity, one approach to selecting a confidence threshold would be to choose the point at which precision reaches unity, ensuring no false positives. 
However, this threshold is highly unstable, as a single misclassified event can cause the threshold to shift dramatically. 
Furthermore, this approach prioritises purity at the potential expense of signal loss, and the threshold derived from mock data may not transfer reliably to real observations if detector or pipeline characteristics differ. 
Nevertheless, we investigate what threshold this method would give. Over 100 permutations of the data, we obtain a conservative 95th percentile of $0.43$ for XGBoost, $0.6$ for \ac{LR} and $0.63$ for \ac{MLP}. 

Another approach would be to use the \ac{ROC} curve. 
The \ac{ROC} curve summarises the trade-off between \ac{TPR} and \ac{FPR} as the classification threshold is varied. 
A common heuristic is to minimise the Euclidean distance to the top-left corner of the \ac{ROC} plane~\citep{perkins2006inconsistency}, $d = \sqrt{(1-\textrm{TPR})^2+\textrm{FPR}^2}$, corresponding to simultaneously high \ac{TPR} and low \ac{FPR}. \citet{Tsukamoto:2025vuu}, for example, employ this approach. 
While this criterion provides a simple and intuitive operating point, it implicitly assumes equal costs for false positives and false negatives and does not account for class imbalance. 
Moreover, the location of the optimal point can be sensitive to statistical fluctuations in finite samples. 
This approach should therefore be used primarily as a diagnostic tool to guide interpretation, rather than as a definitive choice for setting the final detection threshold. 
Using a conservative 95th percentile over 100 permutations of the data, we obtain thresholds of $0.49$ for XGBoost, $0.44$ for \ac{LR}, and $0.39$ for \ac{MLP}. 

Taken together, these threshold estimates cluster around values of approximately $0.5$ across classifiers and selection criteria, despite considerable variability in both approaches. 
The two approaches yield notably different orderings across classifiers: the precision-based method assigns the lowest threshold to XGBoost and the highest to \ac{MLP}, while the \ac{ROC}-based method reverses this, giving XGBoost the highest and \ac{MLP} the lowest. 
This inconsistency illustrates that neither method alone provides a stable or definitive threshold. 
We have therefore chosen to continue using the confidence threshold of $0.5$ throughout the analysis, as it provides a simple, conservative, and broadly supported operating point without relying on unstable data-driven optimisation.

\subsection{Feature importance}\label{sec:shap}
Having established XGBoost as our preferred classifier, we now examine the decision-making process of all classifiers through \ac{SHAP} analysis~\citep{lundberg2017shap}, using XGBoost as the primary example.
\ac{SHAP} analysis quantifies the contribution of each input feature to individual predictions, revealing which pipeline outputs drove a given prediction by how much and in which direction. 
This allows us to move beyond black-box predictions and understand the physical reasoning behind each classification.

A natural starting point for understanding classifier decisions is to examine built-in feature importance metrics, such as coefficient magnitudes for \ac{LR} or impurity-based importance for tree models. 
However, these global metrics reflect the influence of each feature conditional on all others, which can be misleading when features are correlated, and cannot explain how individual predictions are driven by specific feature values. 

\ac{SHAP}~\cite{lundberg2017shap} addresses these limitations by providing a unified framework for feature attribution based on Shapley values from cooperative game theory. 
\ac{SHAP} values quantify the contribution of each feature to individual predictions while accounting for feature dependencies, offering detailed insight into model behaviour and increasing model transparency and interpretability. 
We compute \ac{SHAP} values using the \texttt{shap} Python package~\cite{lundberg2017shap, lundberg2020local}. 

For a given prediction $f(x)$, \ac{SHAP} decomposes the output into a baseline value $E[f(x)]$, the mean model output over the background dataset, plus additive feature contributions:
\begin{equation}
    f(x)=E[f(x)] + \sum_i \phi_i\,,
    \label{eq:shap_sum}
\end{equation}
where the magnitude of each $\phi_i \in \mathbb{R}$ indicates the strength of feature $i$'s influence, and its sign indicates the direction of effect. 
Each $\phi_i$ is computed as the weighted average marginal contribution of feature $i$ across all possible subsets of the remaining features~\citep{lundberg2017shap}. 
Global feature importance can be derived by averaging absolute \ac{SHAP} values across all predictions, while individual \ac{SHAP} values reveal how specific feature values influence particular predictions.
We note that while \ac{SHAP} values account for feature interactions in their computation, the summary plots shown do not make these interactions directly visible; dependence plots would be required to reveal such correlations. 

\begin{figure*}
    \centering
    \subfigure[\ac{MDC}]{\includegraphics[width=0.49\textwidth]{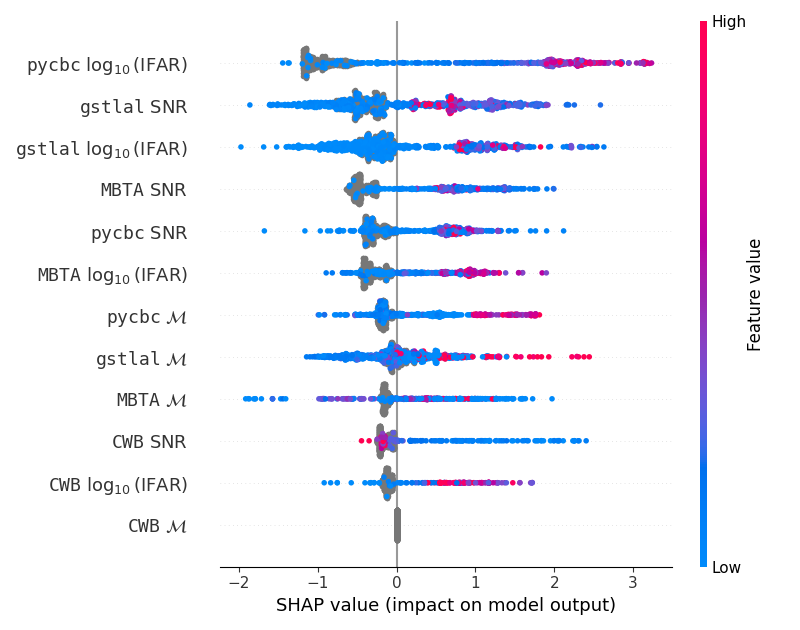} \label{fig:SHAP_xgboost_MDC12}}
    \subfigure[\gwtcTwoPointOne and \gwtcThree]{\includegraphics[width=0.49\textwidth]{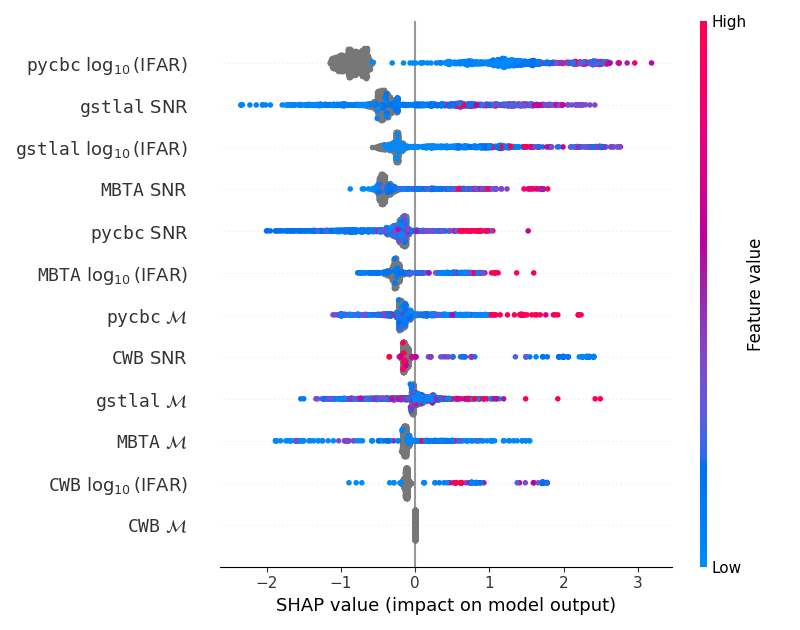}\label{fig:SHAP_xgboost_gwtc3}}
    \caption{Summary plot of the \ac{SHAP} values for the \ac{MDC}-trained XGBoost classifier on the \ac{MDC} and O3 data in \gwtcTwoPointOne and \gwtcThree, respectively. 
    The $x$-axis shows the \ac{SHAP} values: a positive value corresponds to influence toward a signal prediction, while a negative value corresponds to influence toward a noise prediction. 
    The magnitude indicates the strength of influence. 
    The features are ordered by the mean absolute \ac{SHAP} value across all test events, with the most influential feature at the top. 
    The colour scale represents the value of the features, and grey points indicate missing values. 
    Each point represents a single event, plotted at its \ac{SHAP} value for the corresponding feature.} 
    \label{fig:SHAP_xgboost}
\end{figure*}

First, we consider the global \ac{SHAP} values for the XGBoost classifier applied to the \ac{MDC} and O3 data, respectively, using the \ac{MDC} training data as the reference baseline. 
The results are shown in \cref{fig:SHAP_xgboost}, where each point represents the SHAP values for a specific feature and event in the respective dataset, coloured by the feature value. 
For example, high values of \pycbc \logIFAR influence the classifier toward a signal prediction in both datasets, whereas low or missing values (shown in grey) push the prediction toward noise. 
We note similar behaviour in most of the features, and observe that there is little difference between the two datasets. 
This similarity is expected: the \ac{MDC} is constructed to mimic realistic detection scenarios, so a well-trained classifier should respond consistently to equivalent feature values regardless of whether the data are simulated or real.

The \logIFAR features from the templated pipelines dominate the ranking, with \pycbc \logIFAR being the single most important feature in both datasets. 
High values push strongly toward a signal prediction, while missing values, corresponding to non-detections, contribute a small negative score. 
\ac{SNR} features play a secondary but consistent role, while chirp mass features rank lower, indicating that the classifier relies primarily on detection significance rather than source parameters. 
\cwb features are the least influential overall, possibly due to the role as an unmodelled search. 
The close agreement in feature rankings and \ac{SHAP} distributions between the two panels suggests that the classifier generalises well from simulated to real data.

\begin{figure}
    \centering
    \includegraphics[width=0.49\textwidth]{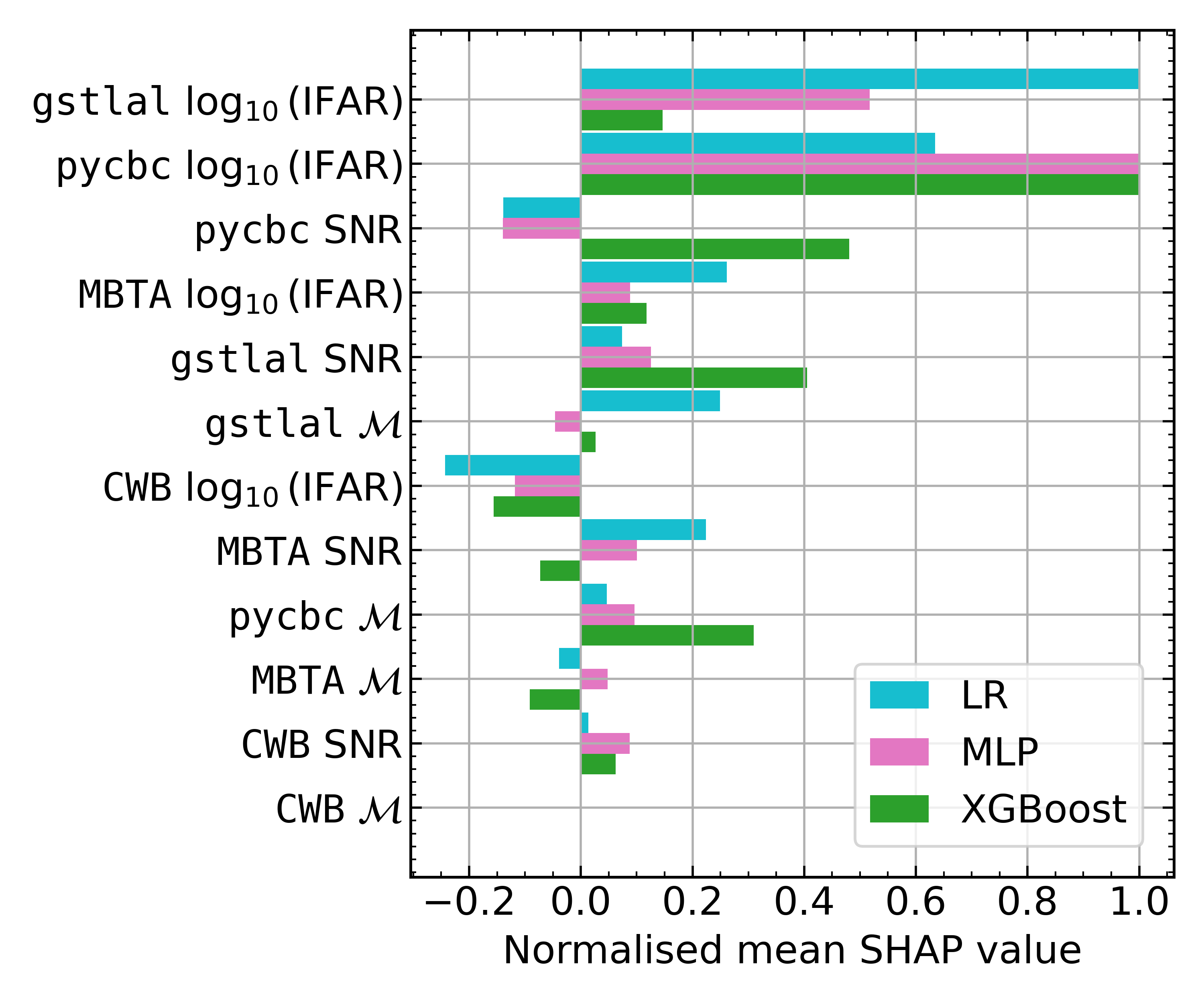}
    \caption{Comparison of mean \ac{SHAP} values for the different classifiers on the \ac{MDC} data. 
    For each classifier, the mean \ac{SHAP} values are normalised by the maximum absolute mean \ac{SHAP} value of that classifier, scaling each to the range $[-1, 1]$ to enable comparison across classifiers with different output scales.} 
    \label{fig:SHAP_classifier_comparison}
\end{figure}

To compare feature importance across classifiers, we show the mean \ac{SHAP} values for \ac{LR}, \ac{MLP}, and XGBoost in \cref{fig:SHAP_classifier_comparison}.
The plot shows the mean \ac{SHAP} value for each feature, normalised by each classifier's maximum absolute mean \ac{SHAP} value for a clearer cross-classifier comparison, and is thus less informative than the distribution of local values shown in \cref{fig:SHAP_xgboost_MDC12} for the same data. 

This also accounts for the slight discrepancy for XGBoost: by default, the features in \cref{fig:SHAP_xgboost} are ranked by the absolute mean \ac{SHAP} value, while \cref{fig:SHAP_classifier_comparison} shows the signed mean. 
All three classifiers agree that \logIFAR features dominate over \ac{SNR} and chirp mass features, confirming that multi-pipeline detection significance is the primary driver of classification. 
\ac{SNR} features play a more prominent secondary role for XGBoost than for \ac{LR} or \ac{MLP}, suggesting that XGBoost exploits non-linear interactions between significance and signal strength. 
\cwb features consistently rank low across all classifiers, and the \cwb \logIFAR carries a negative mean \ac{SHAP} value for all classifiers.
This is driven by non-detections: \cwb produces a trigger for only around 26\% of signals in the \ac{MDC}, so the dominant value of \cwb \logIFAR is zero, which the classifier associates with noise.
A high \cwb \logIFAR would contribute positively (as shown in \cref{fig:SHAP_xgboost_MDC12}), but this is rare in the \ac{MDC}, partly because $41\%$ of injections are \ac{BNS} signals that \cwb is not well suited to detect.

\begin{figure}
    \centering
    \includegraphics[width=0.49\textwidth]{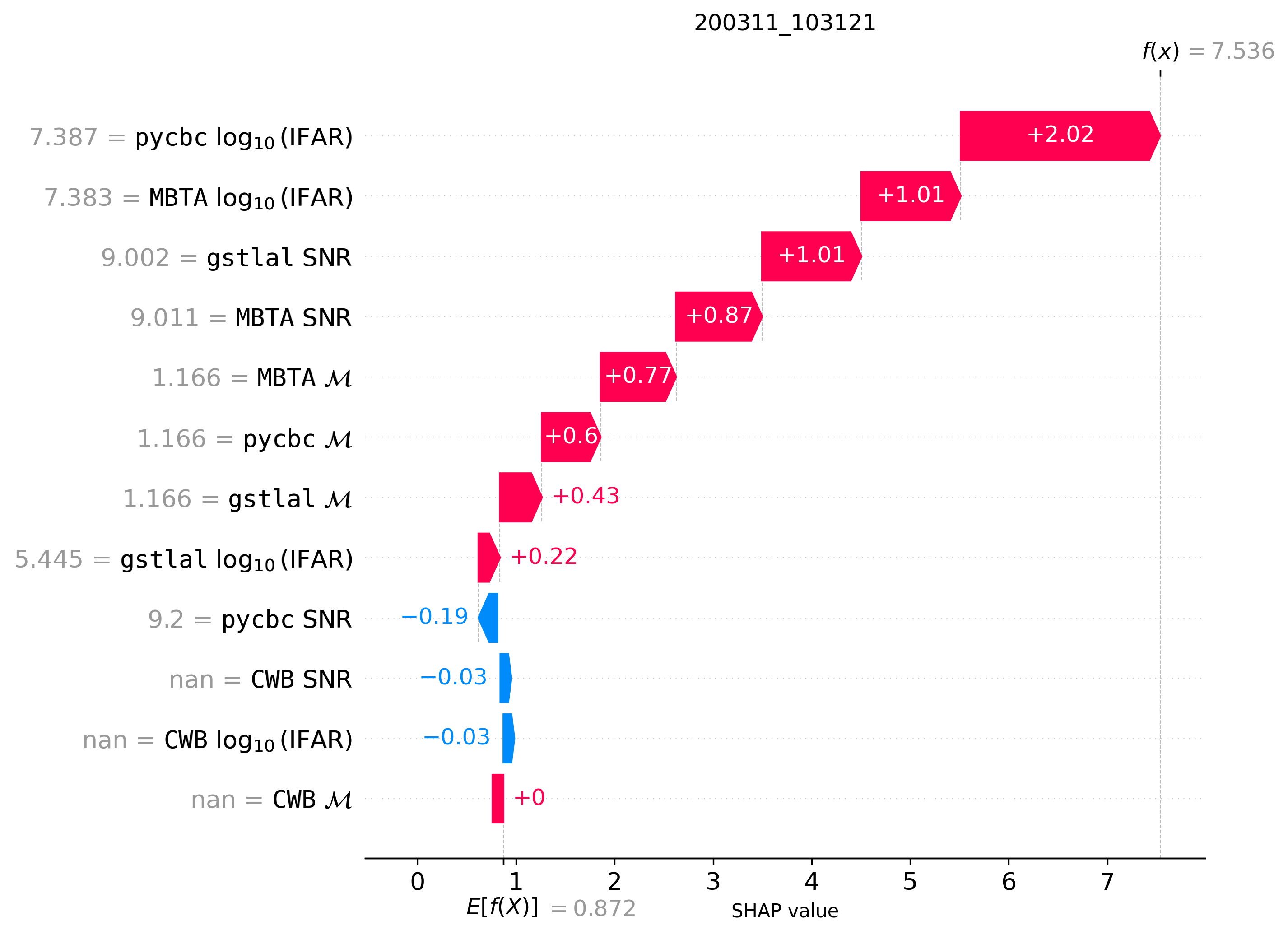}
    \caption{\ac{SHAP} waterfall plot for GW200311\_103121, relative to other O3 data from \gwtcTwoPointOne and \gwtcThree, using XGBoost trained on the \ac{MDC}. 
    The \ac{SHAP} values for each feature are represented by the arrows, where length and colour/direction indicate the magnitude and direction of the feature's influence. 
    The model output $f(x)=7.536$ and expected value $E[f(x)]=0.872$, computed on the raw classifier output log-odds, correspond to probabilities $p\approx1$ and $p\approx0.71$, respectively.}
    \label{fig:waterfall_200311}
\end{figure}

We can also examine feature importance for individual events to understand the reasoning behind a specific prediction.
Focusing on the up-ranked \ac{BNS} candidate GW200311\_103121, the \ac{SHAP} values, computed using \gwtcThree as the reference dataset, are shown in the waterfall plot in \cref{fig:waterfall_200311} for the XGBoost classifier. 
In the waterfall plot, each feature is shown on the $y$-axis alongside its value, with arrows indicating how much it shifts the prediction from the baseline $E[f(x)]$ (the average model output over the \gwtcThree data) toward the final prediction $f(x)$, where the arrow length and colour indicate magnitude and sign respectively. 

The \ac{SHAP} values are computed on the raw \ac{ML} model output, which for the XGBoost classifier corresponds to the log-odds $f(x)=\log\frac{p}{1-p}$. 
While these log-odds can be converted to probabilities $p$, doing so means that the sum of the \ac{SHAP} values no longer exactly equals the difference from the baseline, so the resulting plot provides only an approximate decomposition. 
In \cref{fig:waterfall_200311}, we therefore show the plot in log-odds units. 

The top features in \cref{fig:waterfall_200311} are the \logIFAR values from \pycbc and \mbta, with values just below the threshold of \logIFAR $=7.5$ in units of seconds (corresponding to a \ac{FAR} of 1 per year). 
An \ac{SNR} of $\approx9$ from \gstlal and \mbta also influences the prediction toward being a signal, while a similar \ac{SNR} from \pycbc appears more noise-like to the classifier.
A non-detection from \cwb leads to the \cwb features not affecting the prediction much in any direction. 
The \gstlal \logIFAR value of $5.4$ (sub-threshold) contributes only marginally, with a small positive score.
Additionally, the chirp mass features from all three pipelines, each reporting $\mathcal{M} = 1.166\,M_\odot$, contribute positively to the signal prediction. 
This low chirp mass is consistent with a \ac{BNS} system, and the classifier has evidently learned to associate such values with astrophysical signals.

\begin{figure*}
    \centering
    \subfigure[GW191222\_033537]{\includegraphics[width=0.49\textwidth]{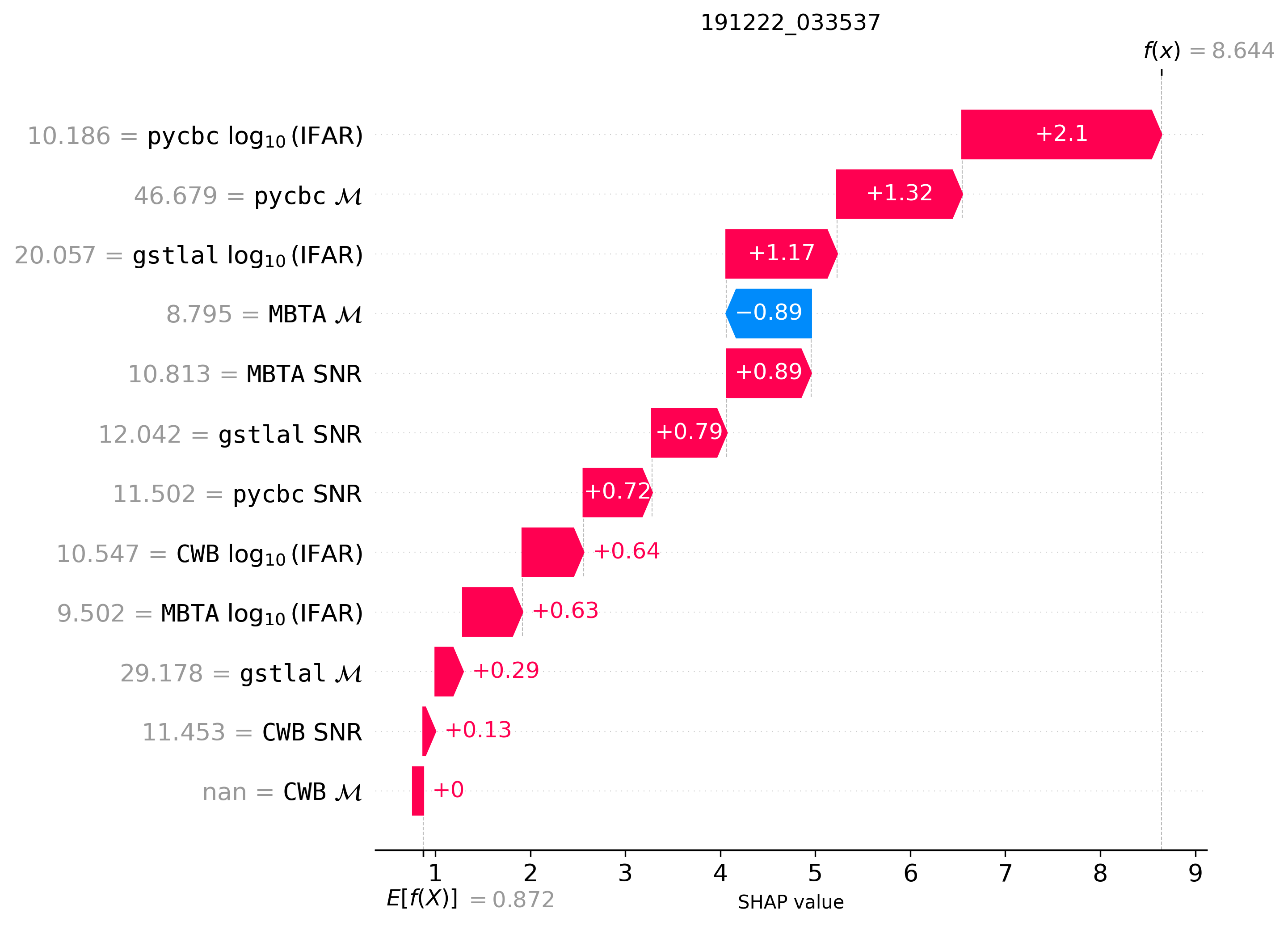} \label{fig:waterfall_191222}}
    \subfigure[GW200324\_081850]{\includegraphics[width=0.49\textwidth]{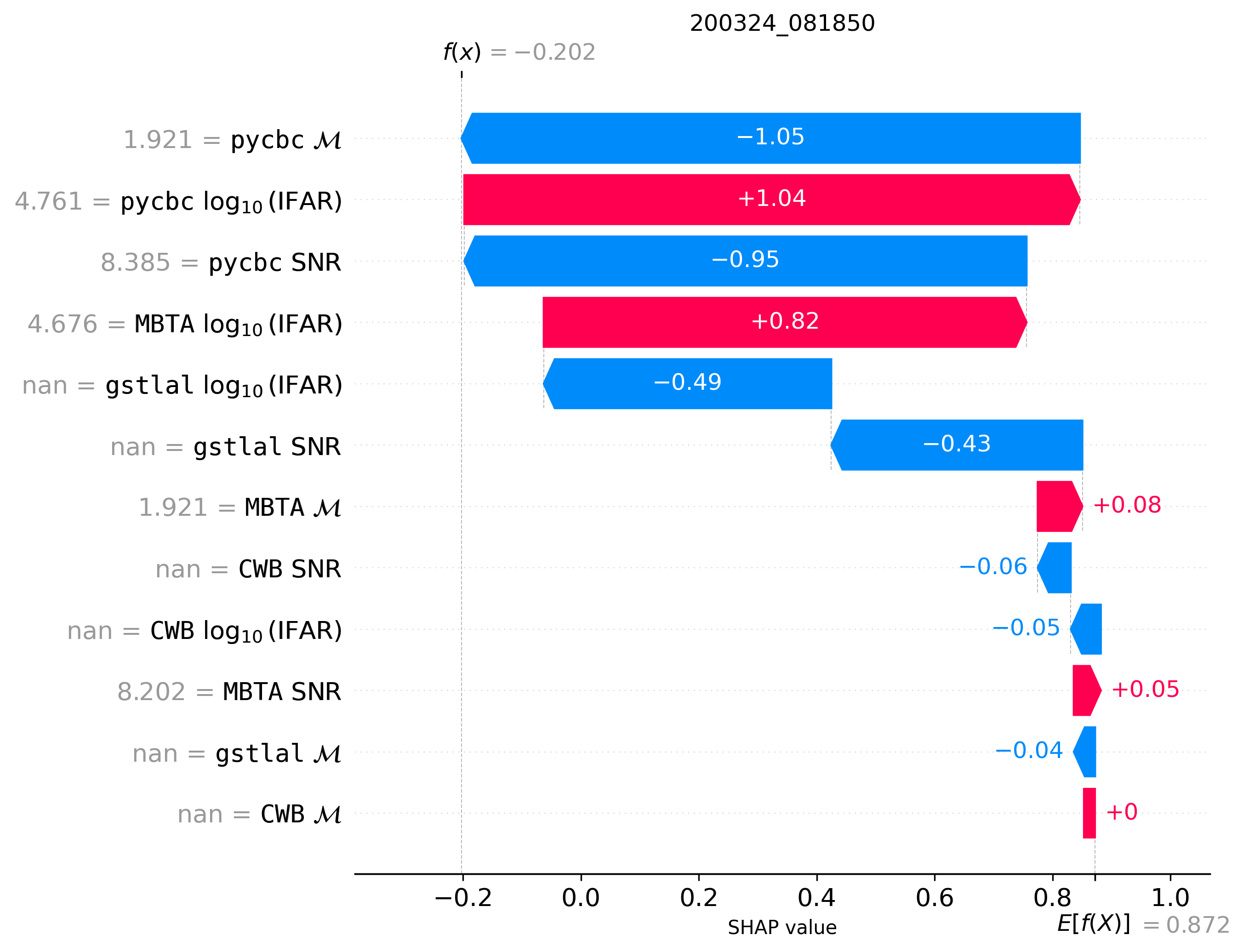} \label{fig:waterfall_200324}}
    \caption{\ac{SHAP} waterfall plot for GW191222\_033537 (signal) and GW200324\_081850 (noise), relative to other O3 data from \gwtcTwoPointOne and \gwtcThree, using XGBoost trained on the \ac{MDC}. 
    The \ac{SHAP} values for each feature are represented by the arrows, where length and colour/direction indicate the magnitude and direction of the feature's influence.} 
    \label{fig:waterfall_191222_200324}
\end{figure*}

For comparison, we also show waterfall plots for a confirmed 4-pipeline signal, GW191222\_033537 from the \gwtcThree catalogue ($\pastro=1$, confidence $0.97$, \cref{fig:waterfall_191222}), and a 2-pipeline noise candidate, GW200324\_081850 ($\pastro=0.001$, confidence $0.13$, \cref{fig:waterfall_200324}).

The \ac{SHAP} decomposition for GW200311\_103121 shares the same qualitative structure as GW191222\_033537, dominated by large positive contributions from multi-pipeline \logIFAR features, supporting its classification as a genuine signal. 
However, the overall prediction is less decisive, reflecting lower \logIFAR values and the absence of a \cwb detection. 
In contrast, the noise candidate GW200324\_081850 shows positive contributions from \pycbc and \mbta \logIFAR, offset by stronger negative contributions from the chirp mass and \ac{SNR} features and the absence of \gstlal, resulting in a clear noise classification.

Overall, the \ac{SHAP} analysis reveals that XGBoost's classification decisions are driven primarily by multi-pipeline detection significance, with source parameters playing a secondary role. 
The consistent feature rankings across classifiers and between simulated and real data demonstrate that the framework is robust and physically interpretable. 
The waterfall analysis for GW200311\_103121 further shows that its high-confidence classification is supported by both strong multi-pipeline \logIFAR values and a chirp mass consistent with a \ac{BNS} system, providing a physically coherent basis for its up-ranking.

\vspace*{\baselineskip}
\section{Tables of conditional confidence scores for candidates from O3, O4a and O4b}\label{app:gwtc_table}
In \cref{tab:gwtc3_up}, \cref{tab:gwtc4_up}, and \cref{tab:gwtc5_up} we tabulate the up-ranked candidates from O3, O4a and O4b studied in this work, enabling a comparison of the \pastro and conditional confidence where they disagree. 

\begin{table*} 
    \centering
    \caption{Up-ranked O3a and O3b candidates from \gwtcTwoPointOne and \gwtcThree. The reported chirp mass is taken from the pipeline with the maximum \pastro value. For \cwb, the chirp mass is not reported, as we do not use this in our analysis.}
    \begin{tabular}{l|l|l|c|c|c|c|c}
        \toprule
        Candidate & Data & Pipeline & $\mathcal{M}$ & \pastro & Confidence LR & Confidence XGBoost & Confidence MLP \\ 
        \midrule
        GW190424\_153213 & O3a & \mbta & 3.5 & 0.076 & 0.600 & 0.510 & 0.480 \\
GW190426\_152155 & O3a & \gstlal & 2.6 & 0.145 & 0.610 & 0.620 & 0.630 \\
GW190531\_023648 & O3a & \gstlal & 2.0 & 0.277 & 0.630 & 0.540 & 0.650 \\
GW190629\_201728 & O3a & \mbta & 3.9 & 0.006 & 0.520 & 0.320 & 0.390 \\
GW190704\_104834 & O3a & \pycbc & 4.6 & 0.175 & 0.560 & 0.590 & 0.480 \\
GW190707\_083226 & O3a & \cwb & - & 0.159 & 0.400 & 0.400 & 0.580 \\
GW190906\_064946 & O3a & \pycbc & 4.4 & 0.099 & 0.400 & 0.380 & 0.500 \\
GW190909\_114149 & O3a & \pycbc & 8.0 & 0.150 & 0.560 & 0.470 & 0.420 \\
GW190910\_012619 & O3a & \pycbc & 9.7 & 0.074 & 0.520 & 0.270 & 0.390 \\
GW190913\_153716 & O3a & \gstlal & 24.1 & 0.004 & 0.130 & 0.270 & 0.630 \\
GW191114\_003412 & O3b & \gstlal & 12.4 & 0.006 & 0.110 & 0.210 & 0.630 \\
GW191126\_115259 & O3b & \pycbc & 11.4 & 0.389 & 0.590 & 0.320 & 0.430 \\
GW191204\_110529 & O3b & \gstlal & 28.5 & 0.070 & 0.480 & 0.830 & 0.380 \\
GW191222\_135709 & O3b & \pycbc & 6.6 & 0.289 & 0.420 & 0.380 & 0.510 \\
GW200105\_162426 & O3b & \gstlal & 3.6 & 0.362 & 0.200 & 0.370 & 0.640 \\
GW200201\_203549 & O3b & \gstlal & 2.1 & 0.116 & 0.500 & 0.420 & 0.540 \\
GW200202\_140019 & O3b & \pycbc & 5.2 & 0.066 & 0.400 & 0.460 & 0.500 \\
GW200221\_142805 & O3b & \gstlal & 40.6 & 0.291 & 0.200 & 0.310 & 0.510 \\
GW200229\_130505 & O3b & \gstlal & 24.1 & 0.001 & 0.230 & 0.270 & 0.880 \\
GW200308\_173609 & O3b & \mbta & 45.2 & 0.236 & 0.460 & 0.640 & 0.440 \\
GW200311\_103121 & O3b & \pycbc & 1.2 & 0.194 & 0.660 & 0.820 & 0.610 \\
GW200316\_235948 & O3b & \pycbc & 5.0 & 0.138 & 0.520 & 0.300 & 0.430 \\
        \bottomrule
    \end{tabular}
    \label{tab:gwtc3_up}
\end{table*}

\begin{table*} 
    \centering
    \caption{Up-ranked O4a candidates from \gwtcFour. The reported chirp mass is taken from the pipeline with the maximum \pastro value. For \cwb, the chirp mass is not reported, as we do not use this in our analysis.}
    \begin{tabular}{l|l|l|c|c|c|c|c}
        \toprule
        Candidate & Data & Pipeline & $\mathcal{M}$ & \pastro & Confidence LR & Confidence XGBoost & Confidence MLP \\ 
        \midrule
        GW230527\_071259 & O4a & \pycbc & 51.2 & 0.204 & 0.450 & 0.350 & 0.520 \\
GW230611\_165753 & O4a & \gstlal & 54.1 & 0.078 & 0.320 & 0.170 & 0.890 \\
GW230622\_160042 & O4a & \mbta & 49.9 & 0.461 & 0.590 & 0.320 & 0.630 \\
GW230625\_063417 & O4a & \pycbc & 38.7 & 0.044 & 0.520 & 0.320 & 0.420 \\
GW230625\_211555 & O4a & \pycbc & 38.7 & 0.225 & 0.560 & 0.300 & 0.460 \\
GW230627\_114138 & O4a & \pycbc & 24.7 & 0.360 & 0.730 & 0.500 & 0.530 \\
GW230719\_070908 & O4a & \gstlal & 55.1 & 0.086 & 0.450 & 0.380 & 0.620 \\
GW230723\_125531 & O4a & \pycbc & 62.6 & 0.113 & 0.450 & 0.380 & 0.580 \\
GW230812\_001904 & O4a & \gstlal & 57.5 & 0.097 & 0.500 & 0.270 & 0.440 \\
GW230813\_144914 & O4a & \gstlal & 34.5 & 0.250 & 0.610 & 0.480 & 0.550 \\
GW230824\_030342 & O4a & \pycbc & 16.3 & 0.229 & 0.530 & 0.610 & 0.360 \\
GW230824\_134554 & O4a & \gstlal & 19.4 & 0.135 & 0.520 & 0.500 & 0.340 \\
GW230826\_031533 & O4a & \pycbc & 45.7 & 0.138 & 0.440 & 0.350 & 0.510 \\
GW230827\_143829 & O4a & \gstlal & 21.5 & 0.098 & 0.520 & 0.520 & 0.370 \\
GW230830\_043238 & O4a & \cwb & - & 0.118 & 0.610 & 0.630 & 0.620 \\
GW230830\_055405 & O4a & \pycbc & 38.7 & 0.040 & 0.530 & 0.240 & 0.410 \\
GW230906\_100145 & O4a & \pycbc & 45.7 & 0.218 & 0.600 & 0.550 & 0.570 \\
GW230907\_064153 & O4a & \cwb & - & 0.134 & 0.610 & 0.400 & 0.610 \\
GW230910\_021701 & O4a & \pycbc & 78.7 & 0.495 & 0.380 & 0.410 & 0.520 \\
GW230911\_134712 & O4a & \gstlal & 44.7 & 0.381 & 0.460 & 0.470 & 0.520 \\
GW230915\_133509 & O4a & \pycbc & 51.2 & 0.156 & 0.590 & 0.360 & 0.490 \\
GW230918\_050515 & O4a & \pycbc & 32.0 & 0.038 & 0.520 & 0.340 & 0.470 \\
GW230925\_090300 & O4a & \gstlal & 80.4 & 0.052 & 0.440 & 0.410 & 0.590 \\
GW231003\_101642 & O4a & \gstlal & 57.5 & 0.049 & 0.430 & 0.280 & 0.560 \\
GW231008\_011953 & O4a & \pycbc & 62.6 & 0.099 & 0.450 & 0.350 & 0.580 \\
GW231009\_081300 & O4a & \gstlal & 29.5 & 0.429 & 0.570 & 0.480 & 0.420 \\
GW231029\_003952 & O4a & \pycbc & 62.6 & 0.278 & 0.460 & 0.410 & 0.580 \\
GW231109\_175405 & O4a & \gstlal & 80.2 & 0.231 & 0.470 & 0.380 & 0.650 \\
GW231110\_122926 & O4a & \gstlal & 48.6 & 0.303 & 0.470 & 0.440 & 0.570 \\
GW231110\_171731 & O4a & \gstlal & 112.3 & 0.431 & 0.450 & 0.500 & 0.500 \\
GW231128\_143157 & O4a & \cwb & - & 0.371 & 0.610 & 0.420 & 0.440 \\
GW231222\_175548 & O4a & \pycbc & 62.6 & 0.034 & 0.430 & 0.280 & 0.550 \\
GW231222\_234123 & O4a & \pycbc & 17.0 & 0.371 & 0.570 & 0.580 & 0.430 \\
GW231229\_133732 & O4a & \gstlal & 39.4 & 0.053 & 0.700 & 0.360 & 0.640 \\
GW240101\_061317 & O4a & \gstlal & 42.5 & 0.399 & 0.520 & 0.530 & 0.420 \\
GW240114\_225929 & O4a & \mbta & 70.2 & 0.361 & 0.590 & 0.400 & 0.610 \\
        \bottomrule
    \end{tabular}
    \label{tab:gwtc4_up}
\end{table*}

\begin{table*} 
    \centering
    \caption{Up-ranked O4b candidates from \gwtcFive. The reported chirp mass is taken from the pipeline with the maximum \pastro value. For \cwb, the chirp mass is not reported, as we do not use this in our analysis.}
    \begin{tabular}{l|l|l|c|c|c|c|c}
        \toprule
        Candidate & Data & Pipeline & $\mathcal{M}$ & \pastro & Confidence LR & Confidence XGBoost & Confidence MLP \\ 
        \midrule
        GW240420\_210120 & O4b & \pycbc & 27.2 & 0.075 & 0.530 & 0.470 & 0.390 \\
GW240422\_224206 & O4b & \pycbc & 38.7 & 0.056 & 0.520 & 0.350 & 0.450 \\
GW240426\_100722 & O4b & \mbta & 26.4 & 0.471 & 0.670 & 0.610 & 0.650 \\
GW240426\_151803 & O4b & \cwb & - & 0.410 & 0.780 & 0.540 & 0.710 \\
GW240430\_111634 & O4b & \cwb & - & 0.227 & 0.460 & 0.500 & 0.530 \\
GW240503\_233347 & O4b & \pycbc & 29.7 & 0.043 & 0.530 & 0.420 & 0.380 \\
GW240506\_030828 & O4b & \pycbc & 62.6 & 0.125 & 0.470 & 0.460 & 0.610 \\
GW240521\_071417 & O4b & \mbta & 14.1 & 0.126 & 0.500 & 0.210 & 0.460 \\
GW240601\_062707 & O4b & \pycbc & 45.7 & 0.304 & 0.460 & 0.390 & 0.530 \\
GW240602\_001317 & O4b & \cwb & - & 0.291 & 0.490 & 0.410 & 0.620 \\
GW240609\_041100 & O4b & \pycbc & 78.7 & 0.114 & 0.510 & 0.440 & 0.660 \\
GW240610\_154921 & O4b & \gstlal & 52.9 & 0.108 & 0.420 & 0.310 & 0.510 \\
GW240615\_023849 & O4b & \gstlal & 11.4 & 0.112 & 0.520 & 0.260 & 0.370 \\
GW240616\_144806 & O4b & \pycbc & 38.7 & 0.077 & 0.560 & 0.420 & 0.460 \\
GW240623\_100729 & O4b & \pycbc & 62.6 & 0.035 & 0.430 & 0.220 & 0.520 \\
GW240627\_210521 & O4b & \gstlal & 83.5 & 0.470 & 0.470 & 0.450 & 0.670 \\
GW240629\_201552 & O4b & \pycbc & 38.7 & 0.230 & 0.450 & 0.520 & 0.490 \\
GW240708\_081052 & O4b & \mbta & 20.6 & 0.054 & 0.510 & 0.420 & 0.440 \\
GW240716\_012708 & O4b & \pycbc & 62.6 & 0.030 & 0.570 & 0.440 & 0.610 \\
GW240828\_064943 & O4b & \pycbc & 25.1 & 0.148 & 0.400 & 0.500 & 0.410 \\
GW240828\_085522 & O4b & \cwb & - & 0.311 & 0.600 & 0.700 & 0.550 \\
GW240829\_213343 & O4b & \pycbc & 62.6 & 0.317 & 0.480 & 0.480 & 0.610 \\
GW240930\_141549 & O4b & \pycbc & 45.7 & 0.074 & 0.450 & 0.470 & 0.500 \\
GW241001\_113345 & O4b & \gstlal & 35.4 & 0.081 & 0.540 & 0.420 & 0.460 \\
GW241102\_010049 & O4b & \gstlal & 62.5 & 0.143 & 0.440 & 0.450 & 0.560 \\
GW241102\_073135 & O4b & \gstlal & 77.0 & 0.070 & 0.460 & 0.340 & 0.640 \\
GW241102\_084419 & O4b & \mbta & 11.3 & 0.050 & 0.500 & 0.390 & 0.310 \\
GW241109\_114629 & O4b & \pycbc & 62.6 & 0.010 & 0.430 & 0.280 & 0.500 \\
GW241129\_122622 & O4b & \mbta & 59.1 & 0.396 & 0.770 & 0.550 & 0.710 \\
GW241130\_134204 & O4b & \gstlal & 47.6 & 0.018 & 0.550 & 0.490 & 0.590 \\
GW241202\_233002 & O4b & \pycbc & 78.7 & 0.411 & 0.520 & 0.390 & 0.690 \\
GW241219\_151137 & O4b & \mbta & 27.1 & 0.397 & 0.610 & 0.640 & 0.480 \\
GW250119\_112442 & O4b & \mbta & 27.1 & 0.169 & 0.450 & 0.380 & 0.550 \\
        \bottomrule
    \end{tabular}
    \label{tab:gwtc5_up}
\end{table*}

In \cref{tab:gwtc3_down}, \cref{tab:gwtc4_down}, and \cref{tab:gwtc5_down} we tabulate the down-ranked candidates from O3, O4a and O4b studied in this work, enabling a comparison of the \pastro and conditional confidence where they disagree. 

\begin{table*} 
    \centering
    \caption{Down-ranked O3a and O3b candidates from \gwtcTwoPointOne and \gwtcThree. The reported chirp mass is taken from the pipeline with the maximum \pastro value. For \cwb, the chirp mass is not reported, as we do not use this in our analysis.}
    \begin{tabular}{l|l|l|c|c|c|c|c}
        \toprule
        Candidate & Data & Pipeline & $\mathcal{M}$ & \pastro & Confidence LR & Confidence XGBoost & Confidence MLP \\ 
        \midrule
        GW190425\_081805 & O3a & \gstlal & 1.5 & 0.686 & 0.220 & 0.380 & 0.660 \\
GW190527\_092055 & O3a & \gstlal & 32.9 & 0.831 & 0.220 & 0.340 & 0.370 \\
GW190620\_030421 & O3a & \gstlal & 24.0 & 0.986 & 0.250 & 0.320 & 0.570 \\
GW190630\_185205 & O3a & \gstlal & 33.6 & 1.000 & 0.420 & 0.280 & 0.840 \\
GW190708\_232457 & O3a & \gstlal & 15.4 & 1.000 & 0.280 & 0.320 & 0.700 \\
GW190725\_174728 & O3a & \pycbc & 9.6 & 0.958 & 0.560 & 0.300 & 0.500 \\
GW190731\_140936 & O3a & \mbta & 29.1 & 0.778 & 0.410 & 0.440 & 0.500 \\
GW190804\_083543 & O3a & \cwb & - & 0.989 & 0.370 & 0.120 & 0.260 \\
GW190910\_112807 & O3a & \gstlal & 38.7 & 0.996 & 0.280 & 0.320 & 0.590 \\
GW190916\_200658 & O3a & \mbta & 57.6 & 0.624 & 0.310 & 0.430 & 0.270 \\
GW190917\_114630 & O3a & \gstlal & 4.2 & 0.743 & 0.160 & 0.380 & 0.470 \\
GW190926\_050336 & O3a & \gstlal & 11.1 & 0.511 & 0.160 & 0.290 & 0.390 \\
GW190930\_234652 & O3a & \cwb & - & 0.646 & 0.310 & 0.310 & 0.280 \\
GW191103\_012549 & O3b & \pycbc & 10.1 & 0.773 & 0.490 & 0.240 & 0.400 \\
GW191105\_143521 & O3b & \pycbc & 9.4 & 0.998 & 0.770 & 0.480 & 0.660 \\
GW191113\_071753 & O3b & \mbta & 13.0 & 0.683 & 0.380 & 0.140 & 0.230 \\
GW191219\_163120 & O3b & \pycbc & 4.7 & 0.822 & 0.320 & 0.300 & 0.360 \\
GW200112\_155838 & O3b & \gstlal & 28.7 & 1.000 & 0.340 & 0.340 & 0.760 \\
GW200202\_154313 & O3b & \gstlal & 8.3 & 1.000 & 0.370 & 0.320 & 0.810 \\
GW200210\_092254 & O3b & \pycbc & 7.6 & 0.530 & 0.450 & 0.380 & 0.600 \\
GW200214\_224526 & O3b & \cwb & - & 0.909 & 0.350 & 0.120 & 0.230 \\
GW200216\_220804 & O3b & \gstlal & 13.6 & 0.773 & 0.510 & 0.390 & 0.490 \\
GW200220\_124850 & O3b & \mbta & 44.3 & 0.826 & 0.310 & 0.380 & 0.210 \\
GW200302\_015811 & O3b & \gstlal & 33.4 & 0.906 & 0.230 & 0.320 & 0.450 \\
GW200306\_093714 & O3b & \mbta & 27.5 & 0.812 & 0.370 & 0.210 & 0.230 \\
GW200322\_091133 & O3b & \mbta & 24.5 & 0.615 & 0.390 & 0.350 & 0.290 \\
        \bottomrule
    \end{tabular}
    \label{tab:gwtc3_down}
\end{table*}

\begin{table*} 
    \centering
    \caption{Down-ranked O4a candidates from \gwtcFour. The reported chirp mass is taken from the pipeline with the maximum \pastro value. For \cwb, the chirp mass is not reported, as we do not use this in our analysis.}
    \begin{tabular}{l|l|l|c|c|c|c|c}
        \toprule
        Candidate & Data & Pipeline & $\mathcal{M}$ & \pastro & Confidence LR & Confidence XGBoost & Confidence MLP \\ 
        \midrule
        GW230517\_044807 & O4a & \gstlal & 133.3 & 0.620 & 0.260 & 0.350 & 0.330 \\
GW230528\_145129 & O4a & \gstlal & 122.0 & 0.502 & 0.250 & 0.380 & 0.320 \\
GW230531\_141100 & O4a & \pycbc & 51.2 & 0.723 & 0.380 & 0.360 & 0.440 \\
GW230603\_174756 & O4a & \pycbc & 27.2 & 0.789 & 0.470 & 0.340 & 0.400 \\
GW230606\_024545 & O4a & \pycbc & 38.7 & 0.877 & 0.480 & 0.350 & 0.510 \\
GW230606\_065320 & O4a & \gstlal & 141.2 & 0.636 & 0.270 & 0.310 & 0.330 \\
GW230610\_061439 & O4a & \gstlal & 218.5 & 0.529 & 0.320 & 0.340 & 0.440 \\
GW230615\_160825 & O4a & \gstlal & 150.4 & 0.758 & 0.440 & 0.430 & 0.540 \\
GW230618\_102550 & O4a & \pycbc & 78.7 & 0.597 & 0.510 & 0.430 & 0.660 \\
GW230624\_214944 & O4a & \pycbc & 14.9 & 0.737 & 0.440 & 0.590 & 0.500 \\
GW230630\_070659 & O4a & \gstlal & 251.7 & 0.958 & 0.370 & 0.350 & 0.580 \\
GW230702\_162025 & O4a & \mbta & 18.9 & 0.603 & 0.410 & 0.410 & 0.390 \\
GW230706\_104333 & O4a & \pycbc & 15.4 & 0.986 & 0.540 & 0.460 & 0.660 \\
GW230717\_102139 & O4a & \pycbc & 31.8 & 0.673 & 0.620 & 0.520 & 0.470 \\
GW230721\_222634 & O4a & \mbta & 62.2 & 0.726 & 0.400 & 0.250 & 0.370 \\
GW230723\_084820 & O4a & \cwb & - & 0.645 & 0.290 & 0.320 & 0.260 \\
GW230728\_083628 & O4a & \gstlal & 136.8 & 0.891 & 0.290 & 0.310 & 0.380 \\
GW230729\_082317 & O4a & \gstlal & 10.9 & 0.983 & 0.480 & 0.430 & 0.640 \\
GW230807\_205045 & O4a & \gstlal & 93.0 & 0.816 & 0.450 & 0.410 & 0.520 \\
GW230823\_142524 & O4a & \mbta & 42.4 & 0.859 & 0.480 & 0.560 & 0.460 \\
GW230824\_135331 & O4a & \cwb & - & 0.583 & 0.270 & 0.300 & 0.230 \\
GW230830\_064744 & O4a & \gstlal & 34.5 & 0.798 & 0.170 & 0.260 & 0.290 \\
GW230831\_134621 & O4a & \pycbc & 10.5 & 0.800 & 0.630 & 0.440 & 0.520 \\
GW230902\_122814 & O4a & \mbta & 17.7 & 0.838 & 0.420 & 0.240 & 0.320 \\
GW230902\_150325 & O4a & \gstlal & 156.8 & 0.651 & 0.270 & 0.290 & 0.370 \\
GW230902\_172430 & O4a & \pycbc & 10.7 & 0.600 & 0.580 & 0.350 & 0.420 \\
GW230902\_224555 & O4a & \pycbc & 62.6 & 0.514 & 0.490 & 0.400 & 0.620 \\
GW230904\_152545 & O4a & \pycbc & 2.8 & 0.746 & 0.480 & 0.340 & 0.440 \\
GW230920\_064709 & O4a & \pycbc & 4.3 & 0.789 & 0.590 & 0.460 & 0.510 \\
GW230925\_143957 & O4a & \mbta & 18.9 & 0.550 & 0.430 & 0.200 & 0.310 \\
GW231004\_232346 & O4a & \cwb & - & 0.967 & 0.510 & 0.440 & 0.580 \\
GW231005\_144455 & O4a & \pycbc & 51.2 & 0.515 & 0.360 & 0.360 & 0.420 \\
GW231007\_134720 & O4a & \gstlal & 32.8 & 0.626 & 0.150 & 0.230 & 0.240 \\
GW231013\_135504 & O4a & \pycbc & 4.3 & 0.530 & 0.300 & 0.260 & 0.300 \\
GW231018\_233037 & O4a & \mbta & 10.2 & 0.927 & 0.610 & 0.410 & 0.440 \\
GW231024\_023603 & O4a & \gstlal & 83.5 & 0.511 & 0.210 & 0.230 & 0.300 \\
GW231025\_142306 & O4a & \gstlal & 47.6 & 0.713 & 0.390 & 0.480 & 0.430 \\
GW231029\_111508 & O4a & \gstlal & 91.0 & 1.000 & 0.360 & 0.260 & 0.610 \\
GW231102\_052214 & O4a & \mbta & 11.0 & 0.719 & 0.430 & 0.150 & 0.310 \\
GW231102\_232433 & O4a & \mbta & 67.4 & 0.800 & 0.200 & 0.170 & 0.270 \\
GW231113\_112825 & O4a & \gstlal & 58.7 & 0.513 & 0.170 & 0.220 & 0.260 \\
GW231204\_090648 & O4a & \gstlal & 147.1 & 0.784 & 0.270 & 0.410 & 0.380 \\
GW231223\_075055 & O4a & \pycbc & 9.3 & 0.981 & 0.710 & 0.470 & 0.630 \\
GW231230\_170116 & O4a & \cwb & - & 0.965 & 0.470 & 0.410 & 0.480 \\
GW240105\_151143 & O4a & \pycbc & 6.0 & 0.745 & 0.280 & 0.360 & 0.390 \\
        \bottomrule
    \end{tabular}
    \label{tab:gwtc4_down}
\end{table*}

\begin{table*} 
    \centering\caption{Down-ranked O4b candidates from \gwtcFive. The reported chirp mass is taken from the pipeline with the maximum \pastro value. For \cwb, the chirp mass is not reported, as we do not use this in our analysis.}
    \begin{tabular}{l|l|l|c|c|c|c|c}
        \toprule
        Candidate & Data & Pipeline & $\mathcal{M}$ & \pastro & Confidence LR & Confidence XGBoost & Confidence MLP \\ 
        \midrule
        GW240407\_214946 & O4b & \pycbc & 62.6 & 0.793 & 0.380 & 0.360 & 0.480 \\
GW240408\_081753 & O4b & \mbta & 43.1 & 0.613 & 0.210 & 0.150 & 0.220 \\
GW240411\_074140 & O4b & \mbta & 48.2 & 0.678 & 0.420 & 0.350 & 0.390 \\
GW240419\_041206 & O4b & \cwb & - & 0.714 & 0.290 & 0.290 & 0.250 \\
GW240421\_052935 & O4b & \cwb & - & 0.774 & 0.470 & 0.420 & 0.510 \\
GW240426\_085802 & O4b & \mbta & 62.2 & 0.626 & 0.620 & 0.480 & 0.640 \\
GW240509\_102351 & O4b & \pycbc & 14.8 & 0.893 & 0.350 & 0.360 & 0.380 \\
GW240512\_051606 & O4b & \mbta & 6.5 & 0.697 & 0.250 & 0.250 & 0.270 \\
GW240516\_041939 & O4b & \mbta & 9.9 & 0.672 & 0.240 & 0.100 & 0.240 \\
GW240525\_201644 & O4b & \gstlal & 30.4 & 0.806 & 0.470 & 0.530 & 0.580 \\
GW240526\_093944 & O4b & \mbta & 13.0 & 0.971 & 0.280 & 0.190 & 0.300 \\
GW240527\_150907 & O4b & \gstlal & 163.5 & 0.812 & 0.300 & 0.280 & 0.400 \\
GW240613\_011503 & O4b & \gstlal & 30.7 & 0.738 & 0.160 & 0.250 & 0.270 \\
GW240619\_212357 & O4b & \gstlal & 39.4 & 0.508 & 0.380 & 0.530 & 0.420 \\
GW240621\_103457 & O4b & \gstlal & 58.5 & 0.505 & 0.560 & 0.410 & 0.530 \\
GW240625\_073331 & O4b & \mbta & 25.3 & 0.641 & 0.250 & 0.300 & 0.230 \\
GW240630\_212937 & O4b & \pycbc & 32.0 & 0.750 & 0.450 & 0.350 & 0.460 \\
GW240701\_011640 & O4b & \pycbc & 26.5 & 0.514 & 0.330 & 0.360 & 0.350 \\
GW240701\_013826 & O4b & \mbta & 52.6 & 0.642 & 0.230 & 0.350 & 0.300 \\
GW240701\_203807 & O4b & \gstlal & 87.3 & 0.547 & 0.620 & 0.410 & 0.750 \\
GW240703\_191355 & O4b & \gstlal & 36.3 & 1.000 & 0.700 & 0.470 & 0.820 \\
GW240807\_214559 & O4b & \pycbc & 10.0 & 0.660 & 0.400 & 0.310 & 0.440 \\
GW240813\_034548 & O4b & \pycbc & 10.3 & 0.638 & 0.400 & 0.340 & 0.430 \\
GW240814\_145043 & O4b & \mbta & 10.8 & 0.743 & 0.620 & 0.440 & 0.480 \\
GW240828\_102340 & O4b & \mbta & 72.8 & 0.621 & 0.610 & 0.490 & 0.720 \\
GW240907\_121650 & O4b & \mbta & 2.7 & 0.521 & 0.240 & 0.280 & 0.280 \\
GW240908\_174149 & O4b & \mbta & 7.6 & 0.668 & 0.250 & 0.130 & 0.260 \\
GW240913\_111216 & O4b & \gstlal & 80.4 & 0.617 & 0.220 & 0.240 & 0.300 \\
GW240919\_215453 & O4b & \cwb & - & 0.701 & 0.280 & 0.270 & 0.210 \\
GW240923\_000715 & O4b & \mbta & 52.6 & 0.783 & 0.240 & 0.350 & 0.310 \\
GW240923\_110846 & O4b & \pycbc & 23.0 & 0.605 & 0.330 & 0.360 & 0.360 \\
GW240925\_000956 & O4b & \gstlal & 36.3 & 0.605 & 0.160 & 0.260 & 0.270 \\
GW240930\_123408 & O4b & \pycbc & 15.4 & 0.809 & 0.600 & 0.300 & 0.410 \\
GW240930\_234614 & O4b & \cwb & - & 0.976 & 0.340 & 0.190 & 0.250 \\
GW241018\_235402 & O4b & \pycbc & 19.0 & 0.983 & 0.360 & 0.420 & 0.400 \\
GW241106\_110932 & O4b & \gstlal & 121.9 & 0.725 & 0.450 & 0.420 & 0.500 \\
GW241109\_005349 & O4b & \cwb & - & 0.560 & 0.270 & 0.220 & 0.210 \\
GW241129\_140416 & O4b & \pycbc & 11.7 & 0.509 & 0.310 & 0.220 & 0.300 \\
GW241201\_142737 & O4b & \pycbc & 51.2 & 0.661 & 0.490 & 0.430 & 0.580 \\
GW241220\_062146 & O4b & \pycbc & 16.4 & 0.567 & 0.320 & 0.360 & 0.330 \\
GW241230\_084504 & O4b & \gstlal & 50.6 & 1.000 & 0.390 & 0.350 & 0.750 \\
GW241230\_233618 & O4b & \gstlal & 64.7 & 0.947 & 0.490 & 0.390 & 0.570 \\
GW250105\_134047 & O4b & \gstlal & 83.8 & 0.507 & 0.410 & 0.310 & 0.430 \\
GW250109\_083206 & O4b & \pycbc & 9.2 & 0.998 & 0.350 & 0.380 & 0.410 \\
GW250116\_015318 & O4b & \pycbc & 45.7 & 0.992 & 0.700 & 0.390 & 0.620 \\
GW250116\_051426 & O4b & \mbta & 62.2 & 0.630 & 0.640 & 0.410 & 0.720 \\
        \bottomrule
    \end{tabular}
    \label{tab:gwtc5_down}
\end{table*}

\section{Tables of posterior parameters for up-ranked events}\label{app:posterior_tabs}
The following tables list the median and $90\%$ credible intervals of the posterior parameter estimates for up-ranked O4 candidates in the \gwtcFour (\cref{tab:gwtc4_posteriors}) and \gwtcFive (\cref{tab:gwtc5_posteriors}) analyses. Parameter estimation was performed using \bilby with the \texttt{IMRPhenomXPHM} waveform model, following the approach of \gwtcFive~\citep{GWTC5}.

\begin{table*}
    \centering
    \caption{Median and $90\%$ credible intervals of the posteriors for up-ranked O4a candidates in \gwtcFour. 
    The parameters shows are the source-frame chirp mass $\mathcal{M}$, mass ratio $q$, the source-frame component masses $m_1$ and $m_2$, effective inspiral spin $\chi_\mathrm{eff}$, effective precession spin $\chi_\mathrm{p}$, source inclination angle $\theta_{JN}$, and luminosity distance $D_L$. }
    \begin{tabular}{l|c|c|c|c|c|c|c|c}
        \toprule
        Candidate & $\mathcal{M} \ [M_\odot]$ & $m_1\ [M_\odot]$ & $m_2\ [M_\odot]$ & $q$ & $\chi_\mathrm{eff}$ & $\chi_p$ & $\theta_{JN}$ & $d_L\ [\mathrm{Gpc}]$ \\
        \midrule
        GW230627\_114138 & $17.12^{+4.07}_{-3.03}$ & $25.46^{+11.49}_{-6.94}$ & $15.60^{+5.42}_{-4.54}$ & $0.62^{+0.33}_{-0.29}$ & $0.32^{+0.29}_{-0.32}$ & $0.51^{+0.33}_{-0.31}$ & $0.95^{+1.91}_{-0.71}$ & $4.50^{+2.92}_{-2.13}$ \\
GW230824\_030342 & $11.71^{+2.01}_{-1.52}$ & $15.61^{+4.72}_{-2.89}$ & $11.85^{+2.90}_{-2.95}$ & $0.78^{+0.20}_{-0.30}$ & $0.02^{+0.14}_{-0.13}$ & $0.41^{+0.42}_{-0.30}$ & $2.43^{+0.54}_{-2.12}$ & $2.39^{+1.76}_{-1.23}$ \\
GW230824\_134554 & $13.41^{+5.33}_{-1.79}$ & $20.53^{+13.30}_{-5.62}$ & $12.19^{+5.20}_{-3.74}$ & $0.60^{+0.35}_{-0.31}$ & $-0.00^{+0.34}_{-0.27}$ & $0.38^{+0.44}_{-0.28}$ & $1.97^{+0.91}_{-1.68}$ & $3.23^{+3.42}_{-1.45}$ \\
GW230827\_143829 & $22.95^{+7.87}_{-5.24}$ & $32.00^{+14.25}_{-8.75}$ & $22.38^{+9.28}_{-7.61}$ & $0.73^{+0.24}_{-0.34}$ & $0.22^{+0.26}_{-0.30}$ & $0.63^{+0.28}_{-0.40}$ & $1.17^{+1.60}_{-0.84}$ & $4.48^{+3.49}_{-2.34}$ \\
GW230830\_043238 & $32.44^{+7.33}_{-5.77}$ & $44.48^{+14.13}_{-10.08}$ & $31.85^{+9.69}_{-9.15}$ & $0.74^{+0.23}_{-0.29}$ & $0.03^{+0.25}_{-0.26}$ & $0.46^{+0.41}_{-0.34}$ & $1.28^{+1.56}_{-0.97}$ & $4.15^{+2.78}_{-1.99}$ \\
GW230906\_100145 & $20.96^{+6.20}_{-4.41}$ & $30.69^{+14.06}_{-9.14}$ & $19.49^{+8.32}_{-6.47}$ & $0.67^{+0.30}_{-0.33}$ & $0.41^{+0.25}_{-0.34}$ & $0.53^{+0.32}_{-0.31}$ & $0.82^{+2.02}_{-0.62}$ & $6.35^{+4.63}_{-3.07}$ \\
GW231110\_171731 & $43.17^{+17.63}_{-8.55}$ & $60.84^{+25.72}_{-15.51}$ & $42.51^{+20.86}_{-14.89}$ & $0.74^{+0.23}_{-0.35}$ & $0.12^{+0.34}_{-0.37}$ & $0.51^{+0.37}_{-0.36}$ & $2.20^{+0.72}_{-1.95}$ & $9.50^{+5.81}_{-5.54}$ \\
GW231222\_234123 & $12.29^{+4.99}_{-1.54}$ & $21.64^{+8.97}_{-6.51}$ & $9.73^{+6.88}_{-2.36}$ & $0.45^{+0.41}_{-0.18}$ & $-0.13^{+0.46}_{-0.31}$ & $0.32^{+0.38}_{-0.22}$ & $1.11^{+1.63}_{-0.80}$ & $2.44^{+4.68}_{-1.17}$ \\
GW240101\_061317 & $17.31^{+4.63}_{-3.45}$ & $25.76^{+18.51}_{-7.34}$ & $15.47^{+6.81}_{-5.72}$ & $0.63^{+0.33}_{-0.38}$ & $0.35^{+0.32}_{-0.37}$ & $0.45^{+0.36}_{-0.29}$ & $0.68^{+2.11}_{-0.50}$ & $4.93^{+5.38}_{-2.39}$ \\
        \bottomrule
    \end{tabular}
    \label{tab:gwtc4_posteriors}
\end{table*}

\begin{table*}
    \centering
    \caption{Median and $90\%$ credible intervals of the posteriors for up-ranked O4b candidates in \gwtcFive.
    The parameters shows are the source-frame chirp mass $\mathcal{M}$, mass ratio $q$, the source-frame component masses $m_1$ and $m_2$, effective inspiral spin $\chi_\mathrm{eff}$, effective precession spin $\chi_\mathrm{p}$, source inclination angle $\theta_{JN}$, and luminosity distance $D_L$. }
    \begin{tabular}{l|c|c|c|c|c|c|c|c}
        \toprule
        Candidate & $\mathcal{M} \ [M_\odot]$ & $m_1\ [M_\odot]$ & $m_2\ [M_\odot]$ & $q$ & $\chi_\mathrm{eff}$ & $\chi_p$ & $\theta_{JN}$ & $d_L\ [\mathrm{Gpc}]$ \\
        \midrule
        GW240426\_100722 & $26.48^{+10.10}_{-6.98}$ & $36.66^{+18.04}_{-11.04}$ & $25.81^{+11.89}_{-9.31}$ & $0.73^{+0.24}_{-0.35}$ & $0.08^{+0.34}_{-0.34}$ & $0.49^{+0.39}_{-0.35}$ & $1.98^{+0.92}_{-1.72}$ & $11.26^{+8.61}_{-5.54}$ \\
GW240426\_151803 & $27.48^{+10.36}_{-7.00}$ & $37.52^{+19.09}_{-10.55}$ & $26.91^{+11.86}_{-9.24}$ & $0.74^{+0.23}_{-0.34}$ & $0.03^{+0.33}_{-0.34}$ & $0.48^{+0.40}_{-0.36}$ & $1.06^{+1.78}_{-0.82}$ & $6.84^{+6.94}_{-3.55}$ \\
GW240430\_111634 & $80.83^{+23.22}_{-18.44}$ & $109.17^{+35.50}_{-27.99}$ & $80.61^{+30.04}_{-25.80}$ & $0.77^{+0.20}_{-0.32}$ & $0.15^{+0.39}_{-0.39}$ & $0.52^{+0.36}_{-0.36}$ & $1.24^{+1.64}_{-1.00}$ & $6.81^{+4.97}_{-3.39}$ \\
GW240629\_201552 & $23.01^{+5.90}_{-4.20}$ & $30.98^{+10.90}_{-6.74}$ & $22.99^{+7.38}_{-6.58}$ & $0.76^{+0.21}_{-0.31}$ & $0.07^{+0.29}_{-0.30}$ & $0.48^{+0.39}_{-0.33}$ & $2.25^{+0.68}_{-1.96}$ & $7.08^{+4.24}_{-3.25}$ \\
GW240828\_064943 & $15.68^{+16.25}_{-2.81}$ & $22.01^{+32.23}_{-5.54}$ & $15.25^{+12.40}_{-4.52}$ & $0.70^{+0.26}_{-0.35}$ & $0.24^{+0.24}_{-0.33}$ & $0.46^{+0.35}_{-0.31}$ & $2.18^{+0.73}_{-1.86}$ & $4.37^{+12.19}_{-2.08}$ \\
GW240828\_085522 & $27.91^{+7.77}_{-5.65}$ & $37.73^{+14.70}_{-8.60}$ & $27.78^{+9.39}_{-8.65}$ & $0.76^{+0.22}_{-0.32}$ & $0.07^{+0.31}_{-0.34}$ & $0.51^{+0.37}_{-0.35}$ & $0.95^{+1.85}_{-0.72}$ & $6.55^{+4.55}_{-3.19}$ \\
GW241129\_122622 & $28.28^{+10.90}_{-7.74}$ & $38.87^{+18.74}_{-11.76}$ & $27.56^{+13.20}_{-10.40}$ & $0.74^{+0.23}_{-0.36}$ & $0.04^{+0.35}_{-0.36}$ & $0.52^{+0.37}_{-0.37}$ & $1.99^{+0.91}_{-1.72}$ & $11.36^{+8.83}_{-5.99}$ \\
GW241219\_151137 & $20.41^{+5.39}_{-4.32}$ & $28.70^{+10.77}_{-6.78}$ & $19.84^{+6.72}_{-7.44}$ & $0.71^{+0.26}_{-0.35}$ & $0.19^{+0.29}_{-0.28}$ & $0.46^{+0.39}_{-0.33}$ & $0.88^{+1.95}_{-0.66}$ & $6.38^{+4.22}_{-3.20}$ \\
        \bottomrule
    \end{tabular}
    \label{tab:gwtc5_posteriors}
\end{table*}

\end{document}